\documentclass[aps,prx,twocolumn,longbibliography]{revtex4-2}
\usepackage{graphicx}
\usepackage{amsfonts}
\usepackage{amssymb}
\usepackage{amsthm}
\usepackage{mathtools}
\usepackage{physics}
\usepackage{thmtools}
\usepackage{booktabs}
\usepackage{subcaption}
\usepackage{siunitx}
\usepackage{xcolor}
\usepackage{placeins}
\usepackage{afterpage}
\usepackage{pgf}

\usepackage[unicode=true,
 bookmarks=false,
 breaklinks=false,pdfborder={0 0 1},backref=false,colorlinks=false]
{hyperref}
\hypersetup{
 urlcolor=blue,linkcolor=black,citecolor=black
}

\newcommand{\calH}[0]{\mathcal{H}}

\begin{document}

\title{Practical advantage beyond the quadratic speedup limit with fully-quantum walks}%

\author{Massimiliano Incudini}%
\email{mincudin@sissa.it}%
\affiliation{International School for Advanced Studies (SISSA), via Bonomea 265, 34136 Trieste, Italy}%

\author{Guglielmo Mazzola}%
\email{gmazzola@sissa.it}%
\affiliation{International School for Advanced Studies (SISSA), via Bonomea 265, 34136 Trieste, Italy}%

\date{\today}

\begin{abstract}
We introduce a new class of fully-quantum Metropolis walks in which both the proposal and acceptance steps are intrinsically quantum. Unlike standard quantum walks obtained by quantizing classically efficient Markov chains, our algorithm employs Hamiltonian simulation as a quantum-native proposal mechanism,  enlarging the class of quantum walks beyond classical counterparts. We target the problem of sampling from the low-temperature Gibbs distribution of classical dense Ising models, within a fixed error in total variation distance. This approach achieves about a cubic polynomial asymptotic advantage over previous quantum-walks, resulting in a total sixth-degree polynomial queries speedup compared to the best classical walk. This shows that speedups beyond the widely assumed quadratic limit are possible within the quantum walk formalism. We perform a complete fault-tolerant compilation of all algorithmic primitives and benchmark against CPU, GPU, and FPGA implementations of the best classical Markov chain. Under identical hardware assumptions, the resulting advantage runtime crossover is reduced from approximately $10^3$ years for conventional quantum walks to less than one day. These results identify fully-quantum Markov chains as a promising route toward practical quantum advantage.
\end{abstract}

\maketitle

\section{Introduction}
\label{sec:introduction}

Sampling is a fundamental computational task with widespread applications across science and technology, ranging from physics, biology, finance, machine learning, optimization~\cite{kirkpatrick1983optimization,kirkpatrick1983optimization,Lovasz2012,pavlopoulos2011using, geman1984stochastic, mantegna1999hierarchical}. As these applications continue to grow in scale and complexity, the development of efficient graph sampling algorithms~\cite{hastings:1970,leskovec2006sampling} has become increasingly important for enabling large-scale data analysis, simulation, and inference~\cite{hamilton2017representation, libbrecht2015machine, peng2020graph, lecun_deep_2015, HuLau2013, Xiaetal2021}.

Digital quantum computers have been proposed as powerful sampling devices, with the potential to accelerate a wide range of computational tasks beyond the reach of classical algorithms~\cite{szegedy:2004,richter:2007,wocjan:2008,somma:2008,wild:2021prl}. A central example is provided by Szegedy-type quantum walks, which can achieve a quadratic speed-up over their parent classical Markov chains.

Whether a quadratic improvement is sufficient to yield a practical runtime advantage has  been questioned in the context of fault-tolerant quantum computing, where the overhead associated with error correction can overshadow the scaling advantage even at large problem sizes~\cite{babbush:2021,Hoefler2023}.

In this context, this concern is supported by the recent resource assessment of Lemieux \textit{et al.}~\cite{lemieux2020efficient}, which analyzes a local-flips quantum walk algorithm, on sparse graphs and concludes that, under realistic fault-tolerant hardware assumptions, achieving a practical runtime advantage will be extremely challenging. 
All in all, the apparent limitation of quantum walks to quadratic speedups raises the question of whether they can ultimately provide a practical computational advantage. 

Here, we show that this limitation is not fundamental and present a practical route to obtaining large, \textit{super-quadratic} speedups over the best fully-classical Markov-chain algorithms.

The key observation is that existing quantum-walk algorithms have been constructed by quantizing \textit{classical} Markov chains, whose proposal moves are chosen to be efficient on classical hardware.
These include local spin flips or different variants of cluster updates.
A quantum computer, however, allows \textit{also} proposal moves that are themselves efficient to implement only on quantum hardware, thus enlarging the class of Markov chains that can be ``quantized''. Therefore our new quantum Metropolis approach
now leverages the features of quantum computation both in the acceptance and in the proposal steps.
As a result, the resulting Markov chains need not be direct quantum counterparts of classical ones, and their performance can exceed the quadratic speedup associated with prior implementations of quantum walks.

More specifically, we analyze a new family of quantum Markov chains, in which the proposal move is implemented by Hamiltonian simulation. We characterize their complexity both at the query level and in terms of end-to-end runtime by explicitly compiling all circuit primitives into a fault-tolerant architecture.
We consider all-to-all connected graphs, with random couplings and random local fields~\cite{SteinNewman2013,lucas:2014}.

To provide a meaningful comparison, we also perform a detailed study of the best classical alternatives. We consider several proposal strategies and estimate their runtimes on CPU, GPU, and FPGA implementations.

Our analysis shows that replacing classical proposal moves with quantum-native ones dramatically changes the practical outlook for quantum sampling. Compared with the best classical walk, the quantum walks introduced here reduce the crossover runtime by six orders of magnitude, bringing it into a practical regime and placing a practical quantum advantage within reach of fault-tolerant quantum computing.

\section{Methods}
\label{sec:methods}

\subsection{Markov chain Monte Carlo sampling}
\label{sec:mcmc_sampling}

We consider the Markov chain Monte Carlo sampling problem \cite{levin2017markov}: given an Ising Hamiltonian $H$, an inverse temperature $\beta \geq 0$, and a precision parameter $\varepsilon \in (0, 1)$, the task is to produce a sample from a distribution that is $\varepsilon$-close in total variation distance to the Gibbs distribution $\pi_\beta(x)\propto e^{-\beta H(x)}$.
For spin configurations $x\in\{\pm1\}^n$, the target distribution is
\begin{equation}
    \pi_\beta(x)
    =
    \frac{e^{-\beta H(x)}}{Z_\beta},
    \qquad
    Z_\beta
    =
    \sum_{z\in\{\pm1\}^n} e^{-\beta H(z)} .
\end{equation}
The chain starts from an initial distribution $q_0$ that is easy to prepare. At each step, a trial configuration $y$ is proposed from the current configuration $x$ with probability $T_{yx}$ and is accepted with probability $A^{(\beta)}_{yx}$. The transition matrix $P_\beta$ has entries
\begin{equation}
    (P_\beta)_{yx}
    =
    T_{yx}A^{(\beta)}_{yx},
    \qquad
    y\neq x,
\end{equation}
with diagonal entries fixed by normalization. For symmetric trial moves, $T_{yx}=T_{xy}$, we use the Metropolis acceptance probability
\begin{equation}
   A^{(\beta)}_{yx}
    =
    \min\left\{1,\exp\!\left(-\beta (H(y)-H(x))\right)\right\}.
\end{equation}

\subsubsection{Convergence to the Gibbs distribution}

We restrict to chains that are ergodic and reversible with respect to $\pi_\beta$. Ergodicity makes $\pi_\beta$ the unique stationary distribution and guarantees convergence from any initial distribution, while reversibility means that the chain satisfies the detailed balance condition
\begin{equation}
    \pi_\beta(x)(P_\beta)_{yx}
    =
    \pi_\beta(y)(P_\beta)_{xy},
\end{equation}
for all configurations $x,y$. Detailed balance ensures that $\pi_\beta$ is stationary. Further details are given in Appendix~\ref{app:mcmc}.

We measure convergence in total variation distance,
\begin{equation}
    \|q-\pi_\beta\|_{\rm TV}
    =
    \frac{1}{2}\sum_x |q(x)-\pi_\beta(x)| .
    \label{eq:tv_distance}
\end{equation}
The mixing time is defined as
\begin{equation}
\label{eq_target}
    t_{\mathrm{mix}}(\varepsilon)
    =
    \max_{q_0}
    \min\left\{
        t\geq 0:
        \|P_\beta^t q_0-\pi_\beta\|_{\rm TV}
        \leq \varepsilon
    \right\}.
\end{equation}
For a finite reversible chain, convergence is controlled by the absolute spectral gap
\begin{equation}
    \delta_\beta
    =
    1
    -
    \max\left\{
        |\lambda|<1:
        \lambda\in\operatorname{spec}(P_\beta)
    \right\},
    \label{eq:absolute_spectral_gap}
\end{equation}
up to logarithmic factors depending on the initial distribution and on $\varepsilon$. We therefore use $\delta_\beta^{-1}$ as a proxy for the number of Markov chain steps required to reach desired accuracy $\varepsilon$.

\subsubsection{Classical trial moves}
\label{sec:proposal_moves}

The trial move matrix $T$ determines how the chain explores configuration space before the Metropolis acceptance step. Different choices of $T$ induce different spectral gaps and therefore different mixing times~\cite{hastings:1970,levin2017markov}. We compare two choices that are expected to provide strong classical baselines for our models in different temperature regimes, following the numerical evidence reported in Ref.~\cite{layden2023quantum}.

Local moves flip exactly $k$ spins in the current configuration. Let $|x\oplus y|$ denote the Hamming distance between two spin configurations, namely the number of sites on which $x$ and $y$ differ. The local proposal matrix is
\begin{equation}
T_{yx}^{\rm loc}
=
\begin{cases}
\binom{n}{k}^{-1}, & |x\oplus y|=k,\\
0, & \text{otherwise}.
\end{cases}
\end{equation}
Uniform moves propose a configuration uniformly at random, independently of the current one. The corresponding proposal matrix is
\begin{equation}
T_{yx}^{\rm unif}
=
2^{-n}.
\end{equation}
Cluster updates~\cite{zhu:2015} are found to be not effective for this model~\cite{layden2023quantum}, so we do not consider them here.

\subsection{Gibbs sampling with quantum walks}
\label{sec:annealed_state_preparation}

We consider a quantum walk algorithm that prepares the coherent Gibbs state
\begin{equation}
    \ket{\pi_\beta}
    =
    \sum_x \sqrt{\pi_\beta(x)}\ket{x}.
\end{equation}
Measuring this state in the computational basis gives a classical sample from $\pi_\beta$. The construction uses the Szegedy quantum walk associated with $P_\beta$, which maps the spectrum of the Markov chain into eigenphases. Its zero-phase component contains the coherent Gibbs state, and a spectral filter projects onto this component. Since the success probability of the projection depends on the overlap between the initial state and the target state, the filters are applied along an annealing schedule, namely a sequence of intermediate inverse temperatures from $\beta=0$ to the target inverse temperature $\bar\beta$.

We denote the Szegedy quantum walk for $P_\beta$ by $W_\beta$. If $\lambda$ is an eigenvalue of $P_\beta$, then the corresponding walk eigenvalues are $e^{\pm i\arccos(\lambda)}$. The stationary eigenvalue $\lambda=1$ is therefore mapped to phase zero. The absolute spectral gap bounds the nonstationary spectrum as $|\lambda|\leq 1-\delta_\beta$. We use the corresponding phase-separation parameter
\begin{equation}
    \theta_\beta
    =
    \arccos(1-\delta_\beta)
    \simeq
    \sqrt{2\delta_\beta},
    \label{eq:phase_separation}
\end{equation}
for $\delta_\beta\ll1$.

We implement $W_\beta$ following the general construction of Ref.~\cite{lemieux2020efficient}. This realizes the Szegedy walk from coherent access to the proposal rule and a coherent Metropolis acceptance step. The full walk operation acts on the composite Hilbert space $\mathcal H_a\otimes\mathcal H_b\otimes\mathcal H_c$. The register $\mathcal H_a$ stores the current configuration, $\mathcal H_b$ stores the proposed configuration, and $\mathcal H_c$ stores the Metropolis coin. The walk decomposes as
\begin{equation}
    W_\beta
    =
    R_0 V^\dagger B_\beta^\dagger F B_\beta V .
    \label{eq:walk_circuit_decomposition}
\end{equation}
Here $V$ coherently prepares the proposal distribution,
\begin{equation}
    V\ket{x}_a\ket{0}_b
    =
    \ket{x}_a
    \sum_y \sqrt{T_{yx}}\ket{y}_b,
    \label{eq:proposal_unitary}
\end{equation}
while $B_\beta$ implements the Boltzmann coin,
\begin{equation}
    B_\beta\ket{x}_a\ket{y}_b\ket{0}_c
    =
    \ket{x}_a\ket{y}_b
    \left(
    \sqrt{1-A^{(\beta)}_{yx}}\ket{0}_c
    +
    \sqrt{A^{(\beta)}_{yx}}\ket{1}_c
    \right),
    \label{eq:boltzmann_coin_unitary}
\end{equation}
where $A^{(\beta)}_{yx}$ is the Metropolis acceptance probability. The unitary $F$ swaps the two configuration registers conditioned on the accept state, and $R_0$ is the reflection about the subspace in which the proposal and coin registers are initialized to zero. The equivalence with the standard Szegedy formulation is reviewed in Appendix~\ref{app:szegedy_quantum_walk}.

\subsubsection{Spectral filtering and annealed preparation}

The exact projector onto the zero phase component of $W_\beta$ is
$
    \Pi_\beta
    =
    \ketbra*{\pi_\beta}.
$
In practice, it is approximated by a spectral filter $\widetilde{\Pi}_\beta$.

When applied to a state with nonzero overlap with $\ket{\pi_\beta}$, this projection prepares the coherent Gibbs state. The filter can be implemented using quantum phase estimation~\cite{wocjan:2008} or quantum singular value transformation~\cite{gilyen2019quantum}. In both cases, the cost is controlled by the phase-separation parameter $\theta_\beta$ (Eq.~\ref{eq:phase_separation}).

The success probability of a spectral filter is determined by the squared overlap between its input state and the target coherent Gibbs state. Direct filtering from the uniform state at $\beta=0$ to the target state $\ket*{\pi_{\bar\beta}}$ at $\bar\beta$ may therefore be inefficient. Following Ref.~\cite{wocjan:2008}, we adopt an annealing schedule
\begin{equation}
    0=\beta_0<\beta_1<\cdots<\beta_L=\bar\beta,
    \label{eq:schedule}
\end{equation}
chosen so that consecutive coherent Gibbs states have squared overlap bounded below by a constant,
\begin{equation}
    |\braket*{\pi_{\beta_{j+1}}}{\pi_{\beta_j}}|^2
    \geq
    1/e ,
\end{equation}
with the threshold $1/e$ being a convenient choice~\cite{wocjan:2008}. In the regimes studied here, the schedule length is not the dominant contribution to the complexity. In the worst case it scales as $O(\norm{H})$, while the dominant contribution is controlled by the smallest phase gap encountered along the schedule~\cite{wocjan:2008}.

Applying all filters sequentially would give a success probability at least $(1/e)^L$, up to filter errors. To avoid this exponential loss, we use the rewind protocol~\cite{lemieux2020efficient,Temme:2009wa}. Let
\begin{equation}
    \Pi_j
    =
    \ketbra*{\pi_{\beta_j}} .
\end{equation}
At step $j$, we apply $\Pi_j$. If the projection fails, we apply $\Pi_{j-1}$ to return to the previous coherent Gibbs state and then retry $\Pi_j$. Since consecutive states have squared overlap at least $1/e$, this procedure has constant expected overhead. Further details are given in Appendix~\ref{app:annealed_spectral_filtering}.

\subsection{Quantum Hamiltonian walks}
\label{sec:new_quantum_walk}

We are now in the position to introduce the conceptual difference between our approach and previous quantum walk constructions. Rather than adopting the local proposal moves, $T_{yx}$, of a classical Markov chain, we use a \emph{non-local} proposal generated by Hamiltonian simulation. This choice is inspired by the quantum-enhanced Monte Carlo method~\cite{mazzola2021sampling,layden2023quantum}, where  quantum simulation of a suitable Hamiltonian is used solely to generate proposal configurations.
In that setting, however, the proposal is followed by a measurement and a classical Metropolis acceptance step. The resulting short-depth algorithm therefore defines a \textit{classical, sequential} Markov chain assisted by a quantum simulator, rather than a fully coherent quantum walk.

Here, we show how the same Hamiltonian simulation proposal can instead be incorporated into Szegedy's quantum walk framework, in a very powerful fashion. In this way, the proposal generation and the Metropolis acceptance step are both implemented coherently, yielding a genuine quantum Markov chain based on quantum-native, non-local updates.

The initial step is to ``upgrade'' the classical cost function $H$ to a non-diagonal operator,
\begin{equation}
\label{eq:tf_ham} 
    H_{\rm tf}(\gamma)
    =
    H + \gamma H_X =
    H
    -
    \gamma\sum_i X_i,
\end{equation}
namely, the sum of transverse field operator, $H_X = -\sum_i X_i $ and the classical Ising problem Hamiltonian $H$.

We define the proposal by averaging the transition probabilities over randomized choices of evolution time and transverse field strength. For $t\in[t_0,t_1]$ and $\gamma\in[\gamma_0,\gamma_1]$ sampled uniformly, the proposal matrix is
\begin{align}
   \nonumber &  T_{yx}
    =
    \frac{1}
    {(t_1-t_0)(\gamma_1-\gamma_0)} \\
    & \qquad
    \int_{t_0}^{t_1}
    \int_{\gamma_0}^{\gamma_1}
    \left|
    \bra{y}
    \exp\!\left(-it H_{\rm tf}(\gamma)\right)
    \ket{x}
    \right|^2
    \,d\gamma\,dt .
    \label{eq:dynamical_proposal_matrix}
\end{align}
 Each term in the average is a valid transition probability generated by unitary evolution, and the uniform average over the grid is therefore stochastic. Since $H_{\rm tf}(\gamma)$ is real symmetric, the transition probabilities satisfy $T_{yx}=T_{xy}$, so the Metropolis rule above defines a reversible chain with stationary distribution $\pi_\beta$.

The motivation for using this quantum hamiltonian  proposal is that, when used in a sequential way, it yields a larger spectral gap $\delta_\beta$ than purely classical proposals (ranging from local, multi-spins, and cluster updates) \cite{layden2023quantum,christmann2025quantum,nakano2026neural,marshall2026quantumenhancedmarkovchainmonte}.
In Appendix~\ref{app:spectral_gap_queries} we confirm again numerically such observations.

The Szegedy construction converts this gap into a walk phase gap of order $\sqrt{\delta_\beta}$. The method therefore combines two mechanisms: an increase in the initial spectral gap due to the Hamiltonian proposal, and the square root improvement offered by the Szegedy walk.

However, this implies a major rethinking of the standard quantum walks implementation, which now requires coherent access to the proposal rule through the unitary $V$ in Eq.~\ref{eq:proposal_unitary}. For a generic classical proposal, constructing $V$ would require a reversible circuit that prepares the amplitudes $\sqrt{T_{yx}}$, which is generally infeasible for the dynamical proposal in Eq.~\ref{eq:dynamical_proposal_matrix}. In the present case, however, the proposal is generated by time evolution under $H_{\rm tf}(\gamma)$. For each choice of $(t,\gamma)$, define
\begin{equation}
    U(t,\gamma)
    =
    \exp\!\left(-it H_{\rm tf}(\gamma)\right).
\end{equation}
We show that, because $H_{\rm tf}(\gamma)$ is real symmetric, these unitaries have the structure required to define a coherent proposal state compatible with the Szegedy construction. The proof of compatibility with the standard Szegedy walk is given in Appendix~\ref{app:szegedy_quantum_walk}.

Finally let us remark that, while we introduce a non-diagonal Hamiltonian, our method still targets sampling of classical problems, and should not be confused with other quantum sampling algorithms which targets thermal state preparation of native quantum many-body Hamiltonians, see e.g. Refs.~\cite{Temme:2009wa,Yung_2012,Chen2025EfficientThermal, Jiang:2024dzh,hahn2026efficientquantumthermalstate,leng2026acceleratingquantumgibbssampling}.

\section{Benchmark setup and resource estimates}
\label{sec:implementation_resources}

\subsection{Sherrington--Kirkpatrick instances}
\label{sec:sk_instances}

We benchmark on instances of the Sherrington--Kirkpatrick (SK) model with fields~\cite{sherrington1975solvable}.
This is a dense all-to-all connected graphs, with  disordered couplings and local fields.
For each system size $n$, we sample independent Gaussian fields $h_i\sim\mathcal{N}(0,1)$ and independent Gaussian couplings $J_{ij}\sim\mathcal{N}(0,1)$ for every pair of spins. The coefficients are then rescaled by
\begin{equation}
    \alpha
    =
    \sqrt{
    \frac{n}
    {\sum_i h_i^2+\sum_{i<j}J_{ij}^2}
    }.
    \label{eq:sk_alpha}
\end{equation}
We define $\tilde h_i=\alpha h_i$ and $\tilde J_{ij}=\alpha J_{ij}$, and use the classical Hamiltonian
\begin{equation}
    H(x)
    =
    -
    \sum_i \tilde h_i x_i
    -
    \sum_{i<j} \tilde J_{ij} x_i x_j,
    \label{eq:sk_hamiltonian}
\end{equation}
This normalization gives the model the standard volume-scaling of the energy. Details of the model are given in Appendix~\ref{app:sk_model}. 

Note that, compared to Refs.~\cite{farhi2022quantum, basso2022quantum} using quantum algorithms to tackle the optimization of the SK model, i.e. finding a configuration close to the ground state of the model, we focus instead on the sampling problem which is a task of different nature and complexity.

\subsection{Query model}
\label{sec:query_model}

The query model counts the number of algorithmic steps required to reach the target Gibbs distribution with accuracy $\varepsilon$. For a classical walk, one query is one attempted Metropolis proposal. For the quantum algorithm, one query is one application of the Szegedy quantum walk $W_\beta$.
Obviously, each comes with a different time-complexity, and we account for this in the next Sections.

\subsubsection{Numerical simulations}

We first determine how the number of classical queries scales with system size. Fix an inverse temperature $\beta$ and an accuracy $\varepsilon$. We choose the initial distribution as a warm Gibbs distribution $q_0=\pi_{\beta_0}$, with $\beta_0\leq\beta$, such that
\begin{equation}
    \left(
    \sum_x
    \sqrt{\pi_{\beta_0}(x)\pi_\beta(x)}
    \right)^2
    \geq
    1/e .
\end{equation}
This choice matches the overlap condition used later in the annealing schedule. For each classical proposal rule and each size $n=3,\ldots,10$, and each SK instance, we construct the full transition matrix $P_\beta$ and compute
\begin{equation}
    Q_{\rm cl,warm}(n,\beta,\varepsilon)
    =
    \min\left\{
        t\geq 0:
        \left\|
        (P_\beta)^t q_0-\pi_\beta
        \right\|_{\rm TV}
        \leq
        \varepsilon
    \right\}.
    \label{eq:classical_query_definition}
\end{equation}
The values are aggregated over $100$ disorder instances for each $n$. At fixed $\beta$ and $\varepsilon$, we fit
\begin{equation}
    Q_{\rm cl,warm}(n,\beta,\varepsilon)
    =
    a_{\beta,\varepsilon}
    2^{\nu_{\beta,\varepsilon} n}
    \label{eq:classical_query_fit}
\end{equation}
This gives the number of classical queries required for one warm start step at inverse temperature $\beta$. We use the same target accuracy $\varepsilon$ for each step of the classical annealing schedule.

For the quantum query model, the fitted quantity is the spectral gap of the Markov chain used to define the Szegedy walk. We compute the absolute spectral gap $\delta_{\beta}(n)$ (Eq.~\ref{eq:absolute_spectral_gap}).
At fixed $\beta$, the instance averaged gap is fitted as
\begin{equation}
    \delta_{\beta}(n)
    =
    c_{\beta}
    \exp\!\left(
        -\nu_{\beta} n
    \right).
    \label{eq:spectral_gap_query_fit}
\end{equation}
The corresponding Szegedy phase separation is $\theta_\beta(n)=\arccos(1-\delta_{\beta}(n))$. The data are evaluated on a finite grid of inverse temperatures. 

\subsubsection{Cost model}

We then build the annealing schedule. For a target inverse temperature $\bar\beta$, the schedule is $0=\beta_0<\beta_1<\cdots<\beta_L=\bar\beta,$
where $L$ is the number of annealing steps. The schedule is chosen so that consecutive coherent Gibbs states have squared overlap at least $1/e$, and is achieved by the following relation
\begin{equation}
    \Delta\beta(n,\beta_i)
    =
    \frac{f(\beta_i)}{\sqrt n},
    \label{eq:query_schedule_spacing}
\end{equation}
where $f(\beta_i)$ is fitted numerically. The scaling in $1/\sqrt n$ follows the observed concentration of the energy fluctuations along the annealing path. The fitted form of $f(\beta_i)$ and its validation are given in Appendix~\ref{app:annealed_spectral_filtering}. The schedule is generated recursively as
\begin{equation}
    \beta_{j+1}
    =
    \min\left\{
        \bar\beta,
        \beta_j+\Delta\beta(n,\beta_j)
    \right\}.
\end{equation}

The classical query count for the full annealing schedule is the sum of the fitted warm start query counts along the schedule,
\begin{equation}
    Q_{\rm cl}(n,\bar\beta,\varepsilon)
    =
    \sum_{j=1}^{L}
    Q_{\rm cl,warm}(n,\beta_j,\varepsilon).
    \label{eq:classical_annealing_queries}
\end{equation}

For the quantum algorithm, the degree of each spectral filter is set by the corresponding Szegedy phase separation. Since the implemented filter is approximate, we allocate part of the total error budget to the filter approximation and set its (operator norm) error to $\varepsilon_{\rm FLT}=\varepsilon/4$. The rewind protocol gives expected overhead $1+e$ on the number of steps in the annealing sequence. Therefore, 
the error assigned to each individual filter is $\varepsilon_{\rm FLT}/((1+e) L)$. The total quantum query count includes both the filter degrees and the rewind overhead,
\begin{align}
\nonumber
    Q_{\rm q}(n,\bar\beta,\varepsilon)
    & =
    2(1+e)
    \log_2\!\left(
        \frac{4(1+e) L}{\varepsilon}
    \right) \times \\
    & \quad
    \sum_{j=1}^{L}
    \frac{1}
    {\theta_{\beta_j}(n)} .
    \label{eq:quantum_annealing_queries}
\end{align}

\subsection{Cost of a single classical step}
\label{sec:classical_baselines}

We estimate the wall clock cost of one attempted Metropolis proposal for CPU, GPU, and FPGA implementations, which have been proposed to study classical and quantum spin systems \cite{bernaschi2024qisg,baity2014janus}. The CPU and GPU programs use single precision floating point arithmetic, while the FPGA program uses fixed point arithmetic. 
In this Section we provide the general results, while the details of calculation and lower-level implementation are reported in Appendix~\ref{app:classical_hardware_baselines}.

\subsubsection{CPU and GPU implementation}

Since our model is dense, we directly evaluate the simulation cost, without resorting to previous literature, specialized to sparse, regular graphs~\cite{bernaschi2024qisg,baity2014janus}.
The CPU implementation evaluates the dense sums using single precision arithmetic. For the local move, we fit
\begin{equation}
\tau_{\rm C/GPU}^{\rm loc}(n)
=
a_{\rm C/GPU}^{\rm loc}
+
b_{\rm C/GPU}^{\rm loc} n.
\end{equation}
and for the uniform move we fit
\begin{equation}
    \tau_{\rm C/GPU}^{\rm unif}(n)
    =
    a_{\rm C/GPU}^{\rm unif}
    +
    b_{\rm C/GPU}^{\rm unif} n
    +
    c_{\rm C/GPU}^{\rm unif} n^2,
\end{equation}
with parameter's values reported in Table~\ref{table:cpu}.
The GPU implementation uses the same scaling forms. It assigns one CUDA block to one Markov chain and parallelizes the sums through block reductions. The reported GPU timing is a single chain latency model. It does not include higher level GPU spin glass strategies such as many replica execution, parallel tempering, or model specific update schedules.

\subsubsection{FPGA implementation}

The FPGA implementation evaluates all terms contributing to the proposed energy difference in parallel. These terms are reduced by a tree of adders, so the reduction depth grows logarithmically in the number of terms. The Metropolis factor $\exp(-\beta\Delta H)$ for $\Delta H>0$ is evaluated using a truncated Chebyshev polynomial approximation.

For both local and uniform proposal move, we fit
\begin{equation}
    \tau_{\rm FPGA}(n)
    =
    a_{\rm FPGA}
    +
    b_{\rm FPGA}\log_2 n.
\end{equation}

\begin{table}[h!]
\centering
\scalebox{0.95}{\begin{tabular}{llccc}
\toprule
Platform & Proposal & $a$~[\si{\second}] & $b$~[\si{\second}] & $c$~[\si{\second}] \\
\midrule
CPU  & local  & \num{5.959e-9} & \num{1.429e-10} & \num{0} \\
CPU  & uniform & \num{0}        & \num{1.173e-8}  & \num{6.964e-11} \\
GPU  & local  & \num{7.837e-7} & \num{1.459e-9}  & \num{0} \\
GPU  & uniform & \num{0}        & \num{0}         & \num{2.215e-10} \\
FPGA & local  & \num{2.679e-7} & \num{1.800e-9}  & -- \\
FPGA & uniform & \num{2.541e-7} & \num{4.200e-9}  & -- \\
\bottomrule
\end{tabular}}
\caption{
Fitted single-step latency constants for the classical Metropolis implementations.
}
\label{table:cpu}
\end{table}

\subsubsection{Bounded error in the fixed point arithmetic}

This subsection is relevant for both the classical FPGA and the quantum walk implementation.
The fixed point implementation replaces $P$ by an approximate transition matrix $\widetilde P$. If $\widetilde\pi$ is the stationary distribution of $\widetilde P$, then we assume that, for small deviations
\begin{equation}
    \|(\widetilde P)^tq_0-\pi\|_{\rm TV}
    \leq
    \|(\widetilde P)^tq_0-\widetilde\pi\|_{\rm TV}
    +
    \|\widetilde\pi-\pi\|_{\rm TV}.
    \label{eq:error_arithmetic_triangular}
\end{equation}
The first term is the mixing error of the implemented chain, while the second is the stationary bias induced by finite precision arithmetic. We control this bias through two internal approximation budgets: the energy discretization budget $\varepsilon_{\rm discr}$ and the operator approximation budget $\varepsilon_{\rm op}$. 
Let $M = n+\binom{n}{2}$ be the number of Hamiltonian coefficients in the dense worst case, and let $\Lambda$ be an upper bound on the energy difference. 
We show in Appendix~\ref{app:classical_hardware_baselines} that the number of fractional bits $b$ must be chosen such that $|\Delta H-\widetilde{\Delta H}| \leq 2M\Lambda 2^{-b}$. For the SK model considered here, this yields
\begin{equation}
    b
    \approx
    \frac{7}{2}\log_2 n
    +
    \log_2(\varepsilon_{\rm discr}^{-1}).
    \label{eq:choice_of_f}
\end{equation}
The error budget $\varepsilon_{\rm op}$ is assigned to the truncated Chebyshev approximation.

\subsection{Cost of a single logical quantum step}
\label{sec:walk_circuit_implementation}

We now estimate the cost of implementing one application of $W_\beta$. We quantify this cost by the logical non-Clifford depth, namely the number of non-Clifford operations on the critical path of the circuit.

This choice anticipates the fault tolerant model used below, where Clifford operations are negligible and the dominant cost is set by non-Clifford resources. 
We work with the gate set formed by Clifford gates, $\mathsf{T}$ gates, and rotations $\mathsf{R}_z(\theta)$. The $\mathsf{T}$ gate is the basic magic state resource. The rotations $\mathsf{R}_z(\theta)$ are kept as native primitives rather than immediately synthesized into Clifford and $\mathsf{T}$, since their synthesis can be performed offline and supplied as rotation resources. 
Here, we provide a description of all the unitaries that are needed for $W_\beta$ (see Eq.~\ref{eq:walk_circuit_decomposition}).
Table~\ref{tab:resources_compact} summarizes the asymptotic costs. Explicit non-asymptotic formulas are reported in Appendix~\ref{app:quantum_walk_circuit_resources}. We provide all notebooks necessary to better understand the implementation and reproduce the calculations in an online repository~\cite{myrepo2026}.

\paragraph{Reflection unitary $R_0$.}
The reflection $R_0$ is a phase flip on the state where the proposal and coin registers are initialized to zero. We implement it as a multi-controlled phase, using a logarithmic depth multi-controlled operation following Ref.~\cite{khattar2025rise}.

\paragraph{Accept-path unitary $F$.}
The accept-path unitary conditionally swaps the current and proposal registers when the coin is in the accept state. We first compute the accept condition into a flag using a logarithmic depth multi-controlled operation. The flag is then copied by Clifford fanout and used to apply the controlled swaps in parallel. The non-Clifford depth is therefore set by the accept-condition computation and by a constant-depth layer of Toffoli-based controlled swaps.

\paragraph{Proposal unitary $V$ for the uniform move.}
The uniform proposal ignores the current configuration and applies Hadamard gates to the proposal register. This has zero non-Clifford depth.

\paragraph{Proposal unitary $V$ for the local move.}
A local move proposes a configuration at fixed Hamming distance $k$ from the current one. This is implemented by preparing a uniform superposition over all weight-$k$ bit masks and applying the mask to the current configuration,
\begin{equation}
    \ket{x}_a\ket{0^n}_b
    \mapsto
    \binom{n}{k}^{-1/2}
    \ket{x}_a
    \sum_{|z|=k}
    \ket{x\oplus z}_b .
\end{equation}
The mask is a Dicke state on $n$ qubits, prepared using the state preparation circuit in Ref.~\cite{aktar2022divide}. In this work we use $k=1$, corresponding to a uniformly random single spin flip.

\paragraph{Proposal unitary $V$ for the Hamiltonian Simulation move.}
The  proposal copies the current configuration into the proposal register and applies a Trotter approximation to the transverse-field Ising evolution. We use the transverse-field Hamiltonian of Eq.~\ref{eq:tf_ham},
where $H$ is diagonal in the computational basis. We use the symmetric second-order formula
\begin{equation}
    U^{(r)}(t,\gamma)
    =
    \left(
    e^{-it H/(2r)}
    e^{-it\gamma H_X/r}
    e^{-it H/(2r)}
    \right)^r .
\end{equation}
This form has essentially the same asymptotic depth as the first-order formula, because the two half diagonal layers can be merged between consecutive Trotter steps. It also preserves the symmetry of the proposal unitary, since each Trotter step is palindromic and built from symmetric factors. Also crucially, the proposal remains effective also even when the Trotter circuit is not a tight approximation to the exact Hamiltonian evolution \cite{layden2023quantum,christmann2025quantum}.
For simplicity, in all simulations and gap estimations we use $r=50$ for all sizes, and do not scale $r$ with $n$ since the model is all-to-all connected.

\paragraph{Boltzmann coin with phase arithmetic.}

The Boltzmann coin applies the square root Metropolis amplitude, which is a function of the energy difference $\Delta H=H(y)-H(x)$. For sparse Hamiltonians and local moves, $\Delta H$ depends only on a small number of terms and can be computed efficiently~\cite{lemieux2020efficient}. For dense SK instances and nonlocal moves, instead, $\Delta H$  involves all terms in the Hamiltonian. We use the phase arithmetic approach in Ref.~\cite{mcardle2022quantum}. Let $H$ denote the diagonal operator associated with the classical Hamiltonian, $H\ket{x}=H(x)\ket{x}$. On registers $\ket{x}_a\ket{y}_b$ we define the Hamiltonian 
\begin{align}
\label{eq:fully_phase_arith_hamiltonian}
    &H_{\rm ph}
    =
    -H^{(a)}
    +
    H^{(b)}, \nonumber \\
    &H_{\rm ph}\ket{x}_a\ket{y}_b
    =
    \Delta H\,\ket{x}_a\ket{y}_b .
\end{align}
where $H^{(a)}, H^{(b)}$ denotes the SK hamiltonian acting on the register $a, b$ respectively. 
A block encoding of $H_{\rm ph}$ is then transformed by a polynomial approximation to $f_\beta(\Delta H)\approx\sqrt{\min\{1,e^{-\beta\Delta H}\}}$. The depth is dominated by the block encoding of $H_{\rm ph}$ and by the polynomial degree needed to approximate the kink of $f_\beta$ at zero.

\paragraph{Boltzmann coin with hybrid arithmetic.}

The phase arithmetic construction is space efficient, but its depth is limited by the cost of block encoding the dense energy difference and by the non-smooth Metropolis function. To reduce this depth, we use a hybrid construction that computes only the part of the energy difference needed by the acceptance rule. Namely, we compute in fixed point
\begin{equation}
    (\Delta H)_+
    =
    \max\{0,H(y)-H(x)\}.
\end{equation}
This value can be obtained by loading the dense set of Hamiltonian terms, summing them with a Wallace tree \cite{ercegovac2004digital}, and selecting the positive branch. The resulting circuit has logarithmic depth in $n$ and in the fixed-point precision, at the cost of using enough workspace and resulting in a quadratic scaling of the number of qubits with $n$.

Once $(\Delta H)_+$ is stored, the Boltzmann coin only needs to apply the smooth function $\exp(-\beta (\Delta H)_+/2)$. We do this by phase arithmetic on the fixed point energy register. Let
\begin{equation}
    H_{\rm hb}\ket{(\Delta H)_+}
    =
    (\Delta H)_+\ket{(\Delta H)_+}
\end{equation}
be the diagonal Hamiltonian acting on the register that stores $(\Delta H)_+$. A polynomial transformation of a block encoding of $H_{\rm hb}$ then applies an approximation to $f(z)=\exp(-\beta z/2)$ on the allowed fixed point range. After truncating the exponentially small tail, the required polynomial degree is logarithmic in the inverse approximation error.

The hybrid construction therefore trades quadratic space in $n$ for logarithmic arithmetic depth. This is the useful regime for dense instances, where extra workspace can be used to turn the dense energy sum into a shallow reduction. The required fixed point precision and polynomial approximation errors are bounded in Appendix~\ref{app:classical_hardware_baselines}.

\begin{table*}[t]
    \centering
    \begin{tabular}{lcc}
        \hline
        Component & Logical non-Clifford depth & Logical qubits \\
        \hline
        Uniform proposal & $0$ & $2n$ \\
        Local proposal, $k=1$ spin flips & $O(\log n)$ & $2n$ \\
        Ham. Sim. proposal, $r$ Trotter steps & $O(rn)$ & $2n$ \\
        Boltzmann coin, phase arithmetic & $O(n^2)$ & $O(n)$ \\
        Boltzmann coin, hybrid arithmetic & $O(\log n\,\log\varepsilon^{-1})$ & $O(n^2\log(n/\varepsilon))$ \\
        Reflection $R_0$ & $O(\log(n+\log\varepsilon^{-1}))$ & $O(n+\log\varepsilon^{-1})$ \\
        Accept path unitary & $O(\log\log(n/\varepsilon))$ & $O(n+\log(n/\varepsilon))$ \\
        \hline
    \end{tabular}
    \caption{
    Asymptotic logical non-Clifford depth and logical qubit cost of the main components of one Szegedy walk step, for the different proposal moves considered above. Constants and lower order logarithmic factors are suppressed. Explicit formulas are given in Appendix~\ref{app:quantum_walk_circuit_resources}.
    }
    \label{tab:resources_compact}
\end{table*}

\subsection{Cost of the quantum algorithm on a fault tolerant architecture}
\label{sec:fault_tolerant_model}

In contrast with the classical case, where the runtime is the number of classical queries multiplied by the estimated wall clock time per query on the best \textit{available} device, the quantum case is more complex. Assigning a definite duration to an abstract logical operation is more nuanced and  misses the details of quantum error correction overheads.

A useful compromise between a platform independent estimate and a precise physical timing model is obtained by mapping the logical circuit to an idealized fault tolerant architecture. We do this by converting the circuit into Pauli product measurements, which can be implemented as lattice surgery instructions on a surface code based system \cite{litinski2019lattice,litinski2019game, beverland2022assessing, chamberland2022universal}. The remaining platform-dependent uncertainties are accounted for by allowing the physical operation time to vary over two orders of magnitude.

We use the Clifford+$\phi$ framework of Ref.~\cite{litinski2019game}, where Clifford gates are tracked classically in a Pauli frame and only update later Pauli product measurements. The explicit hardware cost is assigned to injected non-Clifford resources: $\ket{T}$ states for $\mathsf{T}$ gates~\cite{gidney2024magic}, and $\ket{\mathsf{R}_z(\phi)}$ states for $\mathsf{R}_z(\phi)$ rotations~\cite{mishra2014resource,campbell2016efficient}. This representation preserves the logical dependencies needed for the critical path.

To proceed further, we fix the error budget. For a desired error $\varepsilon$ in TV distance, let $\mathcal A$ be the ideal quantum algorithm and let $\widetilde{\mathcal A}$ be its fault tolerant implementation, also, let $\rho_{\beta}$ and $\widetilde\rho_{\beta}$ be the corresponding output states before measurement. Since measurement cannot increase trace distance,
\begin{equation}
    \|\widetilde\pi_{\beta}-\pi_{\beta}\|_{\rm TV}
    \leq
    \frac{1}{2}\|\widetilde\rho_{\beta}-\rho_{\beta}\|_1
    \leq
    \|\widetilde{\mathcal A}-\mathcal A\|_{\rm op}.
\end{equation}
We split the operator norm budget as
\begin{equation}
    \varepsilon
    =
    \varepsilon_{\rm QEC}
    +
    \varepsilon_{\rm MS}
    +
    \varepsilon_{\rm FLT}
    +
    \varepsilon_W ,
\end{equation}
corresponding to, respectively, failures in the quantum error correction layer, magic state failures, spectral filter approximation error, and implementation error of the individual walk operators.

The error budget $\varepsilon_{\rm QEC}$ determines the surface code distance. Let $Q$ be the number of quantum walk queries, $D$ the non-Clifford depth of one walk step, and $S$ its logical space cost. We follow the model from~\cite{beverland2022assessing}: given the physical error rate $p_{\rm phys}$ and threshold $p_* = 0.01$, we use
\begin{equation}
    p_L(d)
    =
    0.03
    \left(
        \frac{p_{\rm phys}}{p_*}
    \right)^{(d+1)/2}.
    \label{eq:surface_code_logical_error}
\end{equation}
The distance is chosen so that the spacetime volume $QDS$, multiplied by the logical error probability per location, is below the allocated budget:
\begin{align*}
    d
    & =
    \min_{{\rm odd\,} d' \ge 3}
    \left\{d' :
        QDS\,p_L(d')
        \leq
        \varepsilon_{\rm QEC}
    \right\}.
\end{align*}
The duration of one lattice surgery layer is $d(4t_{\rm op}+t_{\rm mr})$, where $t_{\rm op}$ is the physical gate time and $t_{\rm mr}$ is the physical measurement time plus subsequent qubit reset. The fault tolerant runtime estimate is
\begin{equation}
    QD\,d(4t_{\rm op}+t_{\rm mr}).
    \label{eq:surface_code_runtime}
\end{equation}

The price for remaining platform independent is that some layout dependent costs are accounted for only through assumptions, which are discussed in Appendix~\ref{app:fault_tolerant_model}.

The physical operation and measurement times, as well as the physical error rate, are chosen to be compatible with a foreseeable future solid-state platforms \cite{beverland2022assessing}. Given the current state of quantum hardware, we assume that achieving fault-tolerant computation will require physical gate error rates significantly lower than those available today. 
We set $p_{\rm phys}=10^{-3}$, one order of magnitude below the error rates achieved in current memory experiments \cite{google2025quantum}. For simplicity, we impose $t_{\rm mr}=10t_{\rm op}$. We then sweep the physical operation time from $20\,{\rm ns}$ to $20\,{\rm \mu s}$. This wide range is also used to absorb overheads not captured by our assumptions.

\section{Results}
\label{sec:results}

\subsection{Classical and quantum query counts}
\label{sec:query_results}

\begin{figure}[t]
    \centering
    \input{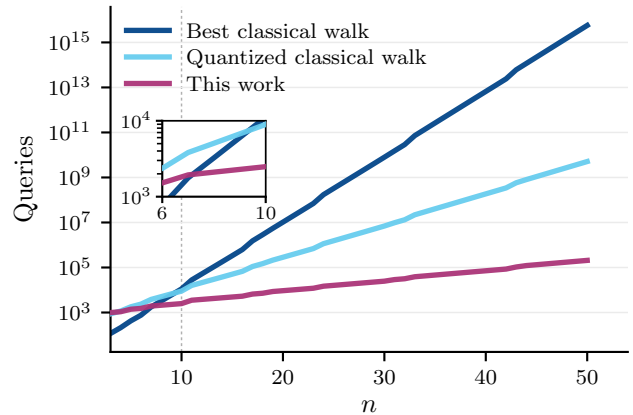}
    \caption{Number of queries as a function of $n$ for $\beta=4$ and $\varepsilon=10^{-2}$. The vertical dashed line separates exact simulations from numerical extrapolations.}
    \label{fig:main_queries}
\end{figure}

We now compare the different methods to sample from dense SK graphs, ranging from purely classical Markov chains, to the fully-quantum walks defined in Sect.~\ref{sec:new_quantum_walk}.
Figure~\ref{fig:main_queries} compares the total query cost as a function of system size, for a reasonably challenging inverse temperature $\beta=4$ and target TV-distance error $\varepsilon=10^{-2}$. The plotted quantity is the cumulative number of queries over the full annealing schedule.

The best classical walk, shown in dark blue, corresponds to the uniform move at this temperature. This is consistent with the low-temperature spectral gap data in Appendix~\ref{app:spectral_gap_queries} and Ref.~\cite{layden2023quantum}.

The quantized classical walk (light-blue line) is obtained by quantizing the same classical proposal. At large $n$'s, its query count is substantially lower than the classical one, as expected from the dependence on the Szegedy phase gap rather than directly on the classical spectral gap. The total annealing cost is not an exact square root of the classical cost, since the schedule, target error, and logarithmic filter factors also contribute; that's why the quantum queries are still larger than the classical ones at small $n$'s. Nevertheless, at $n=50$ the quantized walk improves over the classical baseline by six orders of magnitude.

The quantum Hamiltonian walk, introduced in this work (magenta line), grows much more slowly over the range shown, increasing from about $10^3$ queries at small sizes to about $10^5$ queries at $n=50$. Over the same range, the best classical walk grows by roughly fourteen orders of magnitude, while the quantized classical walk grows by six to seven orders of magnitude. The inset shows that the quantum Hamiltonian walk is already competitive in the numerically accessible regime, and the separation increases rapidly with $n$.

This behavior is consistent with the spectral gap fits in Appendix~\ref{app:spectral_gap_queries:sg}. In the low-temperature regime, the Hamiltonian simulation proposal gives a heuristic cubic improvement in the gap scaling relative to the best classical proposal. Appendix~\ref{app:spectral_gap_queries:cq} isolates this effect in the final annealing step, without summing over the full schedule.
In the main text, we report for simplicity results for 
$\epsilon = 10^{-2}$, corresponding to a  distribution within $1\%$ TVD of the target distribution. This value provides a reasonable representative benchmark for our resource estimates.
Equivalently, a inverse temperature $\beta=4$ is low in the energy scale of the couplings, but not unrealistically low, such that we encounter numerical instabilities in the $\delta_\beta$ calculations.
However,  the same  advantage persists for other low temperatures, starting from $\beta\geq2$, and other target errors, as shown in Appendix~\ref{app:spectral_gap_queries:qq}.

\subsection{Runtime estimates}
\label{sec:runtime_results}

\begin{figure*}[t]
    \centering
    \scalebox{1.5}{\input{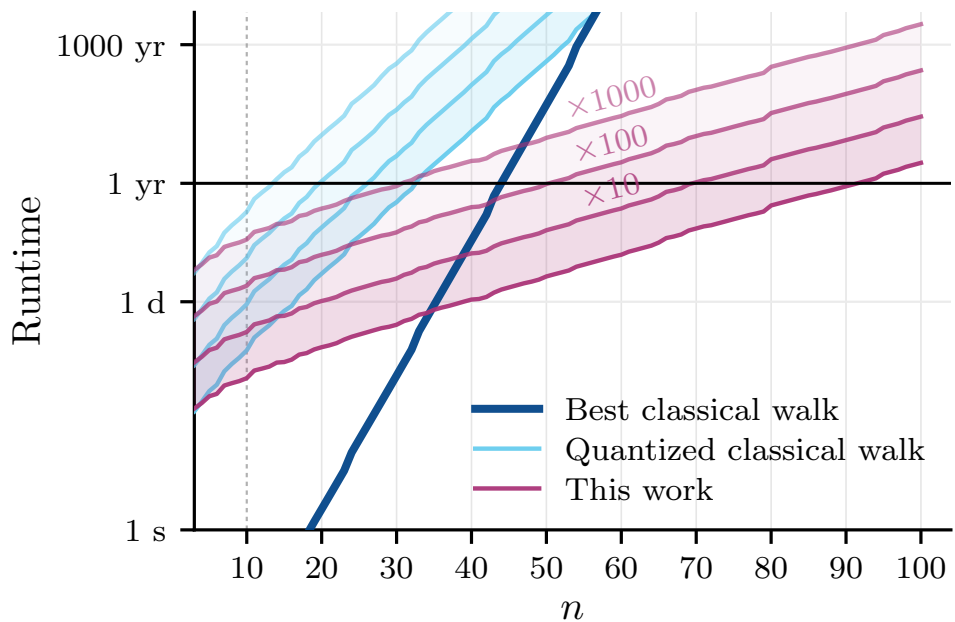}}
    \caption{Runtime versus system size $n$ for $\beta=4$ and TV-distance error $\varepsilon=10^{-2}$. The vertical dashed line separates exact simulations from numerical extrapolations. For the quantum methods, the lowest line assumes a physical operation time of $20\,\mathrm{ns}$, while the subsequent lines increase this time by factors of $10$, $100$, and $1000$. The classical walk is implemented on FPGA hardware.}
    \label{fig:main_runtime}
\end{figure*}

As discussed previously, an asymptotic scaling advantage alone does not necessarily translate into a practical runtime advantage. In particular, the substantial overhead associated with fault-tolerant quantum computing  significantly increases the execution time of quantum algorithms. As a result, the problem size and runtime crossover thresholds may be impractically large.
Here we show that quantum walks are a paradigmatic setting in which the gap between asymptotic and practical performance is evident. The fully quantum walks introduced here bridge this gap by substantially lowering the runtime crossover threshold.

Figure~\ref{fig:main_runtime} shows the runtime estimate as a function of the system size $n$ for the three main walks considered in the text.
These results are obtained by combining the query costs simulations of the previous Sect.~\ref{sec:query_results} with the cost of a single logical classical and quantum steps (see Sect.~\ref{sec:classical_baselines},  \ref{sec:walk_circuit_implementation} and  \ref{sec:fault_tolerant_model}). We choose the hybrid fixed-point arithmetic for the quantum walk operator, which is the most time efficient implementation albeit the space scales about quadratically in the space. 
Additional experiments at different $\beta$ and $\varepsilon$ are shown in Appendix~\ref{app:runtime_plots}. 

The best classical walk, the one with uniform move, is represented by a thick light blue line. Here we have determined that the classical device giving the best performance is a state-of-the-art FPGA hardware thanks to the logarithm depth fixed point arithmetic mitigating the expensive energy calculation routine. The small prefactors for the single step, in hundreds of nanoseconds, gives it the advantage in the small $n$ regime. This represents our classical benchmark.

In contrast to the classical case, it is not possible to provide a single definitive runtime prediction for the quantum algorithms, as large-scale fault-tolerant quantum computers are not yet available. We therefore show the quantized best classical walk in dark blue and the quantum Hamiltonian walk in magenta under a range of hardware assumptions. The lowest line corresponds to an optimistic physical operation time of $20\,\mathrm{ns}$, while the remaining lines increase this time by factors of $10$, $100$, and $1000$; the most pessimistic line therefore corresponds to a physical operation time of $20\,\mu\mathrm{s}$. In our cost model, described in Sect.~\ref{sec:fault_tolerant_model}, these multiplicative factors may also be interpreted as effective overheads arising from the intermediate software, scheduling, control, and quantum error correction layers~\cite{beverland2022assessing}.

The main result is that, as expected, simply quantizing the best classical walk is not competitive: its runtime exceeds one year already for relatively small system sizes (for which also direct classical enumeration is possible). Even under an optimistic hardware scenario, the crossover runtime occurs only after approximately $1000$ years.
This observation is in qualitative agreement with Ref.~\cite{lemieux2020efficient}, although that work is framed in a different and less challenging model instances.

In contrast, the method introduced here achieves substantially lower runtimes owing to its much more favorable scaling. As a result, under the \textit{same} optimistic hardware assumptions, the crossover runtime falls below one day---approximately six orders of magnitude shorter than that of the best previously proposed quantum walk method. Taking one year of wall-clock time as a symbolic practicality threshold, the new fully quantum walk extends the range of tractable simulations from approximately up to $n \simeq 90$ (optimistic).

\section{Discussion}
\label{sec:discussion}

We introduce a new class of \emph{fully quantum} Metropolis walks 
where both the proposal and acceptance steps are genuinely quantum in character.
We target the general problem of sampling from dense graphs, defined through classical disordered spin Hamiltonians.

The main outcomes include: \emph{(i)} the common wisdom that quantum walks are fundamentally limited to quadratic speedups, compared to the best classical walk, is not intrinsic to the framework itself, but rather to the restriction of quantizing classically efficient Markov chains; 
\emph{(ii)} this unlocks a much larger asymptotic polynomial scaling advantage, in the number of queries, compared to the previous best quantum walk, which enables now a reasonable runtime crossover (e.g. 1 day compared to 1000 years, under identical hardware constraints). 
\emph{(iii)} Rather than relying only on asymptotic query complexity, we perform a complete fault-tolerant compilation of all algorithmic primitives and derive end-to-end runtime estimates.
Equally importantly, we benchmark against optimized CPU, GPU, and FPGA implementations of the best  classical walk, allowing for a realistic assessment of the crossover between classical and quantum hardware.

To ensure reproducibility, all resource-estimation notebooks with high and low-level implementation details are provided in an online repository~\cite{myrepo2026}.

Notice that, we choose to include a concrete worst-case resource estimate to provide an initial assessment of the practical feasibility of the algorithm. However, our runtime and memory estimates are necessarily based on the current models of fault-tolerant architectures~\cite{beverland2022assessing}, and states of classical algorithms, and should therefore be regarded as representative rather than definitive. We expect continued advances in quantum hardware, error-correction protocols, compiler and scheduler optimizations, and algorithmic implementations to further reduce the quantum resource requirements. Likewise, also specialized classical implementations may shift the crossover point between classical and quantum approaches. As has been the case for Shor's algorithm, whose resource estimates have been repeatedly revised as the underlying algorithms and fault-tolerant techniques improved~\cite{GidneyEkera2021,Gidney2025RSA2048,cain2026shorsalgorithmpossible10000}, we expect the present estimates to evolve over time and provide a reasonable starting point rather than a final assessment.

Our work presents a different outlook on the practicality of quantum walks than previous resource analyses, most notably that of Lemieux \textit{et al.}~\cite{lemieux2020efficient}. Whereas their study assessed standard quantized local spin-flip walks for optimization on sparse graphs and concluded that fault-tolerant implementations are unlikely to outperform classical algorithms in practice, we show that quantum walks can in fact achieve a practical runtime advantage.

Our method can likely be further improved within the same general framework, for instance by considering different unitary proposals beyond the hamiltonian simulation with the transverse-field mixing operator adopted here. An important aspect of the arithmetic implementation is minimizing its impact on the runtime. This is achieved with logarithmic depth, at the expense of quadratic space in $n$. However, space efficient implementations are also possible, and a study of how arithmetic precision affects the Markov chain could open further possibilities for implementations on early fault-tolerant devices \cite{katabarwa2024early}.
We leave the exploration of these directions for future works. 

In this work we focused on Sherrington--Kirkpatrick dense graph instances as a challenging and controllable benchmark model. It would be interesting to explore how the asymptotic scaling, as well as the crossover system size and runtime, depend on the sparsity and structure of the underlying graph, to understand which classes of real-world problems may benefit first from this acceleration. For instance, dense interaction graphs naturally arise in biological network, gene regulatory and gene co-expression networks~\cite{barabasi2011network,aurora_genomics_2025,hoffmann2024network, dubovitskii2026experimentalimplementationdiscretetimequantum}, as well as in probabilistic graphical models for Bayesian inference~\cite{wainwright2008graphical,lecun_deep_2015}.
In such settings, one should also compare against the best available classical sampling methods at a given target approximation error. Extensions to optimization also represent a natural direction for future works~\cite{somma:2008,lemieux2020efficient,abbas2024quantumoptimization,marshall2026quantum}.

Polynomial quantum advantages beyond the quadratic regime remain comparatively rare, with examples achieving quartic or larger polynomial improvements are known only in a limited number of settings \cite{Hastings2020TensorPCA,Schmidhuber2025QuarticPlantedInference,SchmidhuberZlokapa2025CommunityDetection,Ambainis2016GappedGroupTesting,Ambainis2017PointerFunctions}.

All in all, this new quantum algorithm further reinforces the connection between quantum computing and (classical) Monte Carlo~\cite{szegedy:2004, Mazzola_2024, montanaro2015quantum}, and places quantum sampling among the most promising routes towards practical quantum advantage in the fault-tolerant regime.

\textbf{Acknowledgements.} We acknowledge useful feedback on an earlier version of the manuscript from Alexandru Paler and Scott Aaronson.
We acknowledge Cesare Cozza for providing support for the GPU runtime estimate.
We acknowledge computational resources from the CINECA award under the ISCRA initiative.

\textbf{Data and code availability.} The code and data supporting this work are available at \url{https://github.com/incud/pubmonaqa}~\cite{myrepo2026}.

\bibliography{references} 

\clearpage
\appendix

\onecolumngrid

\section{Background on Markov-chain Monte Carlo}
\label{app:mcmc}

This appendix complements the definitions and properties of Markov chains considered in Section~\ref{sec:mcmc_sampling} and proves the spectral gap bound on convergence. For further background, we refer the reader to Ref.~\cite{levin2017markov}.

\subsection{Ergodicity and reversibility}

A finite Markov chain is irreducible if every configuration can be reached from every other configuration with positive probability after finitely many steps. It is aperiodic if the possible return times to a configuration have greatest common divisor one. A chain that is both irreducible and aperiodic is ergodic. For a finite chain, ergodicity implies convergence to the unique stationary distribution from any initial distribution.

Reversibility with respect to $\pi_\beta$ means detailed balance,
\begin{equation}
    \pi_\beta(x)(P_\beta)_{yx}
    =
    \pi_\beta(y)(P_\beta)_{xy}.
\end{equation}
Summing over $x$ gives
\begin{equation}
    (P_\beta\pi_\beta)(y)
    =
    \sum_x (P_\beta)_{yx}\pi_\beta(x)
    =
    \pi_\beta(y)\sum_x (P_\beta)_{xy}
    =
    \pi_\beta(y),
\end{equation}
where the last equality uses the column-stochastic convention. Thus detailed balance makes $\pi_\beta$ stationary.

\subsection{Spectral-gap control of convergence}

We now justify the use of $\delta_\beta^{-1}$ as a proxy for the number of Markov-chain steps. Define the discriminant matrix
\begin{equation}
    X_\beta
    =
    \operatorname{diag}(\pi_\beta)^{-1/2}
    P_\beta
    \operatorname{diag}(\pi_\beta)^{1/2}.
\end{equation}
For a reversible chain, $X_\beta$ is symmetric, has entries
\begin{equation}
    (X_\beta)_{yx}
    =
    \sqrt{(P_\beta)_{yx}(P_\beta)_{xy}},
\end{equation}
and is similar to $P_\beta$, hence has the same spectrum.

For an initial distribution $q_0$, define
\begin{equation}
    g_t(x)
    =
    \frac{q_t(x)-\pi_\beta(x)}{\sqrt{\pi_\beta(x)}} .
\end{equation}
Since $q_t=P_\beta^t q_0$, one has $g_t=X_\beta^t g_0$. The vector $g_0$ is orthogonal to the stationary eigenvector, so the absolute spectral gap $\delta_\beta$ gives
\begin{equation}
    \|g_t\|_2
    \leq
    (1-\delta_\beta)^t\|g_0\|_2 .
\end{equation}
Using Eq.~\ref{eq:tv_distance},
\begin{align}
    \|P_\beta^t q_0-\pi_\beta\|_{\rm TV}
    &=
    \frac{1}{2}
    \sum_x
    \sqrt{\pi_\beta(x)}\,|g_t(x)| \\
    &\leq
    \frac{1}{2}
    \left(\sum_x \pi_\beta(x)\right)^{1/2}
    \left(\sum_x |g_t(x)|^2\right)^{1/2} \\
    &=
    \frac{1}{2}\|g_t\|_2 \\
    &\leq
    \frac{1}{2}\|g_0\|_2(1-\delta_\beta)^t \\
    &\leq
    \frac{1}{2}\|g_0\|_2 e^{-\delta_\beta t}.
    \label{eq:tv_spectral_gap_bound}
\end{align}
Therefore, reaching total variation error at most $\varepsilon$ is guaranteed whenever
\begin{equation}
    t
    \geq
    \frac{1}{\delta_\beta}
    \log\left(
        \frac{\|g_0\|_2}{2\varepsilon}
    \right).
    \label{eq:mixing_time_gap_bound}
\end{equation}
This is the logarithmic-factor statement used in the main text: the leading dependence is governed by $\delta_\beta^{-1}$, while the initial distribution and target error enter only through the logarithm.


\section{Sherrington--Kirkpatrick model}
\label{app:sk_model}

Here we compute quantities of interest for the Sherrington--Kirkpatrick model defined in Section~\ref{sec:sk_instances}. We refer to the Hamiltonian in Eq.~\ref{eq:sk_hamiltonian} and the normalization factor $\alpha$ in Eq.~\ref{eq:sk_alpha}.

\subsection{Scaling of the coefficient $1$-norm}
\label{app:sk_model:one_norm}

We estimate the coefficient $1$-norm of the normalized SK Hamiltonian. This is used in Appendix \ref{app:fixed_point_errors} to normalize energy differences. If $H=\sum_\ell \tilde c_\ell P_\ell$, the one-norm is $\|\tilde c\|_1 = \sum_\ell |\tilde c_\ell|$. 

Let $c\in\mathbb R^M$ be the vector of unnormalized coefficients, with $M=n+\binom{n}{2}$. Using the normalization in Eq.~\ref{eq:sk_alpha}, the normalized coefficient vector is
\begin{equation}
    \tilde c
    =
    \alpha c
    =
    \frac{\sqrt n}{\|c\|_2} \, c.
\end{equation}
Therefore,
\begin{equation}
    \|\tilde c\|_1
    =
    \alpha\|c\|_1
    =
    \frac{\sqrt n}{\|c\|_2} \, \|c\|_1.
\end{equation}
For a typical Gaussian vector, $\|c\|_1\simeq M\sqrt{2/\pi}$ and $\|c\|_2\simeq\sqrt M$. Hence
\begin{equation}
\label{eq:upper_bound_onenorm_h}
    \|\tilde c\|_1
    \simeq
    \frac{\sqrt n}{\sqrt M} \,M\sqrt{2/\pi} 
    \simeq
    \frac{n^{3/2}}{\sqrt\pi}.
\end{equation}

\subsection{Scaling of the operator norm}

We next estimate the operator norm of the normalized SK Hamiltonian. It is the relevant energy scale for the Gibbs distribution and for the interval on which the Boltzmann coin polynomial is approximated.

\begin{figure}[hbp]
    \centering
    \input{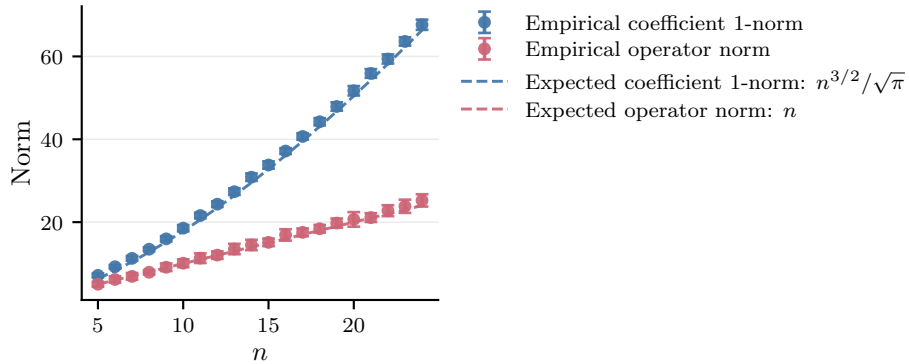}
    \caption{
    Empirical coefficient $1$-norm and operator norm of $30$ independent random instances for each system size $n=5,\ldots,24$, compared with the expected scalings $n^{3/2}/\sqrt{\pi}$ and $n$. Error bars show one standard deviation.
    }
    \label{fig:sk_upper_bounds}
\end{figure}

The operator norm, equivalently the spectral norm for the diagonal Hamiltonian in Eq.~\ref{eq:sk_hamiltonian}, is
\begin{equation}
    \|H\|_{\rm op}
    =
    \max_x |H(x)| .
\end{equation}
For the SK normalization used here, the extremal energies are extensive, so $\|H\|_{\rm op}=O(n)$. We validate this linear scaling numerically on the sampled instances in Fig.~\ref{fig:sk_upper_bounds}.

\subsection{Difference from the  model considered in Lemieux et al.}

The SK model used here is different from the setting considered in Ref.~\cite{lemieux2020efficient}. The construction of Ref.~\cite{lemieux2020efficient} applies to $(k,d)$-local Ising Hamiltonians with bounded locality and bounded degree. In particular, they consider Hamiltonians of the form
\begin{equation}
    H(x)
    =
    \sum_{\ell=1}^{M}
    J_\ell
    \prod_{i\in S_\ell} x_i ,
    \qquad
    |S_\ell|\leq k ,
    \label{eq:kd_local_ising}
\end{equation}
where each spin appears in at most $d$ interaction terms. Importantly, $k$ and $d$ are constants.

This differs from the dense SK Hamiltonian in Eq.~\ref{eq:sk_hamiltonian}. The SK Hamiltonian is still two-local, so $k=2$, but each spin interacts with all the others, hence $d=O(n)$.

This distinction affects the arithmetic cost. In the $(k,d)$-local setting of~\cite{lemieux2020efficient}, the Boltzmann coin for a local move depends only on the neighborhood $\mathcal N_j$ of the flipped spin, i.e. the spins that appear in Hamiltonian terms whose value changes when spin $j$ is flipped. For single-spin flips, $|\mathcal N_j|\leq kd$. This allows an arithmetic construction with depth $2^{|\mathcal N_j|}\leq 2^{kd}$, which is constant in their setting.

Applying the same approach to the SK model, and especially to nonlocal moves, can require a neighborhood containing all spins, making the construction infeasible.

\section{Spectral gap and Monte Carlo query count calculations}
\label{app:spectral_gap_queries}

The spectral gap of a transition matrix $P_\beta$ built from a given proposal move can be used as a proxy for the convergence time of the Markov chain. Here we compute the spectral gap for the proposal moves of interest on an instance set containing $100$ independent instances for each number of spins $n=3,\ldots,10$. We test the inverse-temperature grid $\beta
\in
\{0.01,0.02,\ldots,0.10,
0.20,0.30,\ldots,1,
2,3,\ldots,10,
20,30,\ldots,100\}$. Whenever a fit of the form $f_\beta(n)=C_\beta 2^{-\nu_\beta n}$ is required at a value of $\beta$ not included in the grid, we fit the two available values that bracket it and linearly interpolate $\log C_\beta$ and $\nu_\beta$.

\subsection{Spectral gap plots}
\label{app:spectral_gap_queries:sg}

\begin{figure}[p]
\centering
\scalebox{0.55}{\input{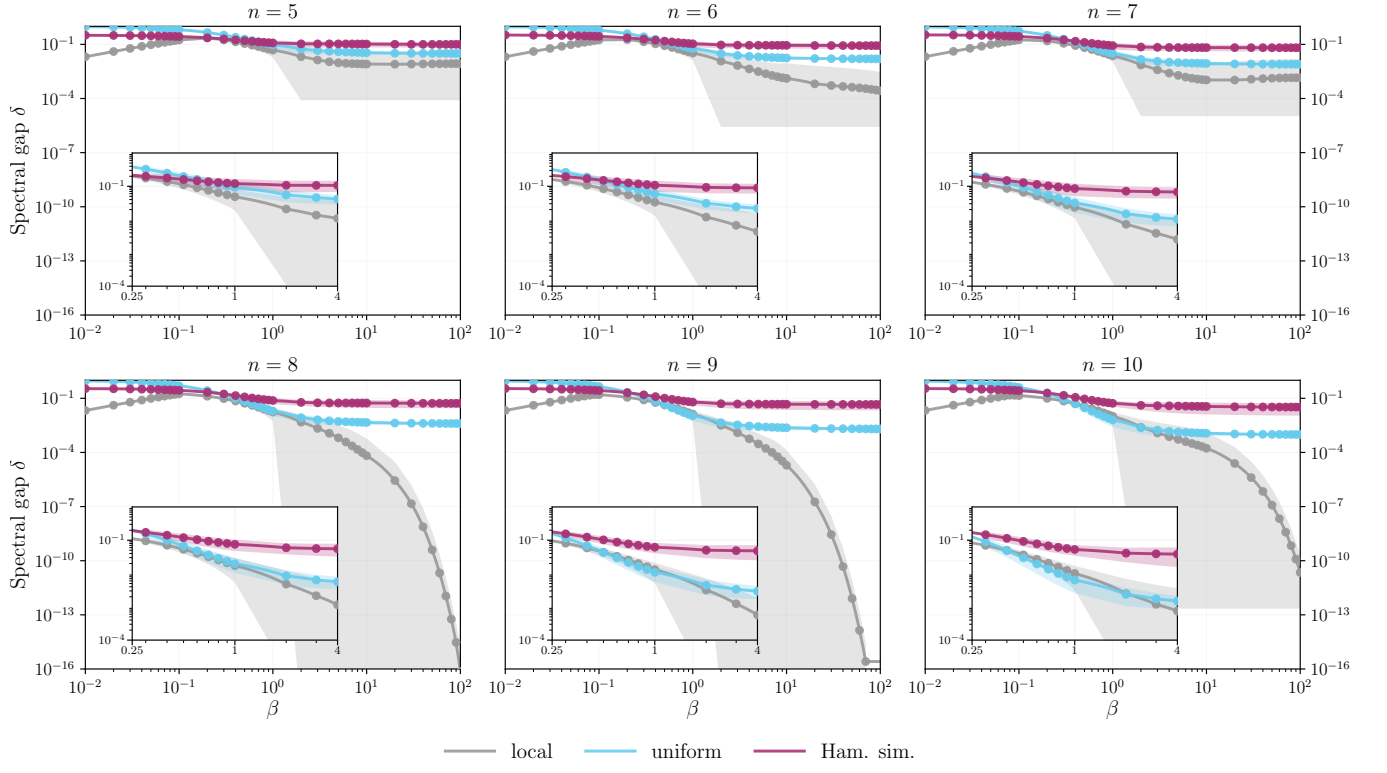}}
\caption{Spectral gap as a function of $\beta$ for a fixed number of spins $n$. Grey, light blue and magenta lines correspond to the local, uniform, and Hamiltonian simulation moves, respectively. The shaded areas show one standard deviation. The statistics shown are the mean and standard deviation.}
\label{fig:spectral_gap_beta_meanstd}
\end{figure}

\begin{figure}[p]
\centering
\scalebox{0.55}{\input{paperimg/spectral_gap_vs_beta_notail.pgf}}
\caption{Same as Figure~\ref{fig:spectral_gap_beta_meanstd}, but with the statistics restricted to the central quartiles.}
\label{fig:spectral_gap_beta_meanstd_notail}
\end{figure}

\begin{figure}[p]
\centering
\scalebox{0.60}{\input{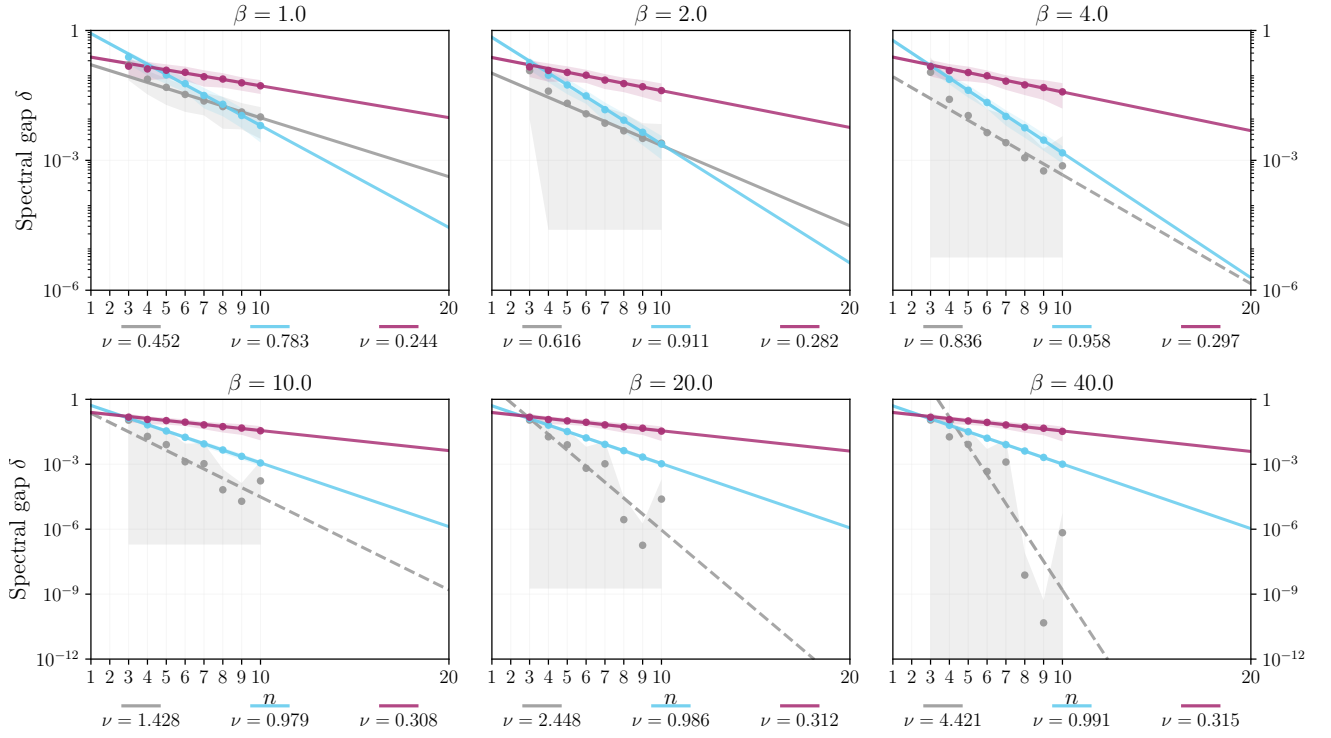}}
\caption{Spectral gap as a function of $n$ at fixed inverse temperature $\beta$. Grey, light blue and magenta lines correspond to the local, uniform, and Hamiltonian simulation moves, respectively. The lines are fitted to $f(n)=C\,2^{-\nu n}$ using the data for $n=5,\ldots,10$, with $\nu$ reported below each line. Points show the mean over instances and shaded regions indicate one standard deviation. Local-move fits are dashed for $\beta\geq4$ because they are affected by the numerical artifact discussed in the text.}
\label{fig:spectral_gap_n_meanstd}
\end{figure}
\afterpage{\clearpage}

\begin{figure}[p]
\centering
\scalebox{0.60}{\input{paperimg/spectral_gap_vs_n_notail.pgf}}
\caption{Same as Figure~\ref{fig:spectral_gap_n_meanstd}, but with the statistics restricted to the central quartiles.}
\label{fig:spectral_gap_n_meanstd_notail}
\end{figure}
\afterpage{\clearpage}

The spectral gap is computed with \texttt{numpy.linalg.eigvalsh} on the discriminant matrix $X$. For reversible chains, $X$ is easy to construct and has the same spectrum as $P_\beta$. Since $X$ is symmetric by construction, its eigenvalues can be computed with a Hermitian eigensolver, which is more numerically stable than a generic eigensolver and avoids small imaginary components that would otherwise have to be truncated.

Figure~\ref{fig:spectral_gap_beta_meanstd} shows the spectral gap of the three trial moves as a function of $\beta$. At high temperature, $\beta \le 0.1$, the best classical baseline is the uniform move: almost all proposals are accepted, and the uniform move mixes faster because it allows long-range jumps, whereas the local move changes only one spin at a time. Around $\beta=1$, the local move outperforms the uniform move on average. At lower temperatures, $\beta \ge 4$, locality becomes a bottleneck: escaping from a local minimum requires uphill moves whose acceptance probabilities are exponentially suppressed in $\beta \Delta H$, creating metastable regions that are hard to leave. Consequently, the spectral gap of the local move decreases again, and the uniform move becomes better. In the regime $\beta>1$, the local move is also affected by a large spread across instances. The Hamiltonian simulation move surpasses the best classical approach above the critical threshold. 

Figure~\ref{fig:spectral_gap_beta_meanstd_notail} shows the same plots, with the average and spread computed as the mean and standard deviation over the central quartiles only.

Figure~\ref{fig:spectral_gap_n_meanstd} shows the scaling of the spectral gap with $n$ at fixed inverse temperature $\beta$. The lines are fitted to $f(n)=C\,2^{-\nu n}$ using the data for $n=5,\ldots,10$, with the fitted value of $\nu$ reported below each line. The points show the mean over instances, while the shaded regions indicate one standard deviation.

Some unreliable points occur for the local move at large inverse temperature, e.g. at $\beta=4$. In this regime, some hard instances have spectral gaps close to machine precision and therefore numerically resemble reducible chains. These instances must be excluded from the statistics. Consequently, the mean at the largest values of $n$, particularly at $n=10$, is computed only over the remaining, easier instances, whereas the means at smaller $n$ include a broader set of instances. This selection effect causes the spectral gap to plateau artificially, biases the fitted line upward, and can make the local move appear to decay more slowly than the uniform move. For this reason, the local move fits are shown as dashed lines for $\beta\geq4$.

Figure~\ref{fig:spectral_gap_n_meanstd_notail} reports the same analysis with the statistics restricted to the central quartiles. Removing both the hardest and the easiest instances reduces the variance and delays the onset of the numerical artifact to larger values of $\beta$. For example, the behavior at $\beta=2$ becomes more stable, while similar distortions appear only at much lower temperatures than $\beta = 4$. This figure should therefore be interpreted as describing typical instances rather than the average behavior over the full instance distribution.

The plots are compatible with those shown in Ref.~\cite{layden2023quantum} for dense Ising models; see Supplementary Figs.~S2 and S3, and Supplementary Table~S1.

\subsection{Classical queries plot}
\label{app:spectral_gap_queries:cq}

\begin{figure}[p]
    \centering
    \scalebox{0.60}{\input{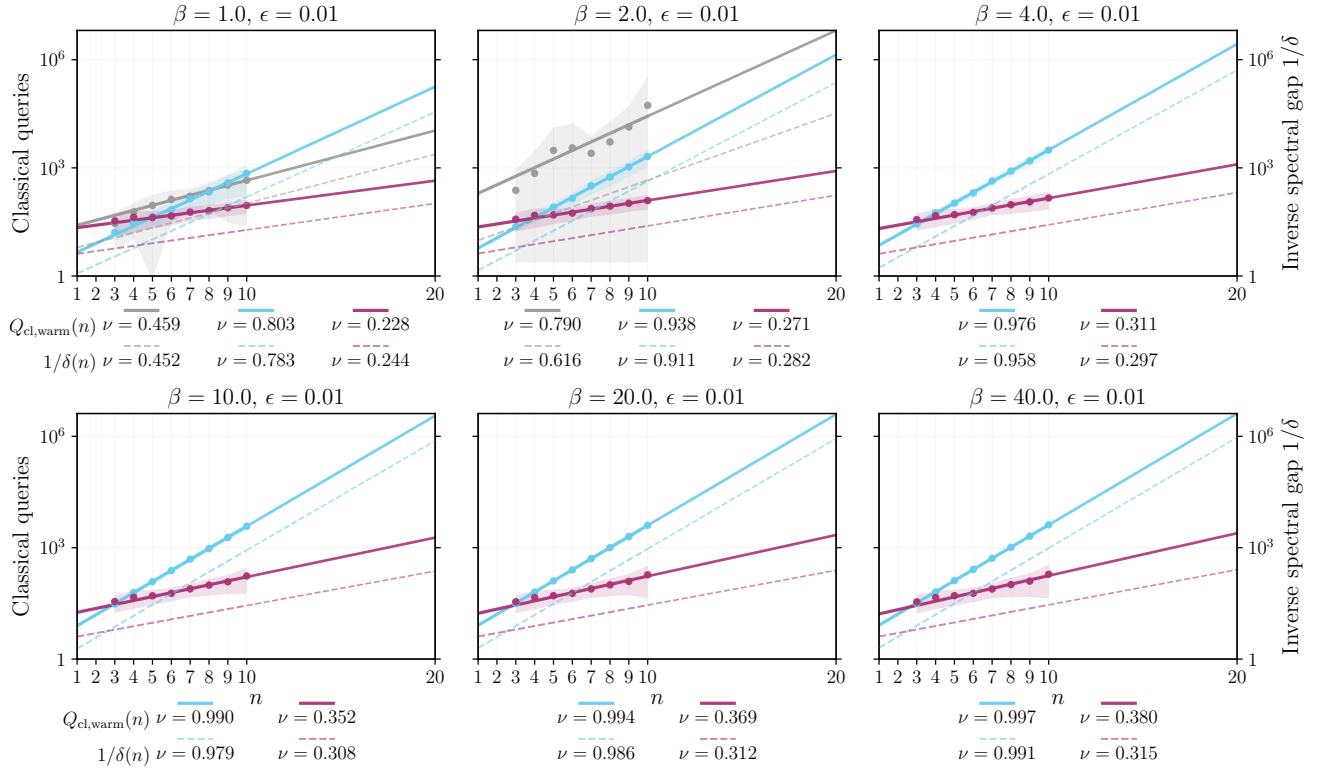}}
    \caption{Classical query count $Q_{\rm cl,warm}(n,\beta,\varepsilon)$ at fixed inverse temperature $\beta$ and TV-distance error $\varepsilon=0.01$. Thick solid lines show the fitted query counts, while thin dashed lines show the corresponding inverse spectral gap fits. Grey, light blue and magenta lines correspond to the local, uniform, and Hamiltonian simulation moves, respectively. Local move are omitted for $\beta\geq4$.}
    \label{fig:classical_queries_vs_n}
\end{figure}

\begin{figure}[p]
    \centering
    \scalebox{0.60}{\input{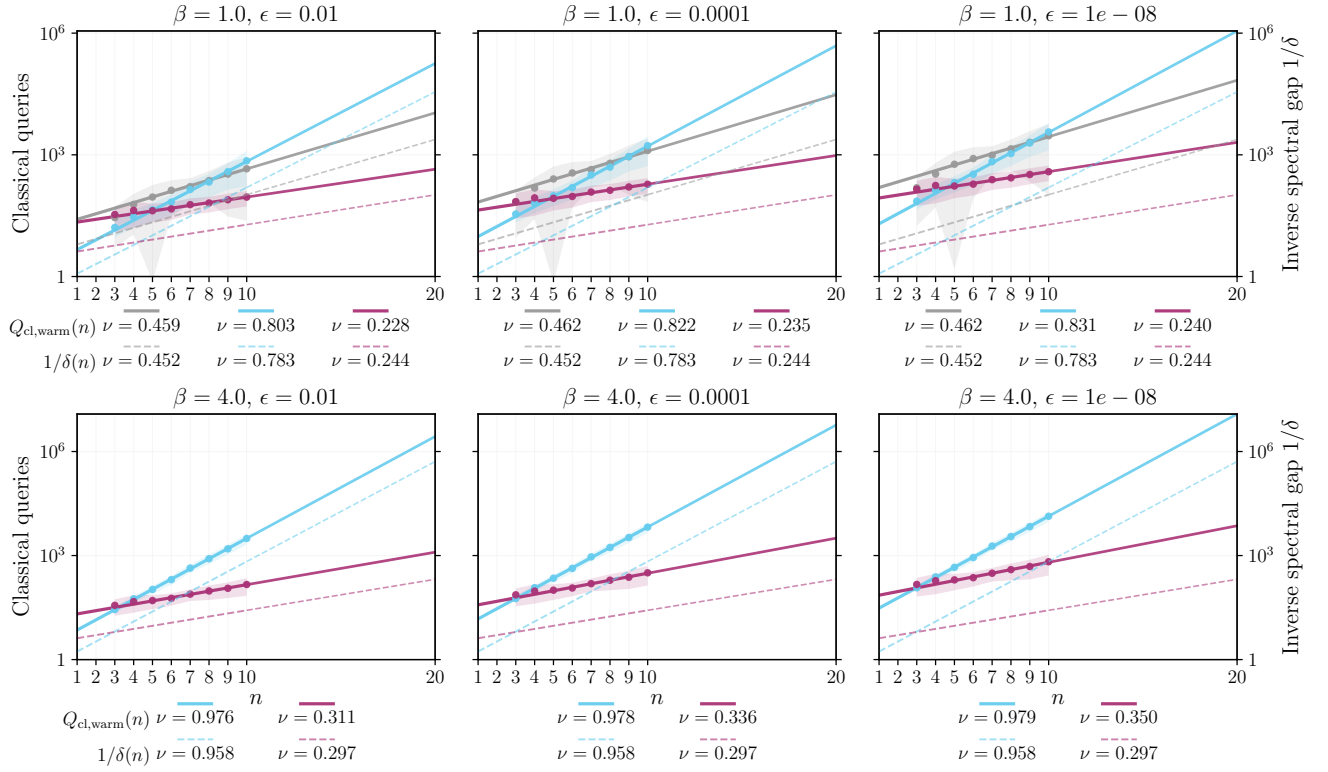}}
    \caption{Same as Figure~\ref{fig:classical_queries_vs_n}, but for several values of the target TV-distance error $\varepsilon$ at fixed inverse temperature $\beta$.}
    \label{fig:classical_queries_vs_n_all_epsilon}
\end{figure}

For each transition matrix $P$, the classical query count is defined as the first time $t$ at which the chain is within TV distance $\varepsilon$ from stationarity. Starting from an initial distribution $q_0$, the direct evolution is $q_t=P^tq_0$, and the convergence time is the smallest $t$ satisfying $\frac{1}{2}\lVert q_t-\pi\rVert_1\leq\varepsilon$. When the stationary distribution is numerically well conditioned, we exploit the reversibility of the chain and instead work with the symmetric discriminant matrix $X$, which has the same spectrum as $P$ and can be diagonalized using a Hermitian eigensolver. In this representation, the weighted deviation $g_0=(q_0-\pi)/\sqrt{\pi}$ evolves as $g_t=X^tg_0$, and the TV distance is computed as $\frac{1}{2}\sum_x\sqrt{\pi_x}\,|g_t(x)|$. If some entries of $\pi$ are too small, this rescaling becomes numerically unstable, and we revert to the direct evolution $q_t=P^tq_0$, with powers computed by repeated squaring using \texttt{numpy.linalg.matrix\_power}.

Figure~\ref{fig:classical_queries_vs_n} shows the number of classical queries required to reach TV distance $\varepsilon=0.01$ from the target distribution. To compare the query count directly with the spectral gap, we consider a single annealing step rather than the full annealing path, which would require summing contributions from different inverse temperatures. The initial distribution is therefore chosen as the preceding Gibbs distribution $\pi_{\beta_0}$, with $\beta_0<\beta$, where $\beta_0$ is the smallest inverse temperature in $[0,\beta)$ such that the Bhattacharyya coefficient $\sum_x\sqrt{\pi_{\beta_0}(x)\pi_\beta(x)}$ is at least $\sqrt{1/e}$.

The points show the mean over instances, while the shaded regions indicate one standard deviation. The query-count data are fitted to $f(n)=C\,2^{\nu n}$ using the points from $n=5$ to $10$. Thick solid lines show these fits, while thin dashed lines show the corresponding inverse spectral-gap fits obtained from the previous analysis. 

This comparison highlights the similar scaling of the classical query count and the inverse spectral gap. The two quantities are not identical: the inverse gap characterizes the slowest possible convergence mode, whereas the observed query count also depends on the overlap of the chosen initial distribution with that mode.

For $\beta\geq4$, the local move are omitted because their large spread and the numerical issues discussed above make the fitted scaling unreliable. This omission does not affect the best classical baseline in this regime, which is given by the uniform move.

The behavior is robust across different target errors. Figure~\ref{fig:classical_queries_vs_n_all_epsilon} reports the same analysis for several values of $\varepsilon$ at fixed $\beta$, showing that the comparison between the query-count and inverse-gap scalings is maintained as the target accuracy is increased.

\subsection{Quantum queries plot}
\label{app:spectral_gap_queries:qq}

The quantum query count is obtained by fitting the spectral gap at each fixed target inverse temperature $\beta$ in the form $\delta_\beta(n)=c_\beta 2^{-\nu_\beta n}$.  The fit is performed as in the previous sections: only the largest sizes are used, $n=5,\ldots,10$. 

\begin{figure}[h]
    \centering
    \scalebox{0.60}{\input{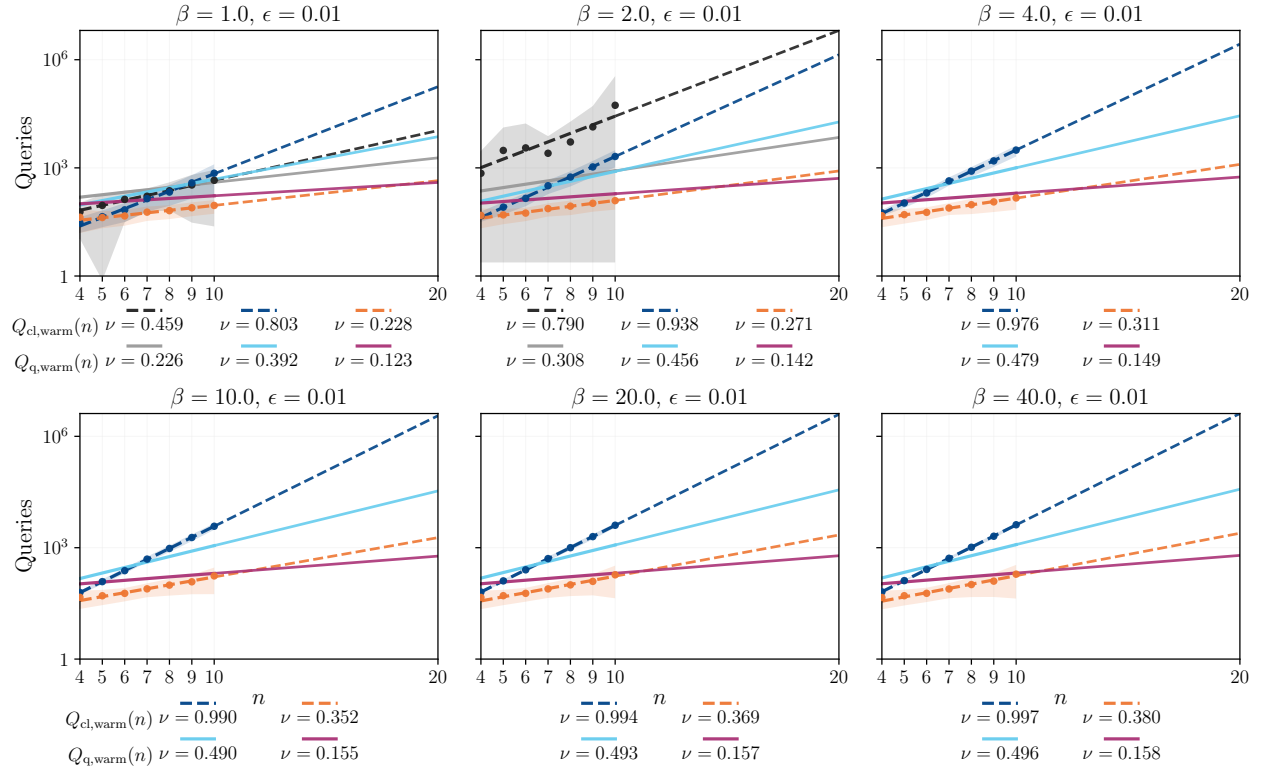}}
    \caption{Classical (dashed) and quantum (solid) query counts, $Q_{\rm cl,warm}(n,\beta,\varepsilon)$ and $Q_{\rm q,warm}(n,\beta,\varepsilon)$, at fixed inverse temperature $\beta$ and TV-distance error $\varepsilon=0.01$. Dashed gray, dark-blue, and orange lines correspond, respectively, to classical Monte Carlo with local, uniform, and Hamiltonian simulation moves; solid gray, light-blue, and magenta lines correspond, respectively, to the quantum-walk-based approach with the same three moves. Local move are omitted for $\beta\geq4$.}
    \label{fig:quantum_queries_vs_n}
\end{figure}

Figure~\ref{fig:quantum_queries_vs_n} compares the number of classical and quantum queries for each proposal move. These plots describe only the \emph{last} step of the annealing schedule, for both the classical and quantum algorithms. The dashed black, dark-blue, and orange lines correspond, respectively, to the classical Monte Carlo approach with local, uniform, and Hamiltonian simulation moves. The latter is applied stepwise following the approach of Ref.~\cite{layden2023quantum}. The solid gray, light-blue, and magenta lines correspond, respectively, to the quantum-walk-based algorithm with local, uniform, and Hamiltonian simulation moves. This help us having a clearer visualization of the number of quantum queries being the approximately the square root of the classical ones. At lower temperatures, moving from the best classical Monte Carlo method with a uniform move (dashed dark blue) to the classical Monte Carlo method with a Hamiltonian simulation move (dashed orange) yields a heuristic cubic speedup. Moving from either classical method to its quantum-walk counterpart, from dark blue to light blue or from orange to magenta, yields a quadratic speedup. Combined, these improvements yields a sextic speedup.

For a single step only the quantum algorithm does not use the annealing rewind procedure. Instead, the query count is the expected number of projection attempts, $1/p$, multiplied by the degree of the spectral filter:
\begin{equation}
    Q_{\rm q,warm}(n,\beta,\varepsilon)
    =
    \frac{1}{p} \times
    \underbrace{
    \frac{
        2\log_2\!\left(1/\varepsilon_{\rm filter}\right)
    }{
        \theta_\beta(n)
    }
    }_{\text{filter degree}}.
\end{equation}
Here $p=1/e$ is the lower bound on the squared overlap between the warm-start input and the target coherent Gibbs state. Note here the error is not the TV-error $\varepsilon$ but instead the portion of such error budgetted for the filtering of the unwanted modes, $\varepsilon_{\rm filter}=\varepsilon/4$, as per the model detailed later in Appendix \ref{app:annealed_spectral_filtering}. Similarly to the previous section, changing target errors does not perturbate the relation between curves, this is shown in Figure~\ref{fig:quantum_queries_vs_n_all_epsilon}.

Figure~\ref{fig:quantum_queries_vs_n_all_epsilon_full_annealing} compares the classical and quantum query counts accumulated over the complete annealing schedule. The schedule $\{\beta_j\}_{j=0}^{L}$ is generated using the ansatz described in Appendix~\ref{app:annealed_spectral_filtering}, with consecutive coherent Gibbs states chosen to have squared overlap at least $p=1/e$. The rewind procedure introduces an expected overhead $1+1/p$ at each annealing step. Moreover, the spectral-filter error budget $\varepsilon/4$ is distributed uniformly over the expected $L(1+1/p)$ filter applications, giving $\varepsilon_{\rm filter}=(\varepsilon/4)/(L(1+1/p))$. The expected total number of calls to $W$ or $W^\dagger$ is therefore
\begin{equation}
    Q_{\rm q}(n,\bar\beta,\varepsilon)
    =
    \underbrace{
    \left(1+\frac{1}{p}\right)
    }_{\text{rewind overhead}}
    \sum_{j=1}^{L}
    \underbrace{
    \frac{
        2\log_2\!\left(1/\varepsilon_{\rm filter}\right)
    }{
        \theta_{\beta_j}(n)
    }
    }_{\text{filter degree at step }j},
    \label{eq:def_Q_q_appendix_c}
\end{equation}
Note that the lines in this figure are less smooth than those in the previous plots because the query counts are summed over the full annealing schedule. An increase in the number of annealing steps therefore produces a sharp increase in the total query count.

\begin{figure}[h]
    \centering
    \scalebox{0.60}{\input{paperimg/classical_and_quantum_queries_all_epsilon.pgf}}
    \caption{Same as Figure~\ref{fig:quantum_queries_vs_n}, but for several values of the target TV-distance error $\varepsilon$ at fixed inverse temperature $\beta$.}
    \label{fig:quantum_queries_vs_n_all_epsilon}
\end{figure}

\begin{figure}[h]
    \centering
    \scalebox{0.60}{\input{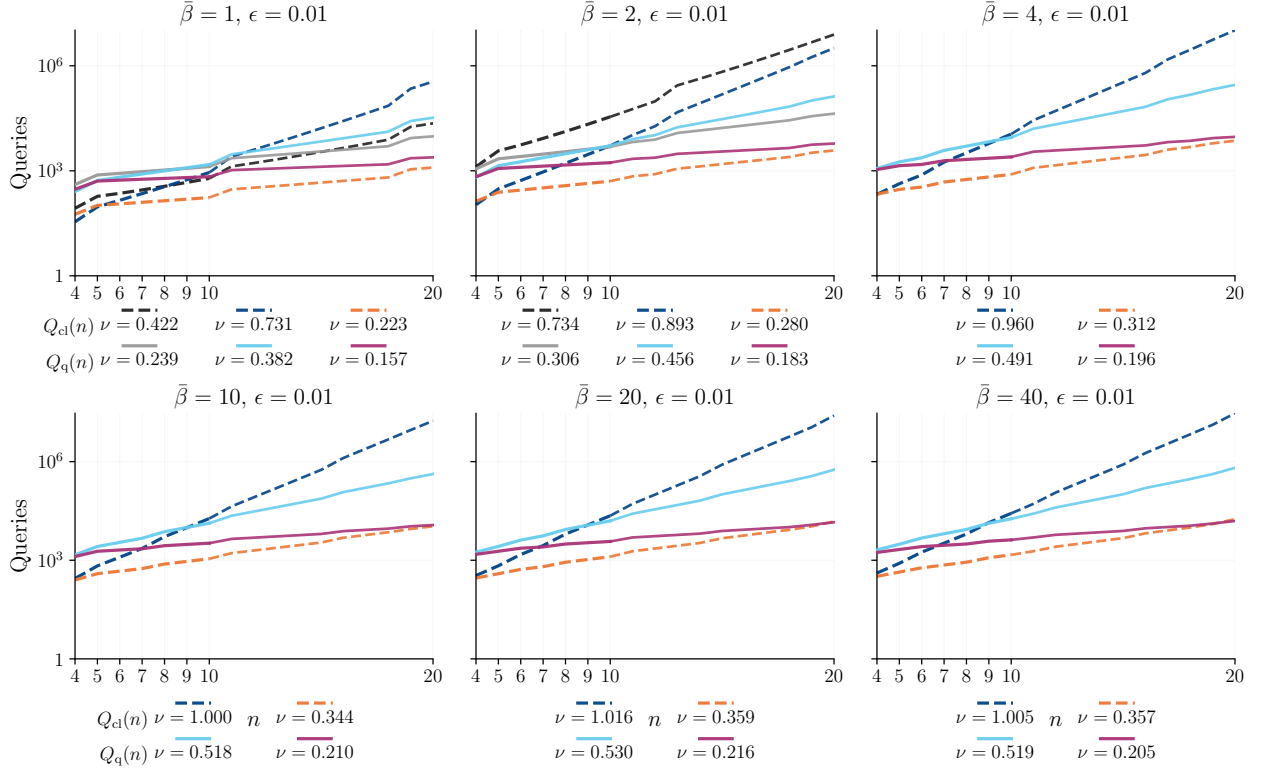}}
    \caption{Same as Figure~\ref{fig:quantum_queries_vs_n_all_epsilon}, but showing the total query counts accumulated over the complete annealing schedule.}
    \label{fig:quantum_queries_vs_n_all_epsilon_full_annealing}
\end{figure}

\clearpage
\section{Classical hardware timing}
\label{app:classical_hardware_baselines}

This appendix derives the hardware baselines used to convert the classical Monte Carlo query counts reported in Appendix~\ref{app:spectral_gap_queries} into wall-clock runtime estimates.

We denote by $\tau_{\rm cpu}(n)$, $\tau_{\rm gpu}(n)$, and $\tau_{\rm fpga}(n)$ the average wall-clock time required to perform one transition of the Markov chain on the CPU, GPU, and FPGA platforms, respectively. The wall-clock time of the full computation, based on an annealing schedule $\vec\beta=(\beta_j)_{j=0}^L$, is estimated as
\begin{equation}
T_{\rm cpu/gpu/fpga}(n,\vec\beta,\varepsilon)=
\sum_{j=1}^{L}
Q_{\rm cl,warm}(n,\beta_j,\varepsilon)
\times
\tau_{\rm cpu/gpu/fpga}(n),
\end{equation}
where $Q_{\rm cl,warm}(n,\beta_j,\varepsilon)$ is the number of Markov-chain transitions required at the $j$-th annealing step to reach the target accuracy as determined in Appendix~\ref{app:spectral_gap_queries:cq}.

These platforms provide three complementary classical hardware baselines. The CPU implementation uses a single core and does not exploit thread-level parallelism. It nevertheless benefits from a high clock frequency and optimized floating-point arithmetic. Energy differences and acceptance probabilities are evaluated using single-precision floating-point arithmetic, avoiding the need to introduce and characterize an additional fixed-point approximation.

The GPU is a throughput-oriented architecture in which many threads execute concurrently. Successive transitions of a single Markov chain are intrinsically sequential and cannot generally be parallelized across steps, unlike methods such as parallel tempering, which can evolve multiple replicas concurrently. In our case, GPU parallelism is instead used within each transition, for example to evaluate and reduce the terms entering the energy difference.

The FPGA provides a reconfigurable hardware baseline. Its circuits are instantiated for the considered transition rule and system size. The hardware resources can therefore increase with $n$, subject to the capacity of the FPGA device, and the measured transition time already includes the resulting circuit-level parallelism.

This resource scaling is important when comparing classical and quantum hardware. Ref.~\cite{ronnow2014defining} argues that directly comparing a fixed-resource CPU with a quantum device whose number of qubits and couplers increases with $n$ gives the quantum device an additional hardware-parallelism advantage. The authors therefore divide the CPU runtime by a factor proportional to $n$, estimating the performance of a hypothetical classical special-purpose device whose hardware resources also scale with the problem size.

We do not apply such a correction to the CPU or GPU runtimes, which are used as direct hardware baselines rather than as resource-matched models of the quantum device. No correction is required for the FPGA baseline in the first place, since the FPGA already represents the type of special-purpose classical architecture considered in Ref.~\cite{ronnow2014defining}, whose resources scale with the system size, and $\tau_{\rm fpga}(n)$ includes the parallel resources instantiated at each system size. 

\begin{figure}[h]
    \centering
    \input{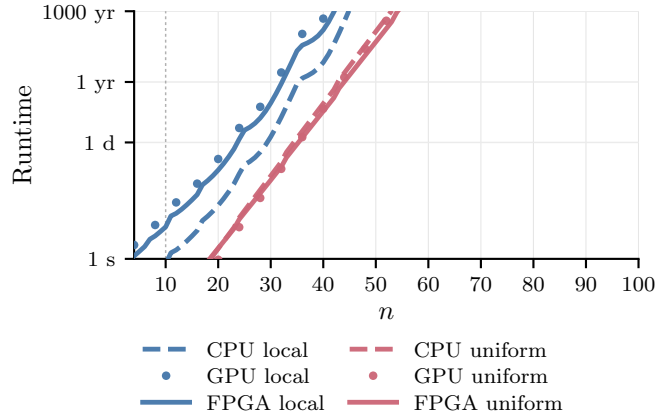}
    \caption{Estimated classical (full) annealing runtime as a function of the system size $n$ for the local and uniform proposal rules on CPU, GPU, and FPGA platforms, at $\beta = 4$ and $\varepsilon=10^{-2}$. Colors distinguish the proposal rule, while line styles and markers distinguish the hardware platform.}
    \label{fig:classical_annealing_runtime_by_device}
\end{figure}

Figure~\ref{fig:classical_annealing_runtime_by_device} compares the classical hardware baselines. For the local proposal, the CPU implementation achieves the lowest runtime over the considered range. For the uniform proposal, the three platforms have comparable runtimes at small system sizes, while the FPGA becomes the fastest platform as $n$ increases.

\subsection{CPU and GPU timing estimation }
\label{app:cpu_baseline}

The CPU benchmarks were performed on the CINECA Leonardo, using a single core of an Intel Xeon Platinum 8358 processor at $2.60\,\mathrm{GHz}$. The code was compiled with GCC using two configurations: a native build with \texttt{-march=native -mtune=native} and a second build with \texttt{-march=sapphirerapids -mprefer-vector-width=512}, both using \texttt{-O3}, \texttt{-ffast-math}, and loop-unrolling optimizations. The GPU benchmarks were performed on a cluster equipped with an NVIDIA H100 NVL GPU using CUDA compilation tools 13.1.115. The CUDA code was compiled with \texttt{nvcc} using \texttt{-O3}, \texttt{--use\_fast\_math}, and \texttt{-arch=sm\_80}, with one block of $128$ threads assigned to each Markov chain.

The benchmarks were performed for $n\in\{64,128,256,512\}$. The CPU measurements used $10$ independent SK instances and $10^5$ attempted transitions per instance. Importantly, we are \emph{not solving} the system for such large $n$, but simply timing a fixed number of steps to estimate the average time per step $\tau$.
The GPU measurements used a single chain of $10^6$ attempted transitions and were averaged over three runs, and therefore measure single chain latency rather than the saturated throughput of the GPU. Pseudorandom number generation was excluded from both measurements, which has been determined a major bottleneck in the calculation \cite{BERNASCHI2026110182}. Instead, random integers and uniform random numbers were generated before the timed region, stored in dedicated pools, and subsequently consumed by the transition kernels.

For the local proposal, each transition evaluates the dense local field of one selected spin and therefore requires $O(n)$ arithmetic operations. Its timing is fitted as $a+bn$. The uniform proposal generates a complete candidate configuration and evaluates its dense SK energy, requiring $O(n^2)$ arithmetic operations, and is fitted as $A+Bn+Cn^2$. The coefficients are obtained by non-negative least squares after dividing the measured total runtime by the number of attempted transitions. Using the Sapphire-Rapids-targeted CPU measurements and the average of the three GPU measurements, the resulting per-transition times are
\begin{align}
\tau_{\rm cpu}^{\rm loc}(n)
&=
\left(
5.959\times10^{-9}
+
1.429\times10^{-10}n
\right)\,\mathrm{s}, 
\\
\tau_{\rm cpu}^{\rm unif}(n)
&=
\left(
1.173\times10^{-8}n
+
6.964\times10^{-11}n^2
\right)\,\mathrm{s},
\\
\tau_{\rm gpu}^{\rm loc}(n)
&=
\left(
7.837\times10^{-7}
+
1.459\times10^{-9}n
\right)\,\mathrm{s},
\\
\tau_{\rm gpu}^{\rm unif}(n)
&=
\left(
2.215\times10^{-10}n^2
\right)\,\mathrm{s}.
\end{align}

\subsection{FPGA timing}
\label{app:fpga_baseline}

The FPGA transition times are estimated from high-level synthesis using AMD Vitis 2025.2. The circuits target an AMD Virtex UltraScale+ \texttt{xcvu19p-fsva3824-2-e} FPGA with a clock period of $3\,\mathrm{ns}$, corresponding to a nominal frequency of $333.3\,\mathrm{MHz}$. We use the fixed-point arithmetic detailed in Appendix~\ref{app:fixed_point_errors}. Specifically, the Hamiltonian coefficients are normalized by twice the Hamiltonian one-norm, which provides an upper bound on the energy difference, as derived in Appendix~\ref{app:sk_model:one_norm}. The Metropolis acceptance probability is evaluated through a piecewise Chebyshev approximation of the exponential, implemented using the Clenshaw recurrence~\cite{Press:2007:NRA}.

For the local proposal, the circuit computes the normalized energy difference associated with a single-spin flip. Only the field term of the selected spin and its $n-1$ interaction terms contribute, giving $O(n)$ terms in total. These terms are generated in parallel and summed using a Wallace-tree implementation~\cite{ercegovac2004digital}. The tree requires $O(n)$ space and has depth $O(\log n)$.

For the uniform proposal, the candidate configuration may differ from the current configuration at every spin. The circuit therefore computes the normalized energy difference from all field and interaction terms. Their number is $n+\binom{n}{2}=n(n+1)/2=O(n^2)$. These terms are generated in parallel and reduced using the same Wallace-tree construction. The corresponding depth is $O\left(\log n\right)$ while the required space scale as $O(n^2)$.

Both transition times are therefore fitted as affine functions of $\log_2 n$. Although the local and uniform reductions contain respectively $O(n)$ and $O(n^2)$ terms, $\log_2(n^2)=2\log_2 n$, so both are described by the same logarithmic fitting form, with different coefficients. Synthesis was performed for $n\in\{8,16,32,64\}$, yielding
\begin{align}
\tau_{\rm fpga}^{\rm loc}(n)
&=
\left(
0.2679
+
0.0018\log_2 n
\right)\,\mu\mathrm{s},
\\
\tau_{\rm fpga}^{\rm unif}(n)
&=
\left(
0.2541
+
0.0042\log_2 n
\right)\,\mu\mathrm{s}.
\end{align}
The logarithmic latency is obtained by increasing the amount of spatially parallel hardware. The uniform-proposal fit therefore assumes that the required $O(n^2)$ resources can be instantiated and is valid only within the capacity of the considered FPGA.

\subsection{Fixed-point precision and approximation errors}
\label{app:fixed_point_errors}

Both the FPGA implementation and the Boltzmann coin quantum circuits rely on fixed-point arithmetic to compute the energy difference between two configurations.

The coefficients $\widetilde h_i$ and $\widetilde J_{ij}$ are encoded in signed fixed-point representation after normalization by $\Lambda$. We choose $\Lambda$ as an upper bound on the largest possible energy difference. This bound is obtained as twice the one-norm of the SK Hamiltonian. For the SK ensemble considered here, it scales as $\Lambda \simeq 2n^{3/2}/\sqrt{\pi}$, as derived in Appendix~\ref{app:sk_model:one_norm}. This choice guarantees that every normalized coefficient, each partial sum, and the final normalized energy difference can be represented using one sign bit, with $0$ and $1$ denoting positive and negative values, respectively. The fixed-point register therefore contains $w=1+b$ bits, where one bit stores the sign and $b$ bits store the fractional part.

The value of $b$ depends on the instance size, the inverse temperature, and the desired accuracy. Given a total error $\varepsilon$ in TV distance in Eq.~\ref{eq:error_arithmetic_triangular}, we assign a negligible fraction of it to arithmetic errors, for example $\varepsilon_{\rm arith}=\varepsilon/50$. Since the arithmetic first computes the energy difference and then evaluates the Metropolis acceptance probability, we divide this budget equally as $\varepsilon_{\rm discr}=\varepsilon_{\rm op}=\varepsilon_{\rm arith}/2$.

The error $\varepsilon_{\rm discr}$ determines the precision with which the Hamiltonian coefficients must be encoded and therefore fixes the choice of $b$. Let $A_\beta(\Delta H)=\min\{1,\exp(-\beta\Delta H)\}$. The energy difference contains contributions from $M=n+\binom{n}{2}$ Hamiltonian coefficients. Since each normalized coefficient is rounded with error at most $2^{-b}$ and its contribution to the energy difference can change by at most a factor of two,
\begin{align}
\left|\Delta H-\widetilde{\Delta H}\right|
&\leq
2M\Lambda 2^{-b},
\\
\left|
A_\beta(\Delta H)
-
A_\beta(\widetilde{\Delta H})
\right|
\leq
\beta
\left|\Delta H-\widetilde{\Delta H}\right|
&\leq
2M\beta\Lambda 2^{-b}
\leq
\varepsilon_{\rm discr}.
\end{align}
It is therefore sufficient to choose
\begin{equation}
b
\geq
\left\lceil
\log_2\left(
\frac{2M\beta\Lambda}{\varepsilon_{\rm discr}}
\right)
\right\rceil.
\end{equation}
Substituting the expressions for $M$ and $\Lambda$, we obtain
\begin{equation}
b
\geq
\left\lceil
\frac{7}{2}\log_2 n
+
\log_2\left(\varepsilon_{\rm discr}^{-1}\right)
+
\log_2\left(\frac{2\beta}{\sqrt{\pi}}\right)
+
\log_2\left(1+\frac{1}{n}\right)
\right\rceil.
\end{equation}
For fixed $\beta$, the constant and subleading terms can be absorbed into the additive constant, giving the asymptotic choice in Eq.~\ref{eq:choice_of_f},
\begin{equation}
b
=
\frac{7}{2}\log_2 n
+
\log_2\left(\varepsilon_{\rm discr}^{-1}\right)
+
O(1).
\end{equation}

To implement $A_\beta$, only the branch with $\Delta H>0$ must be approximated, since the acceptance probability for $\Delta H\leq 0$ is set exactly to one according to the sign of the fixed-point energy difference. Let $x=\Delta H/\Lambda\in[0,1]$. We approximate the truncated function
\begin{equation}
\widetilde g(x)
=
\begin{cases}
e^{-\beta\Lambda x}, & 0\leq x\leq\tau,\\
0, & x>\tau,
\end{cases}
\end{equation}
where $\tau=\min\{1,\ln(\varepsilon_{\rm tail}^{-1})/(\beta\Lambda)\}$ is chosen so that the discarded tail is bounded by $\varepsilon_{\rm tail}$. The operation error budget is divided between the truncated tail and the polynomial approximation as $\varepsilon_{\rm op}=\varepsilon_{\rm tail}+\varepsilon_{\rm poly}$.

Equivalently, defining $u=\beta\Lambda x$, we approximate $e^{-u}$ over the interval $0\leq u\leq\min\{\beta\Lambda,\ln(\varepsilon_{\rm tail}^{-1})\}$. Numerically, we find that the required polynomial degree is well described by $d\approx\log_2(\varepsilon_{\rm poly}^{-1})$. For the target total variation error $\varepsilon=10^{-2}$ used here, the error allocation above gives $\varepsilon_{\rm op}=10^{-4}$, and a polynomial of degree $d=12$ is sufficient to achieve the desired approximation accuracy.

\section{Szegedy quantum walk}
\label{app:szegedy_quantum_walk}

Let $P$ be the transition matrix of an ergodic reversible Markov chain defined on the configuration space $\Omega$, and let $X$ be its discriminant matrix. Since $P$ is stochastic rather than unitary, it cannot be implemented directly as a quantum circuit. The Szegedy quantum walk operator~\cite{szegedy:2004} circumvents this limitation by defining a unitary whose spectrum is tightly related to that of $X$.

In the construction below, the Szegedy walk acts on the Hilbert space $\calH_a^{|\Omega|}\otimes\calH_b^{|\Omega|}$, where registers $a$ and $b$ are two copies of the Markov-chain configuration space. In this work we use the half-step operator $W$, defined explicitly below. With this convention, if $\lambda\in\operatorname{spec}(X)$, then the corresponding walk eigenvalues are $\exp(\pm i\,\arccos(\lambda))$, which is the phase relation underlying the quadratic speedup in the spectral gap.

\subsection{Construction via coherent access to $P$}

Let $V_P$ be a unitary implementing coherent access to $P$ via
\begin{equation}
V_P\ket{x}_a\ket{0}_b
=
\ket{x}_a\sum_y \sqrt{P_{yx}}\ket{y}_b
:=
\ket{\psi_x}
\end{equation}
for all $x\in\Omega$. Let $S$ be the swap operator, $S\ket{x}_a\ket{y}_b=\ket{y}_a\ket{x}_b$. Let $\mathcal A=\mathrm{span}\{\ket{\psi_x}:x\in\Omega\}$ be the image of the zero-ancilla subspace under $V_P$, and let $\mathcal B=S\mathcal A$. The reflection about the zero state of register $b$ is
\begin{equation}
R_0:=2(\mathbb I_a\otimes\ketbra{0}{0}_b)-\mathbb I_{ab}.
\end{equation}
The corresponding reflections about $\mathcal A$ and $\mathcal B$ are $R_{\mathcal A}=V_P R_0 V_P^\dagger$ and $R_{\mathcal B}=S R_{\mathcal A}S$. The full Szegedy walk operator is
\begin{equation}
W_{\mathrm{sz}}=R_{\mathcal B}R_{\mathcal A}.
\end{equation}

The spectral relation between $W_{\mathrm{sz}}$ and $X$ follows from Jordan's lemma applied to $R_{\mathcal A}$ and $R_{\mathcal B}$. Define $\mathcal K:=\mathcal A+\mathcal B$. The Hilbert space decomposes into $\mathcal K^\perp$ and one- or two-dimensional invariant blocks $\mathcal K_j\subseteq\mathcal K$. Since both reflections act as $-\mathbb I_{ab}$ on $\mathcal K^\perp$, the full walk acts trivially there,
\begin{equation}
    W_{\mathrm{sz}}\big|_{\mathcal K^\perp}=\mathbb I_{\mathcal K^\perp}.
\end{equation}

The nontrivial dynamics is therefore contained in the blocks $\mathcal K_j\subseteq\mathcal K$, whose angles are fixed by the overlap between $\mathcal A$ and $\mathcal B$. In fact, it holds that
\begin{equation}
{}_a\!\bra{x}{}_b\!\bra{0}
V_P^\dagger S V_P
\ket{y}_a\ket{0}_b
=
\bra{\psi_x}S\ket{\psi_y}
=
\sqrt{P_{xy}P_{yx}}
=
X_{xy}.
\end{equation}
Thus $V_P^\dagger S V_P$ block-encodes $X$ on the zero state of register $b$:
\begin{equation}
(\mathbb I_a\otimes\bra{0}_b)
V_P^\dagger S V_P
(\mathbb I_a\otimes\ket{0}_b)
=
X.
\end{equation}
Equivalently, if $Q_{\mathcal A}=V_P(\mathbb I_a\otimes\ket{0}_b)$ and $Q_{\mathcal B}=SQ_{\mathcal A}$ are the isometries whose images are $\mathcal A$ and $\mathcal B$, then $Q_{\mathcal A}^\dagger Q_{\mathcal B}=X$.

Since $Q_{\mathcal A}^\dagger Q_{\mathcal B}=X$ and $X$ is real symmetric, the same orthonormal eigenbasis diagonalizes the overlap between the two embedded bases. If $X\ket{v_j}=\lambda_j\ket{v_j}$ and we define $\ket{a_j}=Q_{\mathcal A}\ket{v_j}$ and $\ket{b_j}=Q_{\mathcal B}\ket{v_j}$, then
\begin{equation}
\braket{a_j}{b_k}
=
\bra{v_j}Q_{\mathcal A}^\dagger Q_{\mathcal B}\ket{v_k}
=
\lambda_j\delta_{jk}.
\end{equation}
Thus different values of $j$ do not mix, and each pair $\ket{a_j},\ket{b_j}$ defines one invariant block $\mathcal K_j$ for the two reflections. Writing $\lambda_j=\cos\theta_j$, with $\theta_j\in[0,\pi]$, the angle $\theta_j$ controls the action of $W_{\mathrm{sz}}$ on that block.

For $|\lambda_j|<1$, write
\begin{equation}
\ket{b_j}
=
\cos\theta_j\ket{a_j}
+
\sin\theta_j\ket*{a_j^\perp},
\end{equation}
where $\ket*{a_j^\perp}$ is the normalized state in $\mathcal K_j$ orthogonal to $\ket{a_j}$. Then $\mathcal K_j=\mathrm{span}\{\ket{a_j},\ket*{a_j^\perp}\}$, and in the basis $\{\ket{a_j},\ket*{a_j^\perp}\}$ the full walk is
\begin{equation}
W_{\mathrm{sz}}\big|_{\mathcal K_j}
=
\begin{pmatrix}
\cos(2\theta_j) & -\sin(2\theta_j) \\
\sin(2\theta_j) & \cos(2\theta_j)
\end{pmatrix}.
\end{equation}
Thus $W_{\mathrm{sz}}$ is a rotation by angle $2\theta_j$ on $\mathcal K_j$, with eigenvalues $\exp(\pm 2i\theta_j)=\exp(\pm 2i\arccos(\lambda_j))$. The limiting cases $|\lambda_j|=1$ give one-dimensional invariant blocks.

\subsection{Half-step Szegedy quantum walk}

Using $R_{\mathcal A}=V_P R_0 V_P^\dagger$ and $R_{\mathcal B}=S R_{\mathcal A}S$, the full walk can be written as
\begin{equation}
W_{\mathrm{sz}}
=
R_{\mathcal B}R_{\mathcal A}
=
S R_{\mathcal A}S R_{\mathcal A}.
\end{equation}
We define the half-step walk used in this work as
\begin{equation}
W=SV_P R_0 V_P^\dagger=S R_{\mathcal A},
\end{equation}
so that $W^2=W_{\mathrm{sz}}$. On the block associated with $\lambda_j=\cos\theta_j$, this half-step walk acts as a rotation by $\theta_j$ instead of $2\theta_j$. Hence the corresponding eigenvalues of $W$ are $\exp(\pm i\arccos(\lambda_j))$.

\subsection{Construction via coherent access to $T$}
\label{app:szegedy_quantum_walk:coherent_t}

\textcite{lemieux2020efficient} observed that coherent access to a Metropolis-Hastings transition matrix $P$ is hard to implement directly, because the diagonal entry of $P$ requires coherently summing all rejected off-diagonal transition probabilities out of $x$. This is costly unless the proposal has small support or the sum can be computed analytically.

Instead, one can access only the proposal matrix $T$ coherently. Let
\begin{equation}
V\ket{x}_a\ket{0}_b
=
\ket{x}_a\sum_y \sqrt{T_{yx}}\ket{y}_b.
\end{equation}
The acceptance probability is introduced by a coin register through the Boltzmann coin
\begin{equation}
B\ket{x}_a\ket{y}_b\ket{0}_c
=
\ket{x}_a\ket{y}_b
\left(
\sqrt{1-A_{yx}}\ket{0}_c
+
\sqrt{A_{yx}}\ket{1}_c
\right),
\end{equation}
where $\ket{1}_c$ denotes acceptance and $\ket{0}_c$ rejection. The unitary $F$ then applies the accepted move: it swaps registers $a$ and $b$ when the coin is $\ket{1}_c$, and acts trivially when the coin is $\ket{0}_c$.

The walk acts on $\calH_a\otimes\calH_b\otimes\calH_c$ and is
\begin{equation}
W=R_0 V^\dagger B^\dagger F B V,
\end{equation}
where $R_0=2(\mathbb I_a\otimes\ketbra{0}{0}_{bc})-\mathbb I_{abc}$ is the reflection about the joint zero state of the proposal and coin registers. Writing $U:=V^\dagger B^\dagger F B V$, we have $W=R_0U$, and since $R_0$ acts as $+1$ on the zero subspace,
\begin{equation}
(\mathbb I_a\otimes\bra{0}_{bc})
W
(\mathbb I_a\otimes\ket{0}_{bc})
=
(\mathbb I_a\otimes\bra{0}_{bc})
U
(\mathbb I_a\otimes\ket{0}_{bc}).
\end{equation}

Define
\begin{equation}
\ket{\psi_x}
=
B V\ket{x}_a\ket{0}_b\ket{0}_c
=
\ket{x}_a
\sum_z \sqrt{T_{zx}}\ket{z}_b
\left(
\sqrt{1-A_{zx}}\ket{0}_c
+
\sqrt{A_{zx}}\ket{1}_c
\right).
\end{equation}
Then
\begin{equation}
{}_a\!\bra{x}{}_b\!\bra{0}{}_c\!\bra{0}
U
\ket{y}_a\ket{0}_b\ket{0}_c
=
\bra{\psi_x}F\ket{\psi_y}.
\end{equation}
For $x\neq y$, only accepted branches contribute, and
\begin{equation}
\bra{\psi_x}F\ket{\psi_y}
=
\sqrt{T_{xy}A_{xy}}\sqrt{T_{yx}A_{yx}}
=
\sqrt{P_{xy}P_{yx}}
=
X_{xy}.
\label{eq:szegedy_overlap}
\end{equation}
For $x=y$, both rejected branches and the accepted identity branch contribute, giving
\begin{equation}
\bra{\psi_x}F\ket{\psi_x}
=
\sum_z T_{zx}(1-A_{zx})
+
T_{xx}A_{xx}
=
P_{xx}
=
X_{xx}.
\end{equation}
Thus the zero-block of the walk is the discriminant matrix,
\begin{equation}
(\mathbb I_a\otimes\bra{0}_{bc})
W
(\mathbb I_a\otimes\ket{0}_{bc})
=
X.
\end{equation}

Finally, $W$ is again a product of two reflections. The first is $R_0$, while the second is $U$, an involution because $U=V^\dagger B^\dagger F B V$ and $F^2=\mathbb I_{abc}$. Therefore the same Jordan-lemma argument applies, with the compression of $U$ playing the role of the overlap matrix. Hence, for each $\lambda_j\in\operatorname{spec}(X)$, the corresponding walk eigenvalues are $\exp(\pm i\arccos(\lambda_j))$.

\subsection{Efficient coherent access to the Hamiltonian simulation proposal}
\label{app:szegedy_quantum_walk:hamiltonian_proposal}

For the Hamiltonian simulation proposal, the transition probabilities are not computed by a reversible classical circuit. If $U(t,\gamma)=\exp(-i t H_{\rm tf}(\gamma))$ is the simulated unitary, the proposal kernel is
\begin{equation}
T_{yx}
=
\left|\bra{y}U(t,\gamma)\ket{x}\right|^2.
\end{equation}
Constructing amplitudes proportional to $\sqrt{T_{yx}}=|U_{yx}(t,\gamma)|$ would require coherently removing the phases of the Hamiltonian simulation amplitudes. Hamiltonian simulation instead naturally prepares the phaseful amplitudes $U_{yx}(t,\gamma)$. The point of the construction below is that, when $H_{\rm tf}(\gamma)$ is symmetric in the computational basis, these phases cancel inside the Szegedy overlap of Eq.~\ref{eq:szegedy_overlap}.

Let $a$ and $b$ be two $n$-qubit registers. Consider the unitary loading the state $x$ in the second register initialized to $\ket{0^n}$,
\begin{equation}
C_{a\to b}
=
\sum_x \ket{x}_a\ket{x}_b\,{}_a\!\bra{x}{}_b\!\bra{0^n}
=
\prod_{j=1}^{n}\mathsf{CNOT}_{a_j\to b_j}.
\end{equation}
This operation loads the classical basis label $x$ into register $b$; it is not a copying operation on arbitrary quantum states. Define the unitary
\begin{equation}
V
=
(\mathbb I_a\otimes U(t,\gamma))C_{a\to b}.
\end{equation}
Then
\begin{equation}
V\ket{x}_a\ket{0^n}_b
=
\ket{x}_a U(t,\gamma)\ket{x}_b
=
\ket{x}_a\sum_y e^{i\theta_{yx}(t,\gamma)}|U_{yx}(t,\gamma)| \, \ket{y}_b.
\end{equation}
Thus $V$ is not coherent access to the positive square-root amplitudes $\sqrt{T_{yx}}$, but rather to the phaseful Hamiltonian simulation amplitudes.

Recall that
\begin{equation}
\ket{\psi_x}
=
B V\ket{x}_a\ket{0^n}_b\ket{0}_c
=
\ket{x}_a
\sum_z e^{i\theta_{zx}(t,\gamma)}|U_{zx}(t,\gamma)| \ket{z}_b
\left(
\sqrt{1-A_{zx}}\ket{0}_c
+
\sqrt{A_{zx}}\ket{1}_c
\right).
\end{equation}
The compressed matrix elements are
\begin{equation}
{}_a\!\bra{x}{}_b\!\bra{0^n}{}_c\!\bra{0}
V^\dagger B^\dagger F B V
\ket{y}_a\ket{0^n}_b\ket{0}_c
=
\bra{\psi_x}F\ket{\psi_y}.
\end{equation}

For $x\neq y$, only accepted branches contribute, and therefore
\begin{equation}
\bra{\psi_x}F\ket{\psi_y}
=
e^{i(\theta_{xy}(t,\gamma)-\theta_{yx}(t,\gamma))}
|U_{yx}(t,\gamma)|\,|U_{xy}(t,\gamma)|\,
\sqrt{A_{yx}A_{xy}}.
\end{equation}
Since $H_{\rm tf}(\gamma)$ is symmetric in the computational basis, $U(t,\gamma)^\top=U(t,\gamma)$, and hence $U_{yx}(t,\gamma)=U_{xy}(t,\gamma)$. The phase factor cancels, giving
\begin{equation}
\bra{\psi_x}F\ket{\psi_y}
=
|U_{xy}(t,\gamma)|^2\sqrt{A_{yx}A_{xy}}
=
\sqrt{T_{xy}A_{xy}}\sqrt{T_{yx}A_{yx}}
=
\sqrt{P_{xy}P_{yx}}
=
X_{xy}.
\end{equation}

For $x=y$, first apply $F$ to $\ket{\psi_x}$:
\begin{align*}
F\ket{\psi_x}
&=
\sum_z e^{i\theta_{zx}(t,\gamma)}|U_{zx}(t,\gamma)|
\sqrt{1-A_{zx}}\,
\ket{x}_a\ket{z}_b\ket{0}_c
\\
&\quad+
\sum_z e^{i\theta_{zx}(t,\gamma)}|U_{zx}(t,\gamma)|
\sqrt{A_{zx}}\,
\ket{z}_a\ket{x}_b\ket{1}_c .
\end{align*}
Taking the overlap with $\bra{\psi_x}$ gives
\begin{align*}
\bra{\psi_x}F\ket{\psi_x}
&=
\sum_z
e^{-i\theta_{zx}(t,\gamma)}|U_{zx}(t,\gamma)|\sqrt{1-A_{zx}}\,
e^{i\theta_{zx}(t,\gamma)}|U_{zx}(t,\gamma)|\sqrt{1-A_{zx}}
\\
&\quad+
e^{-i\theta_{xx}(t,\gamma)}|U_{xx}(t,\gamma)|\sqrt{A_{xx}}\,
e^{i\theta_{xx}(t,\gamma)}|U_{xx}(t,\gamma)|\sqrt{A_{xx}}
\\
&=
\sum_z |U_{zx}(t,\gamma)|^2(1-A_{zx})
+
|U_{xx}(t,\gamma)|^2 A_{xx}
\\
&=
\sum_z T_{zx}(1-A_{zx})
+
T_{xx}A_{xx}.
\end{align*}

This is precisely the diagonal entry of the Metropolis-Hastings transition matrix,
\begin{equation}
\bra{\psi_x}F\ket{\psi_x}
=
P_{xx}
=
X_{xx}.
\end{equation}
Therefore
\begin{equation}
(\mathbb I_a\otimes\bra{0^n}_b\bra{0}_c)
V^\dagger B^\dagger F B V
(\mathbb I_a\otimes\ket{0^n}_b\ket{0}_c)
=
X.
\end{equation}

If $H_{\rm tf}(\gamma)$ is not symmetric, then generally $U(t,\gamma)^\top\neq U(t,\gamma)$. The same construction still gives, for $x\neq y$,
\begin{equation}
\bra{\psi_x}F\ket{\psi_y}
=
e^{i(\theta_{xy}(t,\gamma)-\theta_{yx}(t,\gamma))}
|U_{yx}(t,\gamma)|\,|U_{xy}(t,\gamma)|\,
\sqrt{A_{yx}A_{xy}} \neq \sqrt{P_{xy}P_{yx}}.
\end{equation}

\section{Quantum walk circuit construction and resource derivation}
\label{app:quantum_walk_circuit_resources}

This appendix details the resources required to implement the circuit for $W_\beta$. We track the non-Clifford depth and the number of qubits. The Szegedy walk acts on two $n$-qubit system registers and one coin register, according to the structure shown in Appendix~\ref{app:szegedy_quantum_walk}. One application of the walk therefore requires one proposal preparation, one Boltzmann coin, the accept-path unitary, followed by the inverse coin and inverse proposal, and the reflection. If $D_V$, $D_B$, $D_F$, and $D_R$ denote the non-Clifford depths of these blocks, the walk depth is $D_W = 2D_V + 2D_B + D_F + D_R$. An implementation of this walk operator is provided in the accompanying code~\cite{myrepo2026}, written in Python using the Qiskit framework~\cite{javadi2024quantum}.

\subsection{Decomposition of primitives}
\label{app:quantum_walk_circuit_resources:primitives}

\begin{figure}[t!]
    \centering

    \begin{subfigure}[t]{0.68\linewidth}
        \centering
        \includegraphics[width=\linewidth]{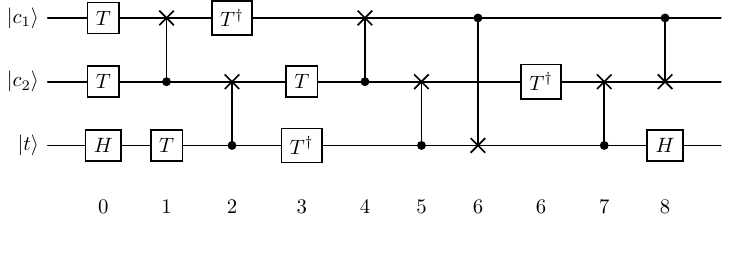}
        \caption{Toffoli gate.}
        \label{fig:primitive_decompositions:ccx}
    \end{subfigure}
    \hfill
    \begin{subfigure}[t]{0.28\linewidth}
        \centering
        \includegraphics[width=\linewidth]{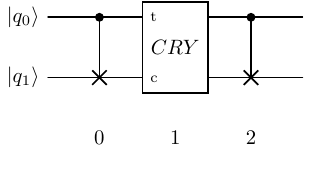}
        \caption{Givens rotation.}
        \label{fig:primitive_decompositions:givens}
    \end{subfigure}
    
    \begin{subfigure}[t]{0.68\linewidth}
        \centering
        \includegraphics[width=\linewidth]{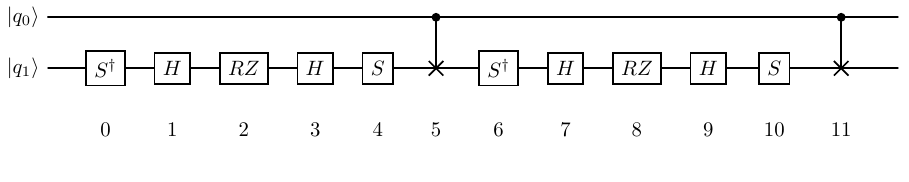}
        \includegraphics[width=\linewidth]{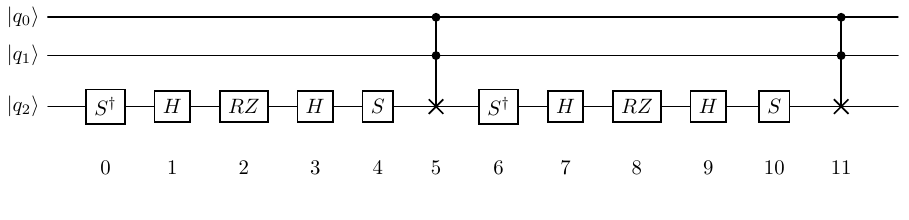}
        \caption{Controlled $\mathsf{R}_y$ rotation.}
        \label{fig:primitive_decompositions:cry}
    \end{subfigure}
    \hfill
    \begin{subfigure}[t]{0.28\linewidth}
        \centering
        \includegraphics[width=\linewidth]{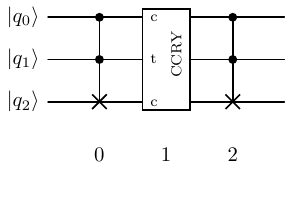}
        \caption{Controlled Givens rotation.}
        \label{fig:primitive_decompositions:controlled_givens}
    \end{subfigure}

    \caption{Explicit decompositions of the primitive gates used in the quantum-walk implementation. The decompositions are written in the Clifford+$\mathsf{T}$+$\mathsf{R}_z$ gate set and are recursively expanded instead of being obtained through transpilation.}
    \label{fig:primitive_decompositions}
\end{figure}

An important operational point is that, in the accompanying notebooks, we provide an explicit implementation of the quantum walk circuit. The decomposition to the Clifford+$\mathsf{T}$+$\mathsf{R}_z$ gate set should not be performed with \texttt{qiskit.transpile}, since transpilation may change the non-Clifford depth of the circuit. We therefore define the required primitive gates directly and give explicit decompositions, which are then unpacked recursively. The primitives used are the Toffoli gate, the controlled and doubly controlled $\mathsf{R}_y$ rotations, and the controlled and uncontrolled Givens rotations,
\begin{align}
    \mathsf{CCX}
    &=
    \left(I-\ketbra{11}{11}\right)\otimes I
    +
    \ketbra{11}{11}\otimes X,
    &
    X
    &=
    \begin{pmatrix}
        0 & 1 \\
        1 & 0
    \end{pmatrix},
    \\
    \mathsf{CR}_y(\theta)
    &=
    \ketbra{0}{0}\otimes I
    +
    \ketbra{1}{1}\otimes \mathsf{R}_y(\theta),
    &
    \mathsf{R}_y(\theta)
    &=
    \begin{pmatrix}
        \cos(\theta/2) & -\sin(\theta/2) \\
        \sin(\theta/2) & \cos(\theta/2)
    \end{pmatrix},
    \\
    \mathsf{CCR}_y(\theta)
    &=
    \left(I-\ketbra{11}{11}\right)\otimes I
    +
    \ketbra{11}{11}\otimes \mathsf{R}_y(\theta),
    &
    G(\theta)
    &=
    \begin{pmatrix}
        1 & 0 & 0 & 0 \\
        0 & \cos\theta & \sin\theta & 0 \\
        0 & -\sin\theta & \cos\theta & 0 \\
        0 & 0 & 0 & 1
    \end{pmatrix},
    \\
    \mathsf{CG}(\theta)
    &=
    \ketbra{0}{0}\otimes I
    +
    \ketbra{1}{1}\otimes G(\theta).
\end{align}
The Givens rotation is used to split an input weight into two branches with non-negative weights $w_\ell$ and $w_r$. It is therefore specified by choosing the angle $\theta$ such that
\begin{equation}
    \cos^2\theta = \frac{w_\ell}{w_\ell+w_r},
    \qquad
    \sin^2\theta = \frac{w_r}{w_\ell+w_r},
    \qquad
    \theta
    =
    \arctan\left(
        \sqrt{\frac{w_r}{w_\ell}}
    \right),
\end{equation}
with $\theta=0$ when $w_\ell+w_r=0$, and $\theta=\pi/2$ when $w_\ell=0$ and $w_r>0$. The decompositions are shown in Fig.~\ref{fig:primitive_decompositions}, and the corresponding resources are reported in Table~\ref{tab:primitive_resources}.

\begin{table}[h!]
    \centering
    \begin{tabular}{lcc}
        \toprule
        Gate & Logical qubits & Non-Clifford depth \\
        \midrule
        $\mathsf{CCX}$ & $3$ & $3$ \\
        $\mathsf{CR}_y(\theta)$ & $2$ & $2$ \\
        $\mathsf{CCR}_y(\theta)$ & $3$ & $8$ \\
        $G(\theta)$ & $2$ & $2$ \\
        $\mathsf{CG}(\theta)$ & $3$ & $14$ \\
        \bottomrule
    \end{tabular}
    \caption{Resource counts for the primitive gates used in the circuit decomposition. The non-Clifford depth counts layers of $\mathsf{T}$, $\mathsf{T}^\dagger$, and arbitrary $\mathsf{R}_z$ rotations after recursive expansion to the Clifford+$\mathsf{T}$+$\mathsf{R}_z$ gate set.}
    \label{tab:primitive_resources}
\end{table}

\subsection{Multi-controlled NOT and reflection}
\label{app:mcx}

Following the approach of Ref.~\cite{khattar2025rise}, we can implement an $m$-controlled NOT can be realized with logarithmic $\mathsf{T}$-depth while requiring only two clean ancilla qubits. The method is already implemented in Qiskit as \texttt{synth\_mcx\_2\_clean\_kg24}. After decomposition into Clifford+$\mathsf{T}$ gates, with each Toffoli assigned $\mathsf{T}$-depth three, its $\mathsf{T}$-depth is upper bounded by $14\left\lceil\log_2 m\right\rceil-10$. The same construction can be used to implement a rotation conditioned on the control register being in $\ket{0^m}$ by applying $X$ gates to the controls, computing their conjunction with the multi-controlled NOT, applying the desired controlled rotation, and then uncomputing the conjunction and the initial $X$ gates.

\subsection{Proposal preparation unitaries}
\label{app:proposal_preparation_unitaries}

The proposal-preparation unitary depends on the choice of trial move.

The uniform proposal has the simplest implementation. It acts on $\ket{x}_a\ket{0^n}_b$ trivially on register $a$ and with a parallel layer of Hadamard gates on register $b$. It therefore requires no qubits beyond the two system registers already used by $W_\beta$, and its non-Clifford depth is zero.

The local proposal also acts on $\ket{x}_a\ket{0^n}_b$. It first prepares on register $b$ the Dicke state of Hamming weight one, using the construction of Ref.~\cite{aktar2022divide}. This gives the uniform superposition over all single-spin flips. The proposal is then obtained by XOR-ing this bitstring with the input configuration $x$ through a layer of parallel $\mathsf{CNOT}$ gates from register $a$ to register $b$. Since this final XOR layer is Clifford, only the Dicke-state preparation contributes to the non-Clifford depth. For the special case of Hamming weight one, the circuit simplifies substantially, and the local proposal has non-Clifford depth upper bounded by $2\lceil \log_2 n\rceil$. 

The Hamiltonian simulation proposal is implemented by Trotterizing the transverse-field Ising evolution used to generate the proposal distribution. As discussed in Sec.~\ref{sec:walk_circuit_implementation}, we use a second-order Trotter formula. Its non-Clifford depth is the same as the corresponding first-order implementation at the level of this estimate, because the two half steps of the same commuting layer can be absorbed into the neighbouring Trotter layers. The dense $Z_iZ_j$ interaction layer contains the edges of the complete graph is scheduled into $n$
parallel matching layers. Each Trotter step therefore contributes $n$ two-qubit interaction layers plus three single-qubit rotation layers. For $r$ Trotter steps, with $r=50$ in our numerical experiments, the resulting non-Clifford depth is $r (n+3)$. 

An important point for the Hamiltonian simulation proposal is that, following Eq.~\ref{eq:dynamical_proposal_matrix}, the proposal is defined by averaging over a range of parameters $t$ and $\gamma$. In practice, we implement this average by assigning a different pair $(t,\gamma)$ to each application of $W_\beta$. In our numerical experiments, we sample $\gamma \in (0.25,0.60)$ and $t \in (2.0,20.0)$. Notice that the simulation time is not scaled with $n$, since the SK model is all-to-all connected and the Hamiltonian is already normalized accordingly.

The resulting resource estimates are summarized in Table~\ref{tab:proposal_unitary_resources}.

\begin{table}[h]
\centering
\begin{tabular}{lcc}
\hline
Proposal block   & Logical qubits & Non-Clifford depth \\
\hline
Uniform proposal                & $2n$ & $0$                     \\
Local proposal                  & $2n$ & $2\lceil\log_2 n\rceil$ \\
Hamiltonian simulation proposal & $2n$ & $r\left(n+3\right)$     \\
\hline
\end{tabular}
\caption{Resource estimates for the proposal-preparation unitary $V$.}
\label{tab:proposal_unitary_resources}
\end{table}

\subsection{Fully phase-based Boltzmann coin}
\label{app:fully_phase_based_coin}

The fully phase-based arithmetic is our most space-efficient implementation of the Boltzmann coin. Instead of computing the energy difference into a binary register and then evaluating the acceptance function arithmetically, we encode the energy difference in the spectrum of the phase Hamiltonian defined in Eq.~\ref{eq:fully_phase_arith_hamiltonian}, and apply the acceptance function through a polynomial transformation. Throughout this subsection and the following hybrid construction, we use $M=n+\binom{n}{2}=n(n+1)/2$ for the number of terms in the original $n$-spin SK Hamiltonian and $\ell_x:=\lceil\log_2 x\rceil$. All circuit depths refer to non-Clifford depth after decomposition into the Clifford+$\mathsf{T}$+$\mathsf{R}_z$ gate set, with each Toffoli gate assigned $\mathsf{T}$-depth $3$.

\subsubsection{Qubitized operator}

Let $H$ be the SK Hamiltonian. Writing the $2n$ system qubits as a single register of size $m=2n$, the coefficient representation of $H_{\rm ph}$ is obtained from the block-diagonal data
\begin{equation}
    h^{\rm ph}=\begin{pmatrix}-\tilde h\\\tilde h\end{pmatrix},\qquad J^{\rm ph}=\begin{pmatrix}-\tilde J&0\\0&\tilde J\end{pmatrix}.
\end{equation}
The first block subtracts the energy of register $a$, while the second block adds the energy of register $b$. We construct a qubitization~\cite{low2019hamiltonian} of $H_{\rm ph}/\Lambda$, where
\begin{equation}
    \Lambda=\sum_{i=1}^{m}|h_i^{\rm ph}|+\sum_{1\leq i<j\leq m}|J_{ij}^{\rm ph}|=2\left(\sum_{i=1}^{n}|\tilde h_i|+\sum_{1\leq i<j\leq n}|\tilde J_{ij}|\right).
\end{equation}
Each copy of the SK Hamiltonian contains $M$ terms, so the phase Hamiltonian $H_{\rm ph}$ contains $2M=n(n+1)$ terms. The qubitized operator takes the form
\begin{equation}
    Q_{\rm ph}=R_0\,\textsc{prepare}^{\dagger}\,\textsc{select}\,\textsc{prepare}.
\end{equation}

The preparation unitary loads the coefficients of the Hamiltonian as amplitudes. A binary selection register would use only $\ell_{2M}$ qubits, but would require a select operation that decodes the binary index and applies one of $O(n^2)$ Pauli terms. Instead, we use a unary factorized encoding with three length-$m$ selection registers $\ket{u_Z}\ket{u_J}\ket{v_J}$. The register $u_Z$ selects one-body $Z_i$ terms, while $u_J$ and $v_J$ select the two endpoints of a two-body $Z_iZ_j$ term. With $\ket{e_i}$ denoting the unary basis state with a single excitation at position $i$, we define $\ket{Z,i}\equiv\ket{e_i}_{u_Z}\ket{0^m}_{u_J}\ket{0^m}_{v_J}$ for $i=1,\ldots,m$, and $\ket{ZZ,i,j}\equiv\ket{0^m}_{u_Z}\ket{e_i}_{u_J}\ket{e_j}_{v_J}$ for $1\leq i<j\leq m$. Thus, the preparation register uses $3m$ qubits. In this notation,
\begin{equation}
    \textsc{prepare}\ket{0^{3m}}=\frac{1}{\sqrt{\Lambda}}\left(\sum_{i=1}^{m}\sqrt{|h_i^{\rm ph}|}\ket{Z,i}+\sum_{1\leq i<j\leq m}\sqrt{|J_{ij}^{\rm ph}|}\ket{ZZ,i,j}\right).
\end{equation}

The select routine acts on the three selection registers and on the $m$-qubit system register,
\begin{equation}
    \textsc{select}=\sum_{i=1}^{m}\operatorname{sgn}(h_i^{\rm ph})\ketbra{Z,i}{Z,i}\otimes Z_i+\sum_{1\leq i<j\leq m}\operatorname{sgn}(J_{ij}^{\rm ph})\ketbra{ZZ,i,j}{ZZ,i,j}\otimes Z_iZ_j.
\end{equation}
Thus, the uncontrolled select acts on $4m$ qubits: $3m$ selection qubits and $m$ system qubits. The signs of the LCU coefficients are implemented inside $\textsc{select}$ as phases on the corresponding selection labels, while the Pauli part acts on the system register. In the two-body case, the Pauli application is factorized over the two selected endpoints.

The GQSP sequence requires the controlled version of this operation. One could implement it on $1+4m$ qubits using a single external control, but then every controlled Pauli operation would share the same control qubit, creating a depth bottleneck. We therefore allocate $m-1$ additional clean qubits and fan out the external control into an $m$-qubit control register. This is the reversible $\mathsf{CNOT}$ fanout $\alpha\ket{0}\ket{0^{m-1}}+\beta\ket{1}\ket{0^{m-1}}\mapsto\alpha\ket{0^m}+\beta\ket{1^m}$, rather than a copy of an arbitrary quantum state. The fanout and unfanout are Clifford operations, so they do not contribute to the non-Clifford depth. The $i$-th control line is then used for the controlled operations associated with the $i$-th unary position. This parallelizes the one-body controlled operations and the endpoint-controlled Pauli applications for the two-body terms. The only pair-dependent part is the sign of $J_{ij}^{\rm ph}$, which is scheduled in at most $m$ matching layers as in the uncontrolled select. Hence, the controlled select uses $1+4m+(m-1)=5m$ qubits.

Finally, $R_0$ is the qubitization reflection associated with the state prepared by $\textsc{prepare}$. Since $\textsc{prepare}$ acts only on the three unary LCU selection registers $u_Z,u_J,v_J$, the reflection is nontrivial only on this $3m$-qubit preparation register and leaves the $m$-qubit system register untouched. It reflects about the all-zero state $\ket{0^{3m}}_{u_Z,u_J,v_J}$, with the sign convention fixed only up to an irrelevant global phase. In our implementation, this multi-controlled reflection uses two additional clean ancillary qubits, so the reflection block acts on $3m+2$ qubits in total. The controlled variant uses one extra control qubit.

The uncontrolled qubitized operator therefore uses $4m+2=8n+2$ qubits, while the controlled qubitized operator uses $5m+2=10n+2$ qubits.

The fully phase arithmetic applies GQSP~\cite{motlagh2024generalized} to implement a polynomial transformation of $Q_{\rm ph}$. For approximation parameter $d_{\rm ph}$, the GQSP sequence uses $2d_{\rm ph}$ applications of the controlled qubitized operator $cQ_{\rm ph}$, $d_{\rm ph}$ applications of the inverse qubitized operator $Q_{\rm ph}^{\dagger}$, and $2d_{\rm ph}+1$ single-qubit $\mathsf{U}_3$ rotations. We count each $\mathsf{U}_3$ rotation as non-Clifford depth $3$. The GQSP layer does not require additional work registers beyond those already needed for $cQ_{\rm ph}$, since the single-qubit rotations act on the same GQSP control qubit and the qubitized calls reuse the same registers.

The necessary resources are detailed in Table~\ref{tab:fully_phase_arithmetic_resources}. The numbers come from the explicit decomposition of the primitives into the Clifford+$\mathsf{T}$+$\mathsf{R}_z$ gate set.

\begin{table}[h!]
\centering
\begin{tabular}{lcc}
\hline
Component & Logical qubits & Non-Clifford depth \\
\hline
$\textsc{prepare}$ & $6n$ & $60n-56$ \\
$\textsc{select}$ & $8n$ & $0$ \\
$c\textsc{-select}$ & $10n$ & $6n+9$ \\
$R_0$ & $6n+2$ & $14\ell_{6n}-10$ \\
$cR_0$ & $6n+3$ & $14\ell_{6n}-10$ \\
$Q_{\rm ph}$ & $8n+2$ & $120n+14\ell_{6n}-122$ \\
$cQ_{\rm ph}$ & $10n+2$ & $126n+14\ell_{6n}-113$ \\
Fully phase arithmetic & $10n+2$ & $3d_{\rm ph}\left(126n+14\ell_{6n}-113\right)+3(2d_{\rm ph}+1)$ \\
\hline
\end{tabular}
\caption{Resource estimates for the fully phase-based Boltzmann coin. Here $d_{\rm ph}$ is the fully phase approximation parameter. The reflection and controlled-reflection bounds both use $\ell_{6n}=\lceil\log_2(6n)\rceil$. These estimates hold for $n\geq2$.}
\label{tab:fully_phase_arithmetic_resources}
\end{table}

\subsubsection{Polynomial approximation}

The GQSP approximation parameter is set by approximating the square-root Metropolis amplitude $f_\beta(\Delta H)=\min\{1,e^{-\beta\Delta H/2}\}$. The only nonsmooth point is the kink at $\Delta H=0$, so we do not require uniform accuracy inside a fixed window $|\Delta H|<\omega$, with $\omega$ a small constant independent of $n$. Writing $\ell_\varepsilon=\lceil\log_2(1/\varepsilon_{\rm ops})\rceil$, we choose
\begin{equation}
    d_{\rm ph}=2+\left\lceil\max\left\{\sqrt{\frac{\beta B}{2}\ell_\varepsilon}+\ell_\varepsilon,\frac{B}{\omega}\ell_\varepsilon\right\}\right\rceil.
\end{equation}
Here $B=\|H_{\rm ph}\|\approx2n$ is the energy scale on which the approximation is required. The first term estimates the degree needed to approximate the exponential branch. The second term estimates the degree needed to resolve the boundary of the excluded window, whose normalized width is $\omega/B$. We take the maximum because both requirements must be satisfied. The additive $2$ is a small safety margin. This choice slightly overestimates the required degree, but matches the target error in our numerical tests. Using $B=\|H_{\rm ph}\|$ avoids the larger LCU normalization $\Lambda$, which would unnecessarily overestimate the approximation interval for the normalized SK instances.

\subsection{Hybrid fixed-point and phase coin}
\label{app:hybrid_coin_resources}

The fully phase-based coin has depth growing as the product of the qubitized-operator depth and the GQSP degree. Since both scale with $n$, the total depth grows as $O(n^2)$ up to logarithmic factors. A lower-depth alternative is to compute the energy difference with fixed-point arithmetic and then use phase arithmetic only for the smooth exponential branch.

\subsubsection{Fixed-point arithmetic part}

Let $w=b+1$ be the signed fixed-point word size and define $s_{\rm SK}=\lceil\log_{3/2}(2M)\rceil$. The word size $w$ is chosen in analogy with the FPGA implementation detailed in Appendix~\ref{app:fixed_point_errors}. The factor $2M$ appears because the circuit loads one copy of every Hamiltonian term for each of the two configurations. The energy-difference block writes $H(y)-H(x)$ into a signed $w$-bit register.

The implementation is as follows.
\begin{enumerate}
    \item For each configuration, the conditional terms loader writes all one-body and two-body contributions into fixed-point rows. A one-body term loads either $+\tilde h_i$ or $-\tilde h_i$, while a two-body term loads either $+\tilde J_{ij}$ or $-\tilde J_{ij}$ depending on the spin parity. This uses fanout, parity computation, and classically controlled coefficient loading, all of which are Clifford operations. Hence, each loader has non-Clifford depth $0$ and uses $Mw+2M-n$ qubits.

    \item The two loaders produce $2M$ signed $w$-bit rows. These rows are summed by a reversible Wallace-tree adder. At each reduction level, disjoint triples of rows are compressed in parallel into two rows. The number of rows changes as $R\mapsto\lceil2R/3\rceil$, so at most $s_{\rm SK}$ levels are sufficient to reduce the initial $2M$ rows to two rows.

    \item Each three-to-two compressor contains three sequential Toffoli gates. Using the depth-$3$ Toffoli decomposition, one compressor has non-Clifford depth $9$. The forward Wallace reduction and its inverse therefore contribute at most $18s_{\rm SK}$. The final carry computation and its inverse contribute at most $18w$. Thus, the Wallace-tree adder has non-Clifford depth bounded by $D_{\rm Wallace}=18s_{\rm SK}+18w$ and uses $6Mw-2w-2M+1$ qubits. This is a convenient uniform upper bound rather than an exact depth formula.

    \item The full energy-difference block applies the loader for one configuration, the loader for the other configuration with inverted coefficients, the Wallace-tree adder, and then uncomputes both loaders. Since the loaders are Clifford-only, its non-Clifford depth remains $18s_{\rm SK}+18w$. Including the two $n$-qubit input registers, the block uses $2n+6Mw-2w-2M+1$ qubits.

    \item After the signed energy difference is computed, the positive-part selector maps $\Delta H$ to $\max\{\Delta H,0\}$. It copies and fans out the sign bit using $\mathsf{CNOT}$ gates, then applies controlled swaps between the delta register and a clean zero register. Each controlled swap is decomposed into one depth-$3$ Toffoli and Clifford gates. Since the controlled swaps are parallel, the selector has non-Clifford depth $3$ and uses $3w$ qubits.

    \item In the active-tail regime, the cutoff-tail block checks whether the positive delta is above the cutoff $2^{-j_{\rm cut}}$. For $j_{\rm cut}\geq3$, the threshold test uses a two-clean-ancilla multi-controlled operation of depth $14\ell_{j_{\rm cut}}-10$. If the value is below the cutoff, the block writes the shifted signal using $w-j_{\rm cut}-1$ sequential Toffoli gates sharing the same tail control, contributing $3(w-j_{\rm cut}-1)$. Therefore, its non-Clifford depth is $D_{\rm cutoff}=14\ell_{j_{\rm cut}}+3w-3j_{\rm cut}-13$, and it uses $2w+3$ qubits. Values above the cutoff are mapped to the saturated signal value $1$, after which the square-root exponential arithmetic is still applied.
\end{enumerate}

The implemented cutoff power is
\begin{equation}
    j_{\rm cut}=\max\left\{0,\min\left[w-1,\left\lfloor\log_2\left(\frac{\beta\Lambda}{2\log(1/\varepsilon_{\rm tail})}\right)\right\rfloor\right]\right\},
\end{equation}
where $\Lambda$ is the coefficient normalization defined in the fully phase-based construction. The resource estimate above assumes the active-tail regime $3\leq j_{\rm cut}\leq w-1$.

\begin{table}[h]
\centering
\begin{tabular}{lcc}
\hline
Component & Logical qubits & Non-Clifford depth \\
\hline
Conditional terms loader    & $Mw+2M-n$         & $0$                                       \\
Three-to-two compressor     & $5$               & $9$                                       \\
Wallace-tree adder          & $6Mw-2w-2M+1$     & $18s_{\rm SK}+18w$                        \\
Energy-difference block     & $2n+6Mw-2w-2M+1$  & $18s_{\rm SK}+18w$                        \\
Positive-part selector      & $3w$              & $3$                                       \\
Cutoff-tail block           & $2w+3$            & $14\ell_{j_{\rm cut}}+3w-3j_{\rm cut}-13$ \\
\hline
\end{tabular}
\caption{Resource estimates for the fixed-point part of the hybrid coin. Here $s_{\rm SK}=\lceil\log_{3/2}(2M)\rceil$, while $j_{\rm cut}$ is the cutoff power. The cutoff-tail estimate assumes the active-tail regime $3\leq j_{\rm cut}\leq w-1$.}
\label{tab:delta_energy_block_resources}
\end{table}

\subsubsection{Phase arithmetic part}

After the fixed-point block has computed the signed energy difference, the circuit first selects the positive branch and then applies the smooth exponential arithmetic. In the active-tail regime, the cutoff block restricts the approximation interval by shifting values below the cutoff and mapping the discarded tail to the saturated signal value $1$.

The square-root exponential is implemented by one-body qubitization and GQSP. The one-body Hamiltonian used for this qubitization is not the SK Hamiltonian. It is a signal Hamiltonian acting on the $w$-qubit fixed-point signal register:
\begin{equation}
    H_{\rm sig}=\frac{1}{2}Z_{w-1}-\sum_{j=0}^{w-2}2^{j-w}Z_j,\qquad H_{\rm sig}\ket{r}=(\eta-r)\ket{r},\qquad \eta=1+3\cdot2^{-w}.
\end{equation}
Here $r$ is the value encoded in the signal register after the optional cutoff rescaling. Thus, approximating $r\mapsto e^{-\beta\Lambda_{\rm eff}r/2}$ is equivalent to applying a polynomial transformation of $H_{\rm sig}$, since $r=\eta-H_{\rm sig}$ on valid signal states. In the active-tail regime, $\Lambda_{\rm eff}=\Lambda2^{-j_{\rm cut}}$ because the cutoff block rescales the input by $2^{j_{\rm cut}}$ before the exponential approximation.

The implementation is as follows.
\begin{itemize}
    \item The one-body $\textsc{prepare}$ is a balanced Givens tree over the $w$ signal coefficients. It has non-Clifford depth $2\ell_w$ and uses $w$ qubits.

    \item The one-body $\textsc{select}$ applies the selected Pauli $Z_j$ and the coefficient sign. It uses only Clifford gates, so its non-Clifford depth is $0$ and it acts on $2w$ qubits.

    \item The controlled one-body $\textsc{select}$ fans out the external GQSP control and applies the selected operation using controlled-controlled-$Z$ gates. Each controlled-controlled-$Z$ is implemented by conjugating a depth-$3$ Toffoli with Hadamard gates. Since these gates are parallel across the $w$ signal positions, the controlled $\textsc{select}$ has non-Clifford depth $3$ and uses $3w$ qubits.

    \item The reflection about the zero state of the one-body selection register uses the same two-clean-ancilla multi-controlled construction as the cutoff test. Both the uncontrolled and controlled reflections have non-Clifford depth bounded by $14\ell_w-10$, and use $w+2$ and $w+3$ qubits, respectively.

    \item The one-body qubitized operator consists of $\textsc{prepare}$, $\textsc{select}$, $\textsc{prepare}^{\dagger}$, and the reflection. Its non-Clifford depth is $18\ell_w-10$. The controlled version replaces $\textsc{select}$ and the reflection by their controlled variants, giving non-Clifford depth $18\ell_w-7$.

    \item The implementation constructs Laurent coefficients indexed by $k=-d_{\rm hyb},\ldots,d_{\rm hyb}$. The resulting GQSP sequence uses $2d_{\rm hyb}$ controlled one-body qubitized calls, $d_{\rm hyb}$ inverse uncontrolled one-body qubitized calls, and $2d_{\rm hyb}+1$ single-qubit $\mathsf{U}_3$ rotations. Counting each $\mathsf{U}_3$ rotation as non-Clifford depth $3$, the square-root exponential arithmetic has depth $D_{\sqrt{\exp}}=d_{\rm hyb}(54\ell_w-18)+3$ and uses $3w+2$ qubits.
\end{itemize}

\begin{table}[h]
\centering
\begin{tabular}{lcc}
\hline
Component & Logical qubits & Non-Clifford depth \\
\hline
One-body $\textsc{prepare}$                 & $w$    & $2\ell_w$                    \\
One-body $\textsc{select}$                  & $2w$   & $0$                          \\
Controlled one-body $\textsc{select}$       & $3w$   & $3$                          \\
One-body reflection                         & $w+2$  & $14\ell_w-10$                \\
Controlled one-body reflection              & $w+3$  & $14\ell_w-10$                \\
One-body qubitized operator                 & $2w+2$ & $18\ell_w-10$                \\
Controlled one-body qubitized operator      & $3w+2$ & $18\ell_w-7$                 \\
Square-root exponential arithmetic          & $3w+2$ & $d_{\rm hyb}(54\ell_w-18)+3$ \\
\hline
\end{tabular}
\caption{Resource estimates for the phase-arithmetic part of the hybrid Boltzmann coin. Here $d_{\rm hyb}$ is the degree of the polynomial used by the hybrid GQSP construction.}
\label{tab:hybrid_coin_resources}
\end{table}

Let $\varepsilon_{\rm ops}$ denote the error budget passed to the complete hybrid arithmetic. We find numerically that a good degree approximation for this phase arithmetic is:
\begin{equation}
    d_{\rm hyb}\approx\left\lceil\log\left(\frac{1}{\varepsilon_{\rm tail}}\right)+\log\left(\frac{1}{\varepsilon_{\rm ops}}\right)+1\right\rceil.
\end{equation}

The complete hybrid arithmetic applies the energy-difference block, the positive-part selector, the cutoff-tail block, and the square-root exponential arithmetic, followed by the inverse cutoff-tail block, the inverse positive-part selector, and the inverse energy-difference block. Its total non-Clifford depth is therefore
\begin{align}
    D_{\rm hybrid}&=d_{\rm hyb}(54\ell_w-18)+3+2(18s_{\rm SK}+18w)+2\cdot3+2(14\ell_{j_{\rm cut}}+3w-3j_{\rm cut}-13)\\
    &=d_{\rm hyb}(54\ell_w-18)+36s_{\rm SK}+42w-6j_{\rm cut}+28\ell_{j_{\rm cut}}-17.
\end{align}
The complete block uses $2n+6Mw-2M+2$ logical qubits.

\subsection{Reflection and accept-path unitary}
\label{app:reflection_accept_path_resources}

\begin{table}[h!]
\centering
\begin{tabular}{lcc}
\hline
Component & Logical qubits & Non-Clifford depth \\
\hline
Reflection & $2n+w+3$ & $14\ell_{n+w}-10$ \\
Accept-path unitary & $3n+w+1$ & $28\ell_{w+1}-11$ \\
\hline
\end{tabular}
\caption{Reflection and accept-path resources for the hybrid Boltzmann coin. The persistent coin register contains the GQSP control qubit and the $w$-qubit one-body selection register.}
\label{tab:reflection_accept_hybrid_resources}
\end{table}

For the hybrid Boltzmann coin, the temporary fixed-point registers, the rescaled signal register, the controlled-\textsc{select} fanout register, and the reflection ancillas are returned to $\ket{0}$ within the coin unitary. The persistent coin register therefore consists only of the GQSP control qubit and the $w$-qubit one-body selection register, giving $q_{\rm coin}=w+1$.

The reflection acts on the proposed-state register and the persistent coin register,
\begin{equation}
    R_0=2\left(I_A\otimes\ket{0^n}\!\bra{0^n}_B\otimes\ket{0^{w+1}}\!\bra{0^{w+1}}_{\rm coin}\right)-I.
    \label{eq:walk_reflection_circuit}
\end{equation}
The all-zero condition involves $n+w+1$ qubits, so the selective phase flip has $n+w$ controls. Using the multi-controlled construction of Appendix~\ref{app:mcx}, the reflection has non-Clifford depth $14\ell_{n+w}-10$. Including the two $n$-qubit system registers and the two clean ancillas required by the multi-controlled operation, it uses $2n+w+3$ logical qubits.

The accept-path unitary swaps the two system registers only when the persistent coin register is in the accepted state,
\begin{equation}
    F\ket{x}_A\ket{y}_B\ket{z}_{\rm coin}=\begin{cases}\ket{y}_A\ket{x}_B\ket{z}_{\rm coin},&z=0^{w+1},\\\ket{x}_A\ket{y}_B\ket{z}_{\rm coin},&z\neq0^{w+1}.\end{cases}
    \label{eq:accept_path_unitary_circuit}
\end{equation}
The circuit computes the zero-coin condition into one flag using the construction of Appendix~\ref{app:mcx}, fans the flag out using Clifford gates, applies the $n$ controlled swaps in parallel, and then uncomputes the flag. Each controlled swap is implemented by three sequential Toffoli gates. Since the swaps act on disjoint system-qubit pairs, they are parallel across the $n$ pairs, while the swap stage has non-Clifford depth $9$. Computing and uncomputing the flag contribute $2(14\ell_{w+1}-10)$, giving a total non-Clifford depth of $28\ell_{w+1}-11$. For $n\geq3$, the circuit uses $3n+w+1$ logical qubits.

\section{Details in the Gibbs sampling with quantum walks}
\label{app:annealed_spectral_filtering}

Section \ref{sec:annealed_state_preparation} has introduced the Gibbs sampling algorithm with quantum walks. This appendices integrates details regarding the choice of the annealing path and how to implement each step of the annealing.

\subsection{Annealing schedule}

A direct spectral projection onto $\Pi_{\beta}=\ketbra{\pi_\beta}{\pi_\beta}$ starting from the uniform coherent state $\ket{\pi_{\beta_0}}$, with $\beta_0=0$, may have exponentially small success probability because the overlap $|\braket{\pi_\beta}{\pi_{\beta_0}}|^2$ can decrease exponentially with the system size. Instead, we use instead an annealing schedule $0=\beta_0<\beta_1<\cdots<\beta_L=\bar\beta$ and sequentially apply the projectors $\Pi_{\beta_j}$. Each $\beta_{j+1}$ is chosen so that consecutive coherent Gibbs states satisfy $|\braket{\pi_{\beta_{j+1}}}{\pi_{\beta_j}}|^2\geq 1/e$, ensuring that every projection succeeds with constant probability.

It remains to determine how many intermediate inverse temperatures are required and how the values $\beta_j$ should be chosen for a given instance. Following~\cite{wocjan2008speedup}, a simple sufficient condition follows from the bound $\left|\braket{\pi_{\beta}}{\pi_{\beta+\Delta\beta}}\right|^2\geq\exp\!\left(-\norm{H}\Delta\beta\right)$. Therefore, for the choice $p_0=1/e$, it is sufficient to choose
\begin{equation}
    \Delta\beta
    \leq
    \frac{\log(1/p_0)}{\norm{H}}
    =
    \frac{1}{\norm{H}}.
\end{equation}

This is a general bound valid for any Ising Hamiltonian. For the SK model, a less conservative local estimate can be obtained from the energy variance. Writing $Z(\beta)$ for the partition function, the overlap between two consecutive coherent Gibbs states is
\begin{equation}
\left|\braket{\pi_{\beta}}{\pi_{\beta+\Delta\beta}}\right|^2
=
\frac{Z\left(\beta+\frac{\Delta\beta}{2}\right)^2}{Z(\beta)Z(\beta+\Delta\beta)}.
\end{equation}
Taking the logarithm and Taylor-expanding $\log Z(\beta)$ symmetrically around the midpoint $\beta+\Delta\beta/2$ gives
\begin{align}
\log\left|\braket{\pi_{\beta}}{\pi_{\beta+\Delta\beta}}\right|^2
&=
2\log Z\left(\beta+\frac{\Delta\beta}{2}\right)
-\log Z(\beta)
-\log Z(\beta+\Delta\beta)
\nonumber\\
&=
-\frac{(\Delta\beta)^2}{4}
\frac{d^2}{d\beta^2}
\log Z\left(\beta+\frac{\Delta\beta}{2}\right)
+
O\left((\Delta\beta)^4\right).
\end{align}
The odd-order terms cancel because the expansion is symmetric around the midpoint. Since the second derivative of the logarithm of the partition function is the Gibbs energy variance, we obtain
\begin{equation}
\log\left|\braket{\pi_{\beta}}{\pi_{\beta+\Delta\beta}}\right|^2
=
-\frac{(\Delta\beta)^2}{4}
\operatorname{Var}_{\pi_{\beta+\Delta\beta/2}}[H]
+
O\left((\Delta\beta)^4\right).
\end{equation}
Therefore, to obtain squared overlap approximately equal to $p_0=1/e$, the local step size is
\begin{equation}
\Delta\beta
\simeq
2\sqrt{\frac{\log(1/p_0)}{\operatorname{Var}_{\pi_{\beta+\Delta\beta/2}}[H]}}
=
\frac{2}{\sqrt{\operatorname{Var}_{\pi_{\beta+\Delta\beta/2}}[H]}}.
\end{equation}
We denote by $\sigma_H^2(n,\beta)=\mathbb{E}_{H\sim\mathrm{SK}_n}\!\left[\operatorname{Var}_{\pi_\beta}[H]\right]$ the disorder-averaged Gibbs energy variance of an $n$-spin SK Hamiltonian at inverse temperature $\beta$, where the expectation is taken over the ensemble of SK instances. For the normalization considered here, the energy variance is extensive in the system size, so we separate its temperature-dependent contribution via:
\begin{equation}
    \sigma_H^2(n,\beta)=n\,v(\beta).
\end{equation}

\begin{figure}[h!]
    \centering
    \includegraphics[width=0.9\linewidth]{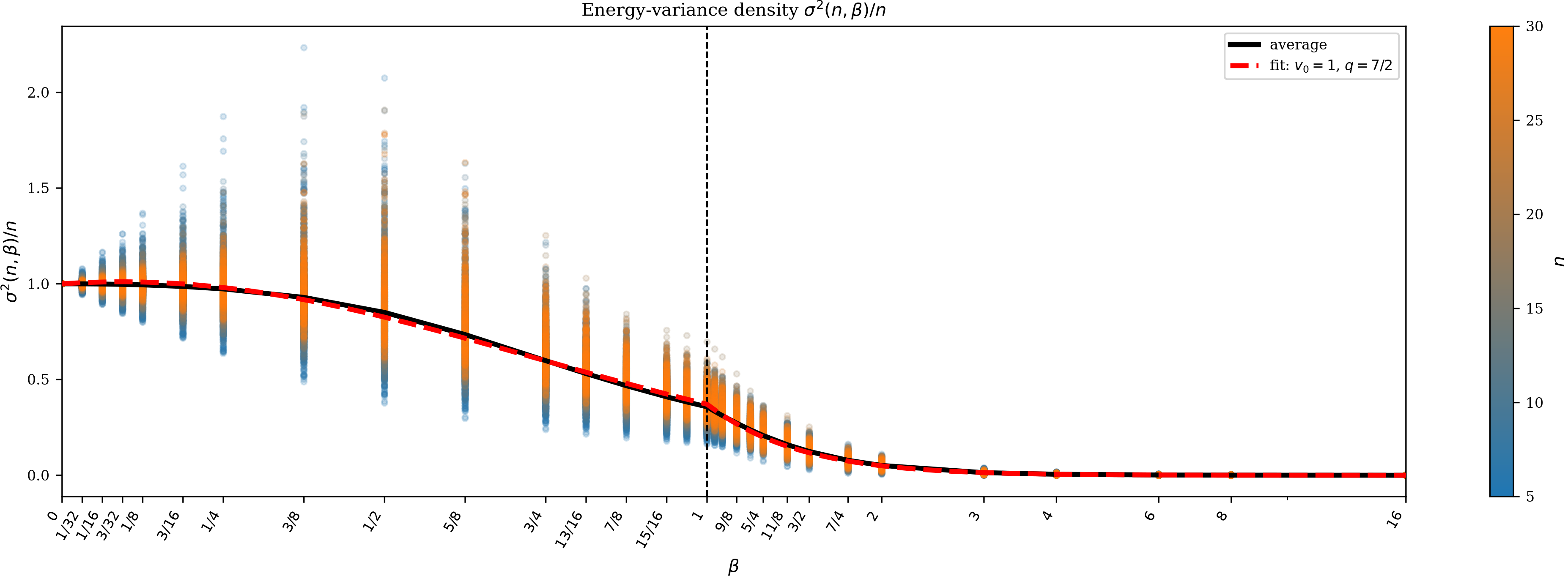}
    \caption{Variance density $v(\beta)$ estimated from 100 SK instances for each system size $n=5,\ldots,30$, with colors ranging from blue to orange as $n$ increases, and inverse temperature $\beta\in[0,16]$. The black line denotes the average across instances and system sizes. The increasing concentration of the data with $n$ supports the extensive scaling $\sigma_H^2(n,\beta)\simeq n\,v(\beta)$. The red dashed line representing the fitted ansatz $v(\beta)$. The vertical dashed line marks the critical inverse temperature $\beta=1$.}
    \label{fig:annealing_extensibility}
\end{figure}

Figure~\ref{fig:annealing_extensibility} shows the instance-wise estimates of the variance density $v(\beta)$ for 100 SK instances at each system size $n=5,\ldots,30$ and inverse temperature $\beta$ sampled between $0$ and $16$. For each instance, the plotted quantity is $\operatorname{Var}_{\pi_\beta}[H]/n$, whose disorder average defines $v(\beta)$ through $\sigma_H^2(n,\beta)\approx n\,v(\beta)$. As $n$ increases, the points concentrate around the average curve, indicating self-averaging and supporting the extensive scaling of the energy variance. The curve also exhibits a change in behavior around $\beta=1$, consistent with the finite-size signature of the spin-glass transition of the SK model.

This suggests that we can fit a numerical ansatz to the disorder-averaged data underlying the black line to estimate the variance density $v(\beta)$ and hence determine the local annealing step $\Delta\beta(n,\beta)$. We first fit the generic piecewise ansatz
\begin{equation}
v(\beta)=
\begin{cases}
v_0\exp\left[\log\left(\dfrac{v_1}{v_0}\right)\beta+\kappa\beta(1-\beta)\right], & \beta<1,\\[2mm]
v_1\left(1+\dfrac{\beta-1}{\tau}\right)^{-q}, & \beta\geq1,
\end{cases}
\end{equation}
where all five parameters $v_0$, $v_1$, $\kappa$, $\tau$, and $q$ are fitted by nonlinear least squares in logarithmic space. This unrestricted fit gives $v_0=0.9908$, $v_1=0.3637$, $\kappa=1.2731$, $\tau=1.3042$, and $q=3.5210$, with logarithmic RMSE $0.14575$. Since the fitted values are very close to the expected value $v(0)=v_0=1$ and to the exponent $q=7/2$, we then fix these two parameters and refit the remaining ones. The reduced ansatz is
\begin{equation}
v(\beta)=
\begin{cases}
\exp\left[\log(v_1)\beta+\kappa\beta(1-\beta)\right], & \beta<1,\\[2mm]
v_1\left(1+\dfrac{\beta-1}{\tau}\right)^{-7/2}, & \beta\geq1,
\end{cases}
\end{equation}
and the second fit gives $v_1=0.36509$, $\kappa=1.23536$, and $\tau=1.28646$, with logarithmic RMSE $0.14582$. The negligible change in the fitting error supports fixing $v_0=1$ and $q=7/2$.

To obtain a conservative schedule that also accounts for the observed instance-to-instance fluctuations, we multiply the fitted variance density by $4$ for $\beta<1$ and by $16$ for $\beta\geq1$. Using the rounded fitted values, the resulting annealing step is
\begin{equation}
\Delta\beta(n,\beta)
=
\frac{1}{\sqrt{n}}
\begin{cases}
\exp\left(-0.111\beta+0.615\beta^2\right), & \beta<1,\\[2mm]
0.533(\beta+0.286)^{7/4}, & \beta\geq1.
\end{cases}
\end{equation}
In the code, we provided the method  \texttt{tight\_schedule\_annealing} that constructs the schedule iteratively as $\beta_{j+1}=\min\{\beta,\beta_j+\Delta\beta(n,\beta_j)\}$ and always includes the point $\beta=1$.

\subsection{Spectral projections}
\label{app:spectral_projection}

Each annealing step at inverse temperature $\beta_j$ is intended to apply the ideal spectral projector $\Pi_{\beta_j}=\ketbra{\pi_{\beta_j}}{\pi_{\beta_j}}$, which removes all components orthogonal to the coherent Gibbs state. In practice, this projector can only be approximated, yielding a filter $\widetilde{\Pi}_{\beta_j}=\Pi_{\beta_j}+R_{\beta_j}$, where the residual leakage satisfies $\norm{R_{\beta_j}}_{\mathrm{op}}\leq\varepsilon_j$ for some predefined error upper bound $\varepsilon_j$.

An efficient way to implement such a filter is through quantum singular value transformation~\cite{gilyen2019quantum} applied to the Szegedy walk $W_{\beta_j}$. Let $\delta_{\beta_j}$ be the spectral gap of the discriminant matrix associated with the Markov chain, and let $T_k$ denote the Chebyshev polynomial of the first kind of degree $k$. For an even degree $d$, the polynomial implemented by the quantum singular value transformation can be chosen as
\begin{equation}
    p_d(x)
    =
    \frac{
    T_{d/2}\left(\dfrac{2x^2}{(1-\delta_{\beta_j})^2}-1\right)
    }{
    T_{d/2}\left(\dfrac{2}{(1-\delta_{\beta_j})^2}-1\right)
    }.
\end{equation}
This polynomial satisfies $p_d(1)=1$ and suppresses the unwanted spectral region $|x|\leq1-\delta_{\beta_j}$. Since $p_d(x)$ depends only on $x^2$, it has even parity and can be implemented directly by alternating applications of $W_{\beta_j}$ and $W_{\beta_j}^{\dagger}$, without combining even- and odd-parity transformations through an additional linear-combination construction.

The required polynomial degree is inversely proportional to the phase-separation parameter $\theta_{\beta_j}=\arccos(1-\delta_{\beta_j})$ and depends logarithmically on the target leakage. Based on the numerical validation of the filter, in the resource estimates we use
\begin{equation}
    d_j
    =
    \frac{
    2\log_2\left(\varepsilon_j^{-1}\right)
    }{
    \theta_{\beta_j}
    }.
\end{equation}

\subsection{Zeno-rewind protocol and total number of walk calls}
\label{app:total_walk_calls}

Let $L$ be the number of annealing steps and define $p_j=\left|\braket{\pi_{\beta_{j-1}}}{\pi_{\beta_j}}\right|^2\geq p_0$. If the filters were applied sequentially and the complete procedure were restarted after any failure, the probability of preparing the final state in a single run would be
\begin{equation}
    p_{\rm succ}
    =
    \prod_{j=1}^{L}p_j
    \geq
    p_0^L.
\end{equation}
For $p_0=1/e$, this lower bound is $e^{-L}$, so naive postselection may require an exponentially large number of repetitions.

We instead use the Zeno-rewind protocol~\cite{wocjan2008speedup}, which corrects a failed transition rather than restarting the complete annealing path. For the transition from $\ket*{\pi_{\beta_{j-1}}}$ to $\ket*{\pi_{\beta_j}}$, let $p_j=\left|\braket{\pi_{\beta_j}}{\pi_{\beta_{j-1}}}\right|^2$ and write
\begin{equation}
    \ket*{\pi_{\beta_{j-1}}}
    =
    \sqrt{p_j}\ket*{\pi_{\beta_j}}
    +
    \sqrt{1-p_j}\ket*{\pi_{\beta_j}^{\perp}}.
\end{equation}
The protocol proceeds as follows:
\begin{enumerate}
    \item The filter $\Pi_{\beta_j}$ is first applied to $\ket*{\pi_{\beta_{j-1}}}$.
    \item If it succeeds, the state $\ket*{\pi_{\beta_j}}$ is obtained and the transition is complete.
    \item If it fails, the state $\ket*{\pi_{\beta_j}^{\perp}}$ is obtained. Since
    \begin{equation}
        \ket*{\pi_{\beta_j}^{\perp}}
        =
        \sqrt{1-p_j}\ket*{\pi_{\beta_{j-1}}}
        -
        \sqrt{p_j}\ket*{\pi_{\beta_{j-1}}^{\perp}},
    \end{equation}
    the filter $\Pi_{\beta_{j-1}}$ is then applied:
    \begin{enumerate}
        \item if it succeeds, the state $\ket*{\pi_{\beta_{j-1}}}$ is recovered and the transition can be retried;
        \item if it fails, the state $\ket*{\pi_{\beta_{j-1}}^{\perp}}$ is obtained.
    \end{enumerate}
    \item Since
    \begin{equation}
        \ket*{\pi_{\beta_{j-1}}^{\perp}}
        =
        \sqrt{1-p_j}\ket*{\pi_{\beta_j}}
        -
        \sqrt{p_j}\ket*{\pi_{\beta_j}^{\perp}},
    \end{equation}
    the filter $\Pi_{\beta_j}$ is applied again:
    \begin{enumerate}
        \item if it succeeds, the state $\ket*{\pi_{\beta_j}}$ is obtained and the transition is complete;
        \item if it fails, the state $\ket*{\pi_{\beta_j}^{\perp}}$ is recovered, and the alternating sequence of filters $\Pi_{\beta_{j-1}}$ and $\Pi_{\beta_j}$ is repeated.
    \end{enumerate}
\end{enumerate}

Let $d_j$ be the number of walk calls required to implement the filter at $\beta_j$. Denoting by $N_j^{(0)}$, $N_j^{(1)}$, and $N_j^{(2)}$ the expected remaining numbers of walk calls when starting from $\ket*{\pi_{\beta_{j-1}}}$, $\ket*{\pi_{\beta_j}^{\perp}}$, and $\ket*{\pi_{\beta_{j-1}}^{\perp}}$. We have
\begin{align}
    N_j^{(0)}
    &=
    d_j+(1-p_j)N_j^{(1)},
    \nonumber\\
    N_j^{(1)}
    &=
    d_{j-1}+(1-p_j)N_j^{(0)}+p_jN_j^{(2)},
    \nonumber\\
    N_j^{(2)}
    &=
    d_j+p_jN_j^{(1)}.
\end{align}
Solving these equations gives
\begin{equation}
    N_j^{(0)}
    =
    \left(1+\frac{1}{2p_j}\right)d_j
    +
    \frac{1}{2p_j}d_{j-1}.
    \label{eq:zeno_rewind_transition_cost}
\end{equation}

Since $p_j\geq p_0$, each transition incurs only a constant expected overhead. Taking $d_0=0$, since the initial uniform state is prepared directly, and summing over all annealing steps gives an \emph{expected} number of quantum queries equals to:
\begin{align}
    \sum_{j=1}^{L}N_j^{(0)}
    \leq
    \sum_{j=1}^{L}d_j
    +
    \frac{1}{2p_0}
    \left(
    \sum_{j=1}^{L}d_j
    +
    \sum_{j=1}^{L}d_{j-1}
    \right)
    \leq
    \left(1+\frac{1}{p_0}\right)
    \sum_{j=1}^{L}d_j := Q_{\rm q}(n,\bar\beta,\varepsilon)
    \label{eq:total_walk_queries}
\end{align}
This matches the definition in Eqs.~\ref{eq:quantum_annealing_queries} and~\ref{eq:def_Q_q_appendix_c} once substituted the degree term. For $p_0=1/e$, the expected Zeno-rewind overhead is therefore bounded by the constant factor $1+e$, rather than growing exponentially with the number of annealing steps.

\section{Fault-tolerant implementation model}
\label{app:fault_tolerant_model}

This appendix explains how the logical circuit implementing the quantum walk-based Gibbs sampling algorithm is converted into a wall-clock runtime estimate. The model is designed to provide credible estimates while remaining as platform-independent as possible, rather than being tailored to a specific hardware technology. We nevertheless adopt the surface code as the error correction model because of its technological maturity relative to more recent proposals. This choice makes the model most directly applicable to solid-state platforms, such as superconducting qubits and silicon spin qubits, and we therefore use physical operation times compatible with these technologies.

\subsection{Logical model}
\label{app:logical_model}

The walk operator is first expressed in the standard quantum circuit formalism. In Appendix~\ref{app:quantum_walk_circuit_resources}, we derive upper bounds on its non-Clifford depth and use this quantity as a proxy for the logical execution time. To connect this circuit level description with a surface code implementation, we then translate the circuit into a logically equivalent representation that is more convenient to run on the quantum error correction level, which in our case are expected to be lattice surgery instructions running on a surface code \cite{horsman2012surface}. Following Ref.~\cite{litinski2019game}, we use the Clifford+$\varphi$ formalism, in which the computation is expressed in terms of Pauli product measurements and non-Clifford Pauli rotations.

Within this formalism, Clifford corrections are absorbed into the Pauli products measured at later stages and tracked classically through updates of the Pauli frame, so their execution cost is considered negligible. The operations contributing to the logical runtime are therefore the Pauli product measurements required to implement non-Clifford rotations. In particular, a $\mathsf{T}$ gate consumes the magic state $\ket{T}=\mathsf{T}\ket{+}=(\ket{0}+e^{i\pi/4}\ket{1})/\sqrt{2}$, while an arbitrary $\mathsf{R}_z(\theta)$ rotation consumes the corresponding resource state $\ket{\mathsf{R}_z(\theta)}=\mathsf{R}_z(\theta)\ket{+}$. A layer of non-Clifford gates in the original circuit is thus mapped to a layer of Pauli product measurements consuming the required magic states.

Turning this logical description into an explicit lattice surgery computation requires choosing a spatial layout. Each logical data qubit is represented by a surface code patch, while auxiliary routing regions connect the data patches and make it possible to measure nonlocal Pauli products. Once a particular layout is fixed, several measurements may compete for the same routing space or for access to the same magic state factories. Operations that are parallel in the abstract circuit may consequently need to be serialized, and the critical path of the routed computation may become longer than the non-Clifford depth of the original circuit. A compiler that maps large logical circuits to layout dependent lattice surgery instructions is presented in Ref.~\cite{watkins2024high}. A particularly restrictive layout places the data qubits and routing bus on two parallel lines. In this geometry, Pauli product measurements with intersecting routing regions must be serialized, which can introduce an $O(n)$ scheduling overhead in the worst case, while more connected layouts can nullify this slowdown.

This layout dependence is particularly relevant in our setting because the logical space required by the walk operator scales quadratically with $n$. For sufficiently large problem sizes, the complete computation may no longer fit on a single chip and may instead require several interconnected modules. The topology, bandwidth, and latency of these interconnects would then impose additional constraints on the implementation of nonlocal Pauli product measurements. These effects cannot be quantified without committing to a particular physical architecture and modular connectivity.

We therefore do not model the routing and scheduling layer explicitly. Instead, we assume that all Pauli product measurements associated with the same non-Clifford layer can be executed in parallel, so that the logical execution time is determined directly by the non-Clifford depth. To account for the architectural overhead omitted by this assumption, together with the additional uncertainties introduced in the next section, we vary the physical operation and measurement times over two orders of magnitude.

The conversion from logical resources to code distance and wall clock runtime follows the surface code model of Ref.~\cite{beverland2022assessing}, summarized in Eq.~\ref{eq:surface_code_logical_error} and Eq.~\ref{eq:surface_code_runtime} of the main text. The spacetime volume $QDS$ includes the algorithmic logical qubits and ancillas, but excludes layout dependent routing patches and magic state factories.

\subsection{Arbitrary angle rotations}
\label{app:arbitrary_angle_rotations}

One advantage of retaining the Clifford+$\varphi$ formalism, rather than immediately compiling the circuit into the more commonly considered Clifford+$\mathsf{T}$ gate set, is that arbitrary rotations can be implemented using angle specific resource states prepared offline. Alternatively, each $\mathsf{R}_z(\theta)$ rotation can be approximated by a Clifford+$\mathsf{T}$ circuit to precision $\varepsilon_{\rm rot}$. Near optimal synthesis requires $O(\log(1/\varepsilon_{\rm rot}))$ $\mathsf{T}$ gates and a comparable non-Clifford depth for a single qubit rotation~\cite{ross2016optimal}. Although this dependence is logarithmic, replacing every arbitrary rotation by a sequence of $\mathsf{T}$ gates can increase the cost of the quantum walk considerably, as also illustrated in Ref.~\cite[c.f. Table~2]{lemieux2020efficient}.

The required resource states must themselves be prepared fault tolerantly. Magic states for $\mathsf{T}$ gates can be produced through cultivation or magic state distillation, while dedicated preparation and distillation protocols can be used for the angle specific states $\ket{\mathsf{R}_z(\theta)}=\mathsf{R}_z(\theta)\ket{+}$. Our model includes the allowed failure probability of these states in the total error budget, but does not estimate the additional qubits or time required for their preparation. The number, capacity, and position of the corresponding factories are part of the layout and routing problem discussed in the previous section.

The injection of a resource state is probabilistic. For a $\mathsf{T}$ gate, the associated Pauli product measurement applies either $\mathsf{R}_z(\pi/4)$ or $\mathsf{R}_z(-\pi/4)$, each with probability $1/2$. The undesired result can be corrected by applying $\mathsf{R}_z(\pi/2)$, which is a Clifford operation and can therefore be absorbed into the Clifford frame without adding another non-Clifford operation.

The same sign ambiguity occurs for an arbitrary rotation. If the first injection applies $\mathsf{R}_z(-\theta)$ instead of $\mathsf{R}_z(\theta)$, the required correction is $\mathsf{R}_z(2\theta)$. This correction is itself probabilistic and, if it has the wrong sign, must be followed by $\mathsf{R}_z(4\theta)$, then by $\mathsf{R}_z(8\theta)$, and so forth. Taking the intended rotation as the zeroth injection, the resource state for the angle $2^k\theta$ is required with probability $2^{-k}$. The expected number of Pauli product measurements is therefore $\sum_{k=0}^{\infty}2^{-k}=2$, corresponding to the intended rotation and, on average, one corrective rotation.

This expectation assumes that all resource states required by the correction sequence are ready when needed. Preparing every possible correction in advance would consume resources for states that are rarely used, while preparing them only after the corresponding measurement outcome introduces a stochastic delay. Rotations outside the critical path may hide part of this delay behind other operations, whereas rotations on the critical path require the corrective states to be supplied more promptly. Since the relevant critical path is determined only after the circuit has been mapped to a specific layout and routing schedule, we do not model this effect explicitly. Instead, the uncertainty associated with resource state production, delivery, and probabilistic corrections is absorbed into the range of physical operation times considered in our estimates.

\subsection{Surface code distance}

\begin{figure}[p!]
    \centering
    \scalebox{0.6}{\input{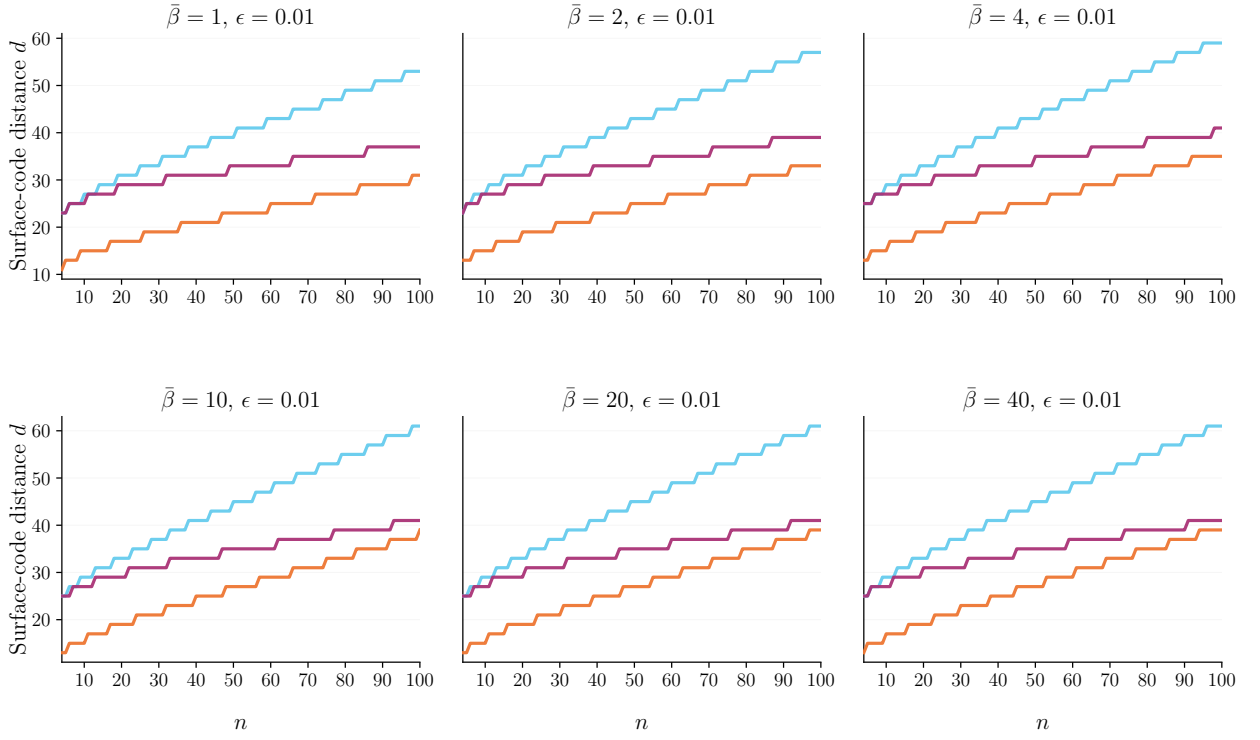}}
    \caption{Surface code distance over the full annealing path as a function of $n$, for $\varepsilon=10^{-2}$ and selected target inverse temperatures $\bar\beta$. The orange line shows classical Monte Carlo with the Hamiltonian-simulation proposal, while the light-blue and magenta lines show the quantized uniform and Hamiltonian-simulation proposals, respectively.}
    \label{fig:surface_code_different_betas}
\end{figure}

\begin{figure}[p!]
    \centering
    \scalebox{0.6}{\input{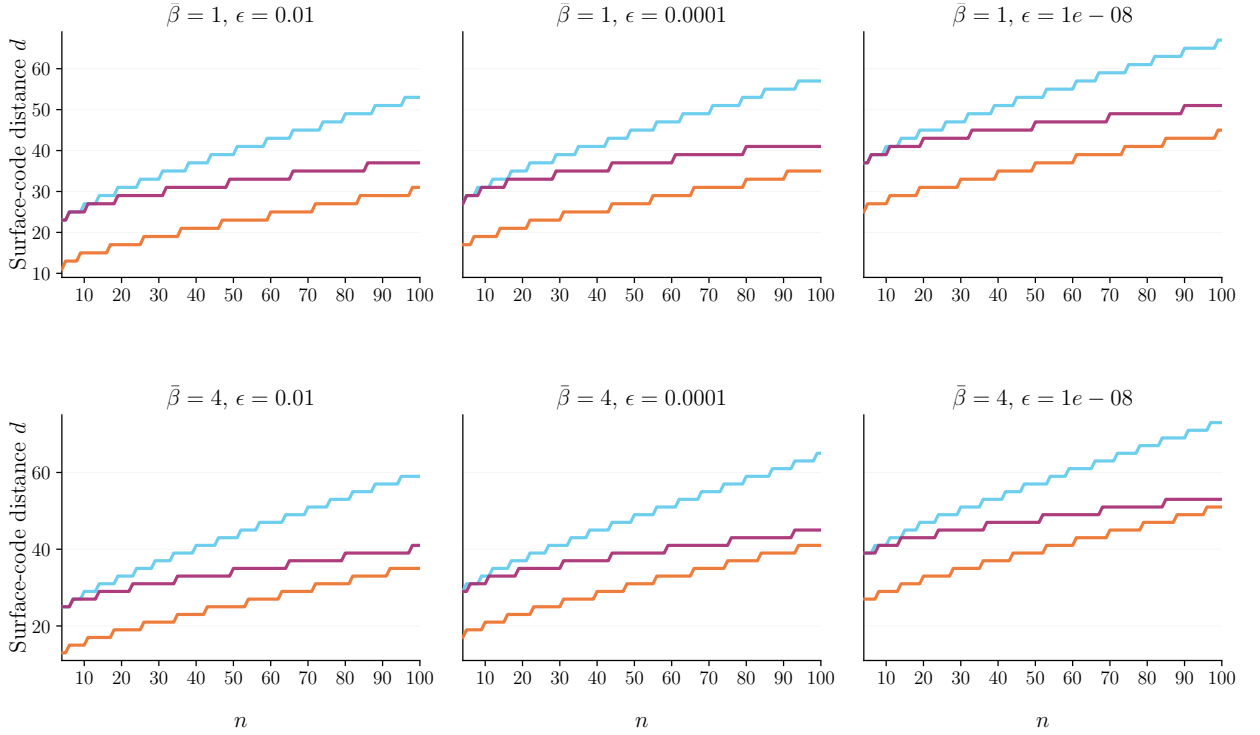}}
    \caption{Same as Figure~\ref{fig:surface_code_different_betas}, but for selected target inverse temperatures $\bar\beta$ and TV-distance errors $\varepsilon=10^{-2}$, $10^{-4}$, and $10^{-8}$.}
    \label{fig:surface_code_different_epsilon}
\end{figure}

We complete the analysis by showing the surface code distance required by our model at fixed physical error rate $p_{\rm phys}=10^{-3}$, according to Eq.~\ref{eq:surface_code_logical_error}. Figure~\ref{fig:surface_code_different_betas} shows the dependence on the target inverse temperature, while Figure~\ref{fig:surface_code_different_epsilon} shows the dependence on the target TV-distance error. Note that, at small $n$, classical Monte Carlo with the Hamiltonian-simulation move requires far fewer resources because it simulates only a single step of the unitary rather than the full annealing path.

\section{Runtime plots}
\label{app:runtime_plots}

This section combines the query-count estimates with the cost of implementing each transition on the corresponding classical or fault-tolerant quantum hardware model. The resulting runtimes therefore include the full annealing schedule rather than only its final step.

\begin{figure}[h!]
    \centering
    \scalebox{0.6}{\input{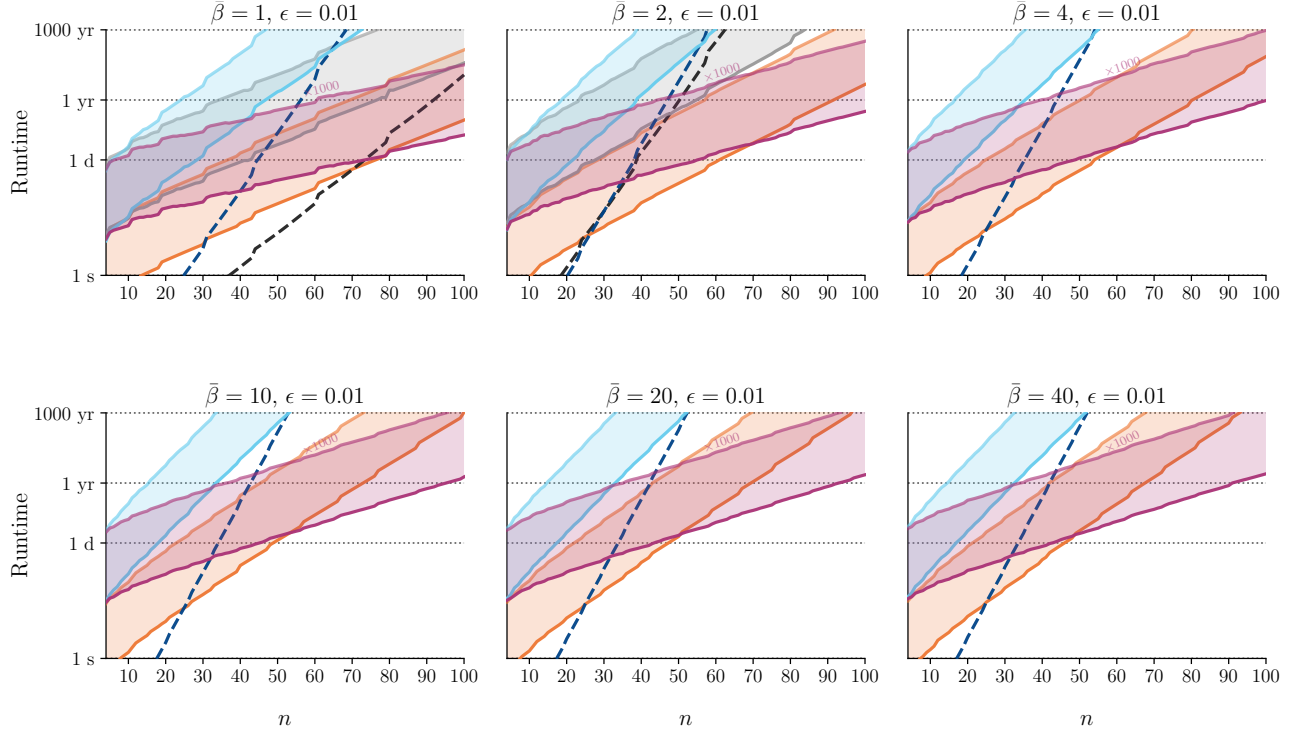}}
    \caption{Annealing runtime as a function of $n$ at fixed TV-distance error $\varepsilon=0.01$ and several final inverse temperatures $\bar\beta$. Dashed curves show classical Monte Carlo with local (dark gray), uniform (dark blue), and Hamiltonian simulation (orange) proposals. Solid curves show the corresponding quantum-walk algorithms with local (gray), uniform (light blue), and Hamiltonian simulation (magenta) proposals. Shaded regions span the optimistic and pessimistic runtime estimates under the physical hardware assumptions. Local move results are omitted for target $\beta\geq4$. Horizontal dotted lines indicate representative time scales.}
    \label{fig:runtime_different_betas}
\end{figure}

Figure~\ref{fig:runtime_different_betas} shows the runtime as a function of $n$ for several final inverse temperatures $\bar\beta$ at fixed target error $\varepsilon=0.01$. Increasing target $\bar\beta$ increases the runtime because sampling at lower temperatures requires more demanding annealing steps and smaller spectral gaps. The Hamiltonian simulation proposal provides a substantially better scaling than the uniform classical proposal (best classical move at lower temperatures), while its quantum-walk implementation further reduces the asymptotic slope. At small $n$, however, the fault-tolerant overhead dominates, and the classical Monte Carlo with Hamiltonian simulation move remain faster. The lower asymptotic scaling of the quantum walk approach leads to a crossover at larger $n$ from $50$ to $60$, beyond which the Hamiltonian simulation quantum walk gives the lowest runtime among the methods considered. Note that at the regime $n \approx 40$ the SK model is still solvable classically by direct enumeration on a classical device. The local proposal is omitted for $\bar\beta\geq4$.

\begin{figure}[h!]
    \centering
    \scalebox{0.6}{\input{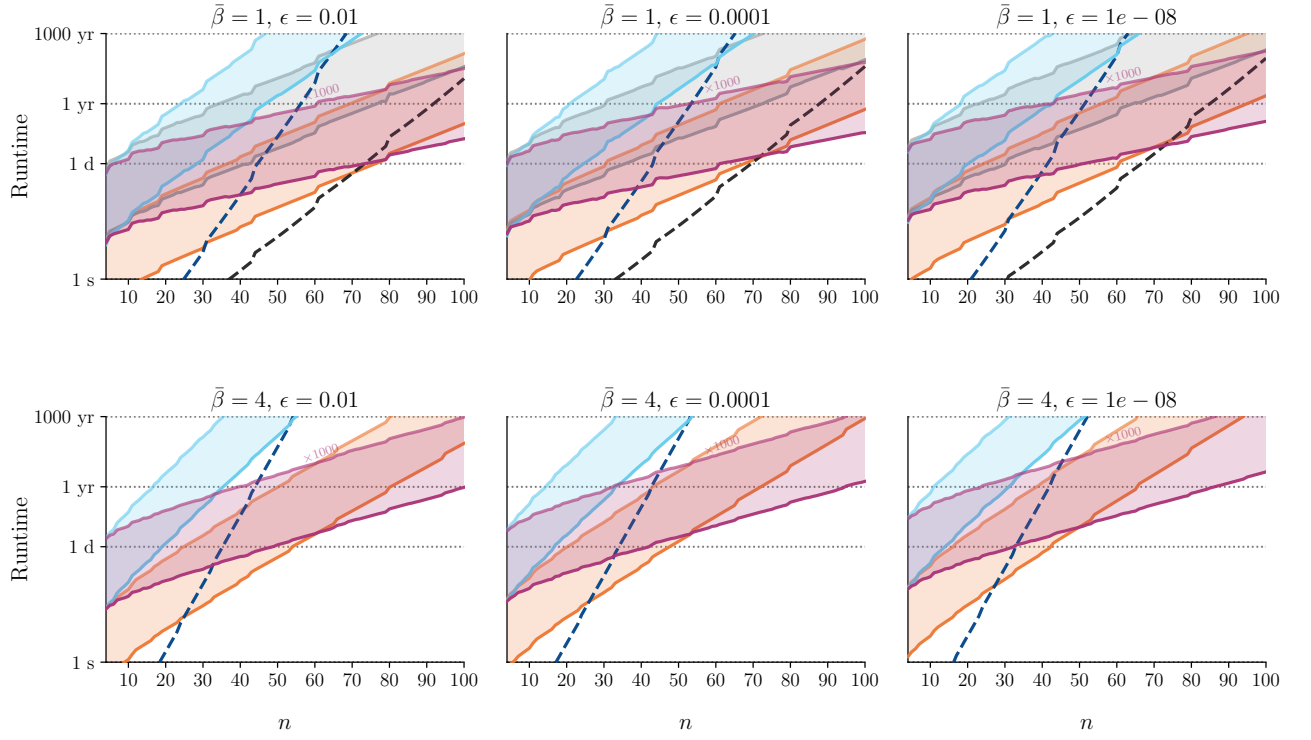}}
    \caption{Same as Figure~\ref{fig:runtime_different_betas}, but for selected $\beta$ and target TV-distance errors $\varepsilon=10^{-2}$, $10^{-4}$, and $10^{-8}$.}
    \label{fig:runtime_different_epsilon}
\end{figure}

Figure~\ref{fig:runtime_different_epsilon} shows the effect of reducing the target TV-distance error from $\varepsilon=10^{-2}$ to $10^{-8}$ at $\bar\beta=1$ and $\bar\beta=4$. A smaller $\varepsilon$ shifts the runtime curves upward, but its effect is weaker than the dependence on $n$ and does not qualitatively change the relative ordering of the methods. In particular, the Hamiltonian simulation proposal retains the most favorable large-$n$ scaling.

The curves are less smooth than those in the single-step plots because the runtime is summed over the full annealing schedule. Whenever the schedule acquires an additional annealing step, the total runtime increases abruptly.

\end{document}